\documentclass[a4paper,11pt]{article}
\usepackage{jheppub}
\usepackage{amsmath}
\usepackage{soul}
\usepackage{color}
\usepackage{float}
\usepackage{dsfont}
\usepackage{tikz}
\usetikzlibrary{arrows.meta, positioning}
\usepackage{amssymb}
\usepackage{physics}
\usepackage{hyperref}
\usepackage{bm,graphicx,graphics,epsfig}
\usepackage{dcolumn}
\usepackage{lineno}
\usepackage{subfigure} 
\usepackage{bbm}
\usetikzlibrary{calc}
\usetikzlibrary{shapes}
\usetikzlibrary{calc,decorations.markings,decorations.pathmorphing}
\tikzset{cross/.style={cross out, draw=black, minimum size=2*(#1-\pgflinewidth), inner sep=0pt, outer sep=0pt}, cross/.default={1pt}}
\usepackage{lmodern}
\usepackage{pgfplots}
\usepackage{tikz}
\pgfplotsset{compat=newest}
\usetikzlibrary{arrows.meta, positioning, decorations.pathreplacing, calligraphy, backgrounds, 3d}

\tikzset{
    axis_style/.style={-Latex, thick},
    grid_style/.style={gray!50, thin},
    surface_style/.style={blue!80!black, thick},
    boundary_curve/.style={red!80!black, very thick},
    boundary_node/.style={draw=red!80!black, fill=red!60, circle, inner sep=2pt, thick},
    internal_node/.style={draw=blue!80!black, fill=blue!60, circle, inner sep=1.5pt, thick},
    integration_line/.style={gray, dashed},
    annotation_text/.style={font=\small},
    info_box/.style={
        draw=gray, 
        rounded corners, 
        inner sep=8pt, 
        text width=12cm, 
        align=center
    }
}

\makeatletter
\gdef\@fpheader{} 
\makeatother

\newcommand{\be}{\begin{equation}}
\newcommand{\ee}{\end{equation}}
\newcommand{\benn}{\begin{equation*}}
\newcommand{\eenn}{\end{equation*}}
\newcommand{\bea}{\begin{eqnarray}}
\newcommand{\eea}{\end{eqnarray}}

\newcommand{\mA}{\mathcal{A}}

\newcommand{\mD}{\mathcal{D}}
\newcommand{\mO}{\mathcal{O}}

\DeclareMathOperator{\im}{Im}

\DeclareMathOperator{\sgn}{sgn}

\title{Constrained Symplectic Quantization II:\\ The Free Scalar Field}

\author[a,b]{Francesco Scardino}
\author[c,d]{Martina Giachello}
\author[e]{Giacomo Gradenigo}

\affiliation[a]{Physics Department, INFN Roma1, Piazzale A. Moro 2, Roma, I-00185, Italy}
\affiliation[b]{Physics Department, Sapienza University, Piazzale A. Moro 2, Roma, I-00185, Italy}
\affiliation[c]{Gran Sasso Science Institute, Viale F. Crispi 7, 67100 L'Aquila, Italy}
\affiliation[d]{INFN-Laboratori Nazionali del Gran Sasso, Via G. Acitelli 22, 67100 Assergi (AQ), Italy}
\affiliation[e]{Physics and Astronomy Department "Galileo Galilei", Universit\`a di Padova, Via Marzolo 8, 35131 Padova, Italy}

\emailAdd{francesco.scardino@uniroma1.it}
\emailAdd{martina.giachello@gssi.it}
\emailAdd{giacomo.gradenigo@unipd.it}

\abstract{
Constrained symplectic quantization is a functional formulation of quantum field theory in which quantum fluctuations are sampled through a deterministic Hamiltonian flow in an auxiliary intrinsic time $\tau$. In this paper we extend the quantum-mechanical framework introduced in \cite{Giachello_Harmonic} to a relativistic scalar quantum field theory in Minkowski space-time. The construction is based on the analytic continuation of fields and action from $\mathbb{R}$ to $\mathbb{C}$ together with constraints that select stable intrinsic-time trajectories and, at the same time, define convergent integration cycles for the corresponding microcanonical functional. We show that, in the continuum limit, the microcanonical generating functional reproduces the Feynman generating functional. For the free scalar field in $1+1$ dimensions we derive the constrained equations of motion, implement the resulting dynamics numerically, and verify real-time two-point correlators, equal-time commutator relations, and Dyson--Schwinger equations including the expected contact terms.
}
\begin{document}
\maketitle
\flushbottom
\newpage

\section{Introduction}
\label{sec:intro}

This paper is the second installment of the {\it constrained symplectic quantization} program. Symplectic quantization (SQ) is a functional formulation of quantum field theory proposed in~\cite{Gradenigo_1,Gradenigo_2,Gradenigo_3} and supported numerically in~\cite{Giachello_JHEP,Giachello_Lattice2024,Giachello_trani}. In the first paper of this series~\cite{Giachello_Harmonic} (CSQ~I), we introduced the constrained SQ framework, proved its equivalence with the Feynman path integral in the continuum limit, and validated it in the $0+1$-dimensional case of the quantum harmonic oscillator.\par
In the present work we extend the same framework of~~\cite{Giachello_Harmonic}to a relativistic quantum field theory, focusing on a real scalar field $\hat\varphi(x)$ in Minkowski space-time. Our goal is to show that constrained SQ provides a stable and practical framework for sampling real-time correlators directly in Minkowski signature.\par
The starting point of SQ is an extended variable space in which quantum fluctuations are parametrized by an additional intrinsic time $\tau$~\cite{Gradenigo_1,Gradenigo_2,Gradenigo_3}. In this formulation the operator $\hat\varphi(x)$ is replaced by a $\tau$-dependent field $\varphi(x,\tau)$ together with a conjugate momentum field $\pi(x,\tau)$. The intrinsic-time evolution is generated by a generalized Hamiltonian $\mathbb{H}_{\rm SQ}$. Under a weak ergodicity assumption,  standard in microcanonical approaches, long-$\tau$ averages are identified with averages in a microcanonical ensemble associated with the conservation of $\mathbb{H}_{\rm SQ}$~\cite{Gradenigo_1,Gradenigo_2,Giachello_JHEP,deAlfaro:1983zz,Strominger:1982xu,Iwazaki:1984kx,Callaway:1983,Duane:1987de}.\par
The realization of SQ previously discussed in~\cite{Giachello_JHEP,Giachello_Lattice2024,Giachello_trani} suffered from a major limitation: it was proved~\cite{Giachello_Harmonic} that in the large-number-of-degrees-of-freedom limit quantum fluctuations had a weight proportional to $\exp\{S/\hbar\}$ rather than to the oscillatory Feynman weight $\exp\{iS/\hbar\}$, thus yielding an ill-defined theory in absence of interactions, due to the non-positivity of the free relativistic action. The key step introduced in CSQ~I is to analytically continue \emph{all} the real fields (and all real components of fields which are already complex in the original formulation of the theory) and the action from $\mathbb{R}$ to $\mathbb{C}$ while keeping the generalized Hamiltonian real and imposing suitable constraints on the intrinsic-time flow. This yields both a stable deterministic dynamics and a direct correspondence with the Feynman path integral measure.\par
The logic of constrained SQ is closely related to modern contour-deformation strategies for the sign problem~\cite{Gattringer:2016kco,Troyer:2004ge,Loh:1990zz,Alexandru:2020wrj,Cristoforetti:2012su,Scorzato:2015qts,Behtash:2015loa,Witten:2010cx}. One complexifies the field space and restricts the dynamics to a suitable stable manifold. The conceptual difference from Lefschetz-thimble methods is that the constraint is imposed not only on the integration paths for fields in the  functional integral, but at a more fundamental level already on the microscopic Hamiltonian flow in intrinsic time which generates the measure. In this paper we show that, for the free scalar field in Minkowski space-time, this construction reproduces the expected real-time correlators and their causal structure. In particular, we compute the two-point function, verify the equal-time commutator structure, and recover the Dyson--Schwinger equations including the correct contact terms.\par
Beyond the numerical results, the main conceptual point is that constrained SQ reconstructs the functional framework of quantum field theory directly from deterministic $\tau$-evolution. The most important theoretical result presented here is the proof that the CSQ microcanonical generating functional is equivalent, in the continuum limit, to the Feynman generating functional for a generic \emph{interacting} scalar quantum field theory, provided it is renormalizable. We also prove, still in full generality for an interacting theory, that the intrinsic-time Hamiltonian flow encodes the quantum equations of motion and, more generally, the full Dyson--Schwinger hierarchy. Numerical validation of the theoretical results is then show for a free scalar field theory because, as it will be clarified through the whole paper, while equivalence between constrained SQ and Feynman path-integral can be demonstrated in general, the particular choice of constraints to guarantee the functional integral convergence and the stability of the Hamiltonian dynamics is model dependent, so that it must be engineered for specific theories. We started from the free one, which is the more tractable, still providing the evidence of several non-trivial results.\\

The paper is organized as follows. In Sec.~\ref{sec:formulation} we set up the complexified field space for the scalar theory and introduce the generalized Hamiltonian $\mathbb{H}_{\rm SQ}$. In Sec.~\ref{sec:equivalence} we prove the equivalence between the corresponding microcanonical generating functional and the Feynman path integral in $d+1$ dimensions, emphasizing the points that are specific to a relativistic field theory and closely following the logic of CSQ~I~\cite{Giachello_Harmonic}. In Sec.~\ref{sec:ds} we derive the SQ version of the quantum equations of motion and the Dyson-Schwinger relations from the intrinsic-time Hamilton equations, including the appearance of the contact terms. In Sec.~\ref{sec:SF-dynamics} we specialize to the free scalar field and identify the class of constrained, stable intrinsic-time flows, making explicit their interpretation as a constructive choice of integration cycles compatible with the Feynman prescription. In Sec.~\ref{sec:numerical-algorithm} we discuss the numerical symplectic integrator and the practical implementation of the constraints, and in Sec.~\ref{sec:numericalresults} we present numerical results
for the free scalar field, including both periodic-boundary measurements and the mixed temporal (initial-value) setup of Sec.~\ref{sec:initial-value-dynamics}.

\section{Fields and Action: analytic continuation from $\mathbb{R}$ to $\mathbb{C}$}
\label{sec:formulation}
We briefly recall the constrained symplectic quantization setup introduced in~\cite{Giachello_Harmonic} and adapt the notation to a real scalar field theory in $d+1$ dimensions. The central requirement is to construct a generalized Hamiltonian functional $\mathbb{H}_{\rm SQ}$ which is {\it real} and can be fixed to the value $\mathbb{H}_{\rm SQ}=\mA=\hbar\,M$, on a $d+1$ lattice with $M$ degrees of freedom  while at the same time generating, in the continuum limit, the same correlation functions of the Feynman path integral.\par
The need for an analytic continuation to $\mathbb{C}$ and for a modified generalized Hamiltonian is already apparent from the ``unconstrained'' ansatz of~\cite{Giachello_JHEP,Giachello_Lattice2024,Giachello_trani} based on
\begin{equation}
  \mathbb{H}[\varphi,\pi]=\mathbb{K}[\pi]-S[\varphi]\,,
\end{equation}
where $S[\varphi]$ is the Minkowskian action and $\mathbb{K}$ is a positive quadratic form in the intrinsic-time momenta. As shown there~\cite{Giachello_JHEP,Giachello_Lattice2024,Giachello_trani}, fixing $\mathbb{H}$ to a real value produces in the large number-of-degrees-of-freedom limit a real canonical weight proportional to $\exp\{S/\hbar\}$ rather than the Feynman weight $\exp\{iS/\hbar\}$, and moreover it leads to instabilities in the free theory because the Minkowskian action is not positive definite.\par
Constrained SQ resolves both issues by complexifying the field space while keeping the generalized Hamiltonian real. Concretely, we promote the operator $\hat\varphi(x)$ to a complexified intrinsic-time field
\begin{align}
  \varphi(x,\tau)\in\mathbb{R}\quad \longrightarrow\quad
  \varphi_R(x,\tau)+ i\,\varphi_I(x,\tau)\in\mathbb{C}\,,
\end{align}
together with a complex intrinsic-time momentum field
\begin{align}
  \pi(x,\tau)\in\mathbb{R}\quad \longrightarrow\quad
  \pi_R(x,\tau)+ i\,\pi_I(x,\tau)\in\mathbb{C}\,.
\end{align}
Incidentally, let us notice that if the original field was already a complex one, as it happens for a theory which contains both particles and antiparticles, consider for instance
\begin{align}
\Psi(x,\tau) = \varphi(x,\tau) + i ~\vartheta(x,\tau),\quad\quad\quad \varphi(x,\tau),\vartheta(x,\tau)~\in~\mathbb{R}
\end{align}
one should consider the analytic continuation from $\mathbb{R}$ to $\mathbb{C}$ of both the real and imaginary part of the \emph{physically} complex field, namely:
\begin{align}
  \varphi(x,\tau)\in\mathbb{R}\quad &\longrightarrow\quad
  \varphi_R(x,\tau)+ i\,\varphi_I(x,\tau)\in\mathbb{C}\,,\nonumber \\
  \vartheta(x,\tau)\in\mathbb{R}\quad &\longrightarrow\quad
  \vartheta_R(x,\tau)+ i\,\vartheta_I(x,\tau)\in\mathbb{C}.
\end{align}
Since there are several important issues to be touched and clarified, with the presentation of corresponding proofs and numerical results, already in the case of a \emph{real} scalar field theory, for the sake of clarity we will discuss elsewhere the case of fields with antiparticles, focusing now on a purely real field.
The analytic continuation extends to the action, so that $S[\varphi]\in\mathbb{C}$, and we treat $\varphi$ and $\bar\varphi$ (and similarly $\pi$ and $\bar\pi$) as independent variables to be integrated along suitable one-dimensional contours in the complex plane at each space-time point.\par
We then introduce the generalized Hamiltonian of constrained symplectic quantization as
\begin{equation}
  \mathbb{H}_{\rm SQ}[\pi,\bar\pi,\varphi,\bar\varphi]
  =
  \int d^{d+1}x\,\bar\pi(x,\tau)\pi(x,\tau)
  +2\,\Im S[\varphi,\bar\varphi]\,,
 \label{eq:csq-hamiltonian}
\end{equation}
where by a slight abuse of notation we have denoted
\begin{equation}
  \Im S[\varphi,\bar\varphi]\equiv \frac{S[\varphi]-\bar S[\bar\varphi]}{2i}\,,
\end{equation}
indeed the fields $\varphi$ and $\bar{\varphi}$ are independent off-shell.
Equation~\eqref{eq:csq-hamiltonian} is the direct field-theory analogue of the ansatz proposed in CSQ~I~\cite{Giachello_Harmonic,Giachello_lattice2025}. Its correctness is justified by the fact that it yields, in the continuum limit and after imposing the appropriate intrinsic-time constraints, a microcanonical generating functional equivalent to the Feynman path integral. The intrinsic-time evolution is generated by Hamilton equations on the complexified phase space,
\begin{align}\label{eq:sq-Ham}
  \partial_\tau \varphi(x,\tau)&=\frac{\delta \mathbb{H}_{\rm SQ}}{\delta \pi(x,\tau)}\,,
  \qquad
  \partial_\tau \pi(x,\tau)=-\frac{\delta \mathbb{H}_{\rm SQ}}{\delta \varphi(x,\tau)}\,, \nonumber \\  \nonumber \\
  \partial_\tau \overline{\varphi}(x,\tau)&=\frac{\delta \mathbb{H}_{\rm SQ}}{\delta \overline{\pi}(x,\tau)}\,,
  \qquad
  \partial_\tau \overline{\pi}(x,\tau)=-\frac{\delta \mathbb{H}_{\rm SQ}}{\delta \overline{\varphi}(x,\tau)}\,,
\end{align}
The additional ingredient of constrained SQ is that this Hamiltonian flow is restricted by constraints that guarantee stability and select the relevant integration paths in the complexified field space. We discuss these constraints explicitly for the free scalar field in Sec.~\ref{sec:SF-dynamics}.

On the basis of a weak ergodicity assumption, which, as thoroughly explained on a recent review on the foundations of statistical mechanics~\cite{Baldovin:2024dpo}, simply amount to the equivalence between dynamical and statistical averages for "generic enough" observables for "generic enough" initial data of the constrained Hamiltonian flow, it is natural to identify long-$\tau$ averages with averages in the corresponding microcanonical ensemble~\cite{Gradenigo_1,Gradenigo_2,Giachello_Harmonic}. This leads to the microcanonical density
\begin{align}\label{eq:micorcanonical-emasure}
\rho_{\text{micro}}[\varphi,\bar{\varphi},\pi,\bar{\pi}]
=
\frac{1}{\Omega[\mA,0]}\,
\delta\!\left(\mA-\mathbb{H}_{\rm SQ}[\varphi,\bar{\varphi},\pi,\bar{\pi}]\right),
\end{align}
with partition function
\begin{align}
\Omega[\mA,0] =
\int_{\boldsymbol{\Gamma}(x)} \mathcal{D}\bar{\pi}\,\mathcal{D}\pi\,\mathcal{D}\bar{\varphi}\,\mathcal{D}\varphi\;
\delta\!\Big(\mA - \mathbb{H}_{\rm SQ}[\pi,\bar{\pi},\varphi,\bar{\varphi}] \Big)\,,
\label{eq:micro-part-func-qft0}
\end{align}
where we call $\mA$ the \textit{generalized action}.\par
For a generic holomorphic observable $\mO[\varphi]$, the ergodicity assumption states that, for generic initial conditions, the long-$\tau$ dynamical average coincides with the corresponding microcanonical average:
\begin{align}
\lim_{\tau \rightarrow \infty} \frac{1}{\tau}\int_{0}^{\tau} d\tau'\,\mO[\varphi(x,\tau')]
=
\int \mD\varphi\,\mD\bar \varphi\,\mD\pi\,\mD\bar\pi\;
\rho_{\text{micro}}[\varphi,\bar{\varphi},\pi,\bar{\pi}]\,\mO[\varphi(x)]\,.
\label{eq:mild-ergodicity}
\end{align}
\section{Microcanonical action and the Feynman Path Integral in $d+1$ dimensions}
\label{sec:equivalence}
In this section we generalize to relativistic quantum field
theory the equivalence proven in the $0+1$-dimensional case in Ref.~\cite{Giachello_Harmonic}.  Our goal
is to show that, once the generalized Hamiltonian of constrained
symplectic quantization (CSQ) is fixed to $\mA=\hbar M$, the
microcanonical generating functional constructed from the conservation
of $\mathbb{H}_{\rm SQ}$ reproduces, in the continuum limit, the
standard Feynman generating functional of a scalar field in Minkowski
space-time. More precisely, introducing an external source $J(x)$, we
prove that
\begin{align}
  \lim_{M\rightarrow\infty}\,
  \frac{1}{\Omega[\hbar M,0]}\,
  \frac{\delta^n\Omega[\hbar M,J]}{\delta J(x_{1})\ldots\delta J(x_{n})}\Bigg|_{J=0}
  \;=\;
  \big\langle \varphi(x_1)\ldots \varphi(x_n)\big\rangle
  \;=\;
  \frac{1}{\mathcal{Z}[\hbar,0]}\,
  \frac{\delta^n\mathcal{Z}[\hbar,J]}{\delta J(x_{1})\ldots\delta J(x_{n})}\Bigg|_{J=0}\,,
  \label{eq:equivalence-sq-fey-qft}
\end{align}
where $\mathcal{Z}[\hbar,J]$ denotes the Feynman path integral of the
holomorphic theory defined by the analytically continued Minkowskian
action $S[\varphi]$. Let us stress that, despite the numerical algorithm and results which we will present in Sec.~\ref{sec:numerical-algorithm} and Sec.~\ref{sec:numericalresults} will be specific of a free field theory,  the following proof is drawn in full generality for a generic field theory with a renormalizable interaction.

We start by defining the microcanonical generating functional of CSQ in
$d+1$ dimensions as
\begin{align}
\Omega[\mA,J] =
\frac{1}{\Omega[\mA,0]}
\int_{\boldsymbol{\Gamma}(x)} \mathcal{D}\bar{\pi}\,\mathcal{D}\pi\,\mathcal{D}\bar{\varphi}\,\mathcal{D}\varphi\;
\delta\!\Big(\mA - \mathbb{H}_{\rm SQ}[\pi,\bar{\pi},\varphi,\bar{\varphi}] + i\, J\!\cdot\!\varphi\Big)\,,
\label{eq:micro-part-func-qft}
\end{align}
where $x\equiv (x_0,\mathbf{x})$ is a Minkowskian space-time point and
\begin{align}
  J\!\cdot\!\varphi \equiv \int d^{d+1}x\, J(x)\,\varphi(x)\,,\qquad
  \bar{\pi}\!\cdot\!\pi \equiv \int d^{d+1}x\,\bar{\pi}(x)\,\pi(x)\,.
\end{align}
The generalized Hamiltonian is the one introduced in
Eq.~\eqref{eq:csq-hamiltonian},
\begin{align}
  \mathbb{H}_{\rm SQ}[\pi,\bar{\pi},\varphi,\bar{\varphi}] =
  \bar{\pi}\!\cdot\!\pi + 2\,\Im S[\varphi,\bar{\varphi}]\,,\qquad
  \Im S[\varphi,\bar{\varphi}] = \frac{S[\varphi]-\bar{S}[\bar{\varphi}]}{2i}\,.
  \label{eq:hsq-qft}
\end{align}
As in Ref.~\cite{Giachello_Harmonic}, $\boldsymbol{\Gamma}(x)$ denotes
the set of one-dimensional integration contours in the complex plane
parametrized by the space-time point $x$, one contour for each of the
fields to be integrated, $\varphi(x)$, $\bar{\varphi}(x)$, $\pi(x)$ and
$\bar{\pi}(x)$. We write
\begin{align}
\boldsymbol{\Gamma}(x) = \Gamma_\varphi(x)\cup \Gamma_{\bar{\varphi}}(x)\cup \Gamma_\pi(x)\cup \Gamma_{\bar{\pi}}(x)\,,
\label{eq:contours-qft}
\end{align}
so that, for instance,
\begin{align}
\Gamma_\varphi(x)=\left\lbrace \gamma_\varphi(x)\in\mathbb{C}: x\in\mathcal{V}_{d+1}\right\rbrace,
\end{align}
with $\mathcal{V}_{d+1}$ a finite space-time volume. Let us stress again
a crucial point, namely that the analytic continuation from $\mathbb{R}$ to
$\mathbb{C}$ does not imply any doubling of integration degrees of
freedom. The integration domain for each field at each space-time point
remains a one-dimensional path and so the only difference is that this path is
embedded in $\mathbb{C}$ rather than being the real line. The additional
variables in Eq.~\eqref{eq:micro-part-func-qft} are the intrinsic-time
momenta $\pi,\bar{\pi}$, which will be integrated out explicitly.

We discretize a finite Minkowskian space-time region $\mathcal{V}_{d+1}$
on a hypercubic lattice with spacing $a$.\footnote{For notational
simplicity we assume an isotropic lattice spacing. The generalization to
anisotropic lattices is immediate and does not affect the large-$M$
analysis.}  Denoting by $x_i$, $i=1,\ldots,M$, the lattice sites, we
have $M = \mathrm{Vol}(\mathcal{V}_{d+1})/a^{d+1}$ degrees of freedom for
each field. In complete analogy with the discussion in
Ref.~\cite{Strominger:1982xu} (see also Ref.~\cite{Giachello_Harmonic} in
the $0+1$ case), one may equivalently represent the discretized field in
a generic orthonormal basis $\{f_n(x_i)\}$, for instance Fourier modes
with a UV cutoff $\Lambda\sim \pi/a$:
\begin{equation}
\varphi(x_i)=\sum_{n=1}^M f_n(x_i)\,c_n\,,\qquad
\bar{\varphi}(x_i)=\sum_{n=1}^M \bar{f}_n(x_i)\,\bar{c}_n\,,
\end{equation}
with orthonormality
\begin{equation}
\sum_{i=1}^M f_n(x_i)\, f_m(x_i)=\delta_{nm}\,,\qquad
\sum_{i=1}^M \bar{f}_n(x_i)\,\bar{f}_m(x_i)=\delta_{nm}\,.
\end{equation}
The functional measure is then understood as the continuum limit of the
finite-dimensional measure
\begin{equation}
\int \mathcal{D}\varphi\,\mathcal{D}\bar{\varphi}
\;\approx\;
\int \mathcal{D}_M\varphi\,\mathcal{D}_M\bar{\varphi}
\;\equiv\;
\prod_{n=1}^M \left(\int_{\Gamma_n} dc_n\, d\bar{c}_n\right)
\;\equiv\;
\prod_{i=1}^M \left(\int_{\Gamma_i} d\varphi(x_i)\, d\bar{\varphi}(x_i)\right)\,,
\end{equation}
and analogously for $\pi,\bar{\pi}$. As in the standard convention, the
measure does not explicitly contain factors of $\hbar$; the latter enters
through the quantization constraint $\mA=\hbar M$ discussed below.

Finally, the relation between $M$ and a UV cutoff can be made explicit.
On a finite $(d+1)$-volume with temporal extent $T$ and spatial volume
$V_d$, one has parametrically
\begin{equation}\label{eq:M-cutoff-qft}
  M \;=\; \frac{T\,V_d}{a^{d+1}}
  \;\sim\; T\,V_d\left(\frac{\Lambda}{\pi}\right)^{d+1}\,,
\end{equation}
so that the large-$M$ limit can be approached by sending $\Lambda\to\infty$
(continuum limit at fixed physical volume) and/or by increasing the
physical volume at fixed cutoff (thermodynamic limit). In what follows,
the large-$M$ limit is understood in the continuum sense, unless stated
otherwise.

We now perform explicitly the integration over the intrinsic-time
momenta $\pi$ and $\bar{\pi}$. As in Ref.~\cite{Giachello_Harmonic}, it is
convenient to represent the delta function as a Fourier integral:
\begin{align}
\Omega[\mA,J]
=
\frac{1}{\Omega[\mA,0]}
\int \mathcal{D}_M\bar{\pi}\,\mathcal{D}_M\pi\,
\mathcal{D}_M\bar{\varphi}\,\mathcal{D}_M\varphi\,
\frac{d\lambda}{2\pi}\;
\exp\!\Big[
-i\lambda\,\bar{\pi}\!\cdot\!\pi
+i\lambda\big(\mA-2\,\Im S[\varphi,\bar{\varphi}]+ i\,J\!\cdot\!\varphi\big)
\Big]\,.
\label{eq:lambda-rep-qft}
\end{align}
The $\pi$-integral is a complex Gaussian integral, factorized over the
$M$ lattice degrees of freedom, and yields (up to multiplicative
constants independent of $J$ and $\varphi$, which cancel in the
normalization)
\begin{align}
\Omega[\mA,J]
=
\frac{1}{\Omega[\mA,0]}
\int \mathcal{D}_M\bar{\varphi}\,\mathcal{D}_M\varphi\,
\mathcal{D}\lambda\;
\lambda^{-M}\;
\exp\!\Big[
i\lambda\big(\mA-2\,\Im S[\varphi,\bar{\varphi}]+ i\,J\!\cdot\!\varphi\big)
\Big]\,.
\label{eq:after-pi-qft}
\end{align}
The $\lambda$ integral can then be performed explicitly, again up to
irrelevant constants. One obtains
\begin{align}
\Omega[\mA,J]
=
\frac{1}{\Omega[\mA,0]}
\int \mathcal{D}_M\bar{\varphi}\,\mathcal{D}_M\varphi\;
\Big(\mA-2\,\Im S[\varphi,\bar{\varphi}]+ i\,J\!\cdot\!\varphi\Big)^{M-1}\,.
\label{eq:Omega-power-qft}
\end{align}
At this point we fix the generalized Hamiltonian to the quantization
condition
\begin{align}
  \mA=\hbar M\,,
  \label{eq:quant-constraint-qft}
\end{align}
which assigns a unit of action $\hbar$ to each degree of freedom. By
plugging Eq.~\eqref{eq:quant-constraint-qft} into
Eq.~\eqref{eq:Omega-power-qft} we get
\begin{align}
\Omega[\hbar M,J]
&=
\frac{1}{\Omega[\hbar M,0]}
\int \mathcal{D}_M\bar{\varphi}\,\mathcal{D}_M\varphi\;
(\hbar M)^{M-1}
\left(
1-\frac{2}{\hbar M}\Im S[\varphi,\bar{\varphi}]
+\frac{i}{\hbar M}J\!\cdot\!\varphi
\right)^{M-1}\nonumber\\
&=
\frac{1}{\Omega[\hbar M,0]}
\int \mathcal{D}_M\bar{\varphi}\,\mathcal{D}_M\varphi\;
\exp\!\left[
(M-1)\ln\!\left(
1-\frac{2}{\hbar M}\Im S[\varphi,\bar{\varphi}]
+\frac{i}{\hbar M}J\!\cdot\!\varphi
\right)\right]\,,
\label{eq:Omega-log-qft}
\end{align}
where the overall factor $(\hbar M)^{M-1}$ cancels in the normalized
ratio and will be henceforth omitted.
We now expand Eq.~\eqref{eq:Omega-log-qft} in powers of the source $J$
to recover the standard source expansion of a generating functional. We
write
\begin{equation}
\left(
1-\frac{2}{\hbar M}\Im S[\varphi,\bar{\varphi}]
+\frac{i}{\hbar M}J\!\cdot\!\varphi
\right)^{M-1}
=
\sum_{n=0}^{\infty}\binom{M-1}{n}
\left(\frac{i}{\hbar M}J\!\cdot\!\varphi\right)^n
\left(
1-\frac{2}{\hbar M}\Im S[\varphi,\bar{\varphi}]
\right)^{M-1-n}\!,
\end{equation}
and we use
\begin{equation}
(J\!\cdot\!\varphi)^n
=
\int d^{d+1}x_1\cdots d^{d+1}x_n\,
J(x_1)\cdots J(x_n)\,
\varphi(x_1)\cdots \varphi(x_n)\,.
\end{equation}
This yields
\begin{align}
\Omega[\hbar M,J]
=&
\frac{1}{\Omega[\hbar M,0]}
\sum_{n=0}^{\infty}\frac{1}{n!}
\left(\frac{i}{\hbar M}\right)^n
\frac{\Gamma(M)}{\Gamma(M-n)}
\nonumber\\
&\int d^{d+1}x_1\cdots d^{d+1}x_n\,
J(x_1)\cdots J(x_n)\,
\Big\langle \varphi(x_1)\cdots \varphi(x_n)\Big\rangle_{M-n}\,,
\label{eq:Omega-series-qft}
\end{align}
where $\binom{M-1}{n}=\Gamma(M)/\big[n!\,\Gamma(M-n)\big]$, and where we
introduced the ``dressed'' correlator
\begin{align}
\Big\langle \varphi(x_1)\cdots \varphi(x_n)\Big\rangle_{M-n}
=
\int \mathcal{D}_M\bar{\varphi}\,\mathcal{D}_M\varphi\;
\varphi(x_1)\cdots \varphi(x_n)\,
\left(
1-\frac{2}{\hbar M}\Im S[\varphi,\bar{\varphi}]
\right)^{M-n}\,,
\label{eq:dressed-corr-qft}
\end{align}
where, as in Ref.~\cite{Giachello_Harmonic}, we neglected the harmless
shift by $-1$ in the exponent since we are interested in the large-$M$
limit.

Our next goal is to evaluate
Eq.~\eqref{eq:dressed-corr-qft} in the continuum (large-$M$) limit and substitute the
result back into Eq.~\eqref{eq:Omega-series-qft}. We proceed exactly as
in Ref.~\cite{Giachello_Harmonic}, expanding in inverse powers of $M$.
Let
\begin{equation}
x \equiv \frac{2}{\hbar}\,\Im S[\varphi,\bar{\varphi}]\,.
\end{equation}
Then
\begin{equation}
\left(1-\frac{2}{\hbar M}\Im S[\varphi,\bar{\varphi}]\right)^{M-n}
=
\left(1-\frac{x}{M}\right)^{M-n}\,.
\end{equation}
We now factor out the exponential term suggested by the notable limit
$\lim_{M\to\infty}(1-\frac{x}{M})^{M-n}=e^{-x}$:
\begin{equation}
\left(1-\frac{x}{M}\right)^{M-n}
=
e^{-x}\times
\left(1-\frac{x}{M}\right)^{M-n}e^{x}\,,
\end{equation}
and expand the second factor by convoluting the two series
\begin{align}
\left(1-\frac{x}{M}\right)^{M-n}
&=
\sum_{k=0}^{\infty}\binom{M-n}{k}\left(\frac{-x}{M}\right)^k\,,
\qquad
e^x=\sum_{\ell=0}^{\infty}\frac{x^\ell}{\ell!}\,.
\end{align}
Using the Cauchy product one obtains
\begin{equation}
\left(1-\frac{x}{M}\right)^{M-n}e^{x}
=
\sum_{j=0}^{\infty}x^j
\sum_{k=0}^{j}
\binom{M-n}{k}\,
\frac{1}{(j-k)!}\,
\frac{(-1)^k}{M^k}\,.
\end{equation}
Since $x=\frac{2}{\hbar}\Im S[\varphi,\bar{\varphi}]$, we define the
coefficients $c_j(M,n)$ as
\begin{equation}\label{eq:coeff-qft}
c_j(M,n)
=
\left(\frac{2}{\hbar}\right)^j
\sum_{k=0}^{j}
\binom{M-n}{k}\,
\frac{1}{(j-k)!}\,
\frac{(-1)^k}{M^k}\,,
\end{equation}
so that
\begin{equation}
\left(
1-\frac{2}{\hbar M}\Im S[\varphi,\bar{\varphi}]
\right)^{M-n}
=
e^{-\frac{2}{\hbar}\Im S[\varphi,\bar{\varphi}]}
\sum_{j=0}^{\infty}
c_j(M,n)\,
\big(\Im S[\varphi,\bar{\varphi}]\big)^j\,.
\label{eq:weight-expanded-qft}
\end{equation}
By inserting Eq.~\eqref{eq:weight-expanded-qft} into
Eq.~\eqref{eq:dressed-corr-qft} we find
\begin{align}
\Big\langle \varphi(x_1)\cdots \varphi(x_n)\Big\rangle_{M-n}
&=
\sum_{j=0}^{\infty}c_j(M,n)
\int \mathcal{D}_M\bar{\varphi}\,\mathcal{D}_M\varphi\;
\varphi(x_1)\cdots \varphi(x_n)\,
e^{-\frac{2}{\hbar}\Im S[\varphi,\bar{\varphi}]}\,
\big(\Im S[\varphi,\bar{\varphi}]\big)^j
\nonumber\\
&=
\sum_{j=0}^{\infty}c_j(M,n)
\int \mathcal{D}_M\bar{\varphi}\,\mathcal{D}_M\varphi\;
\varphi(x_1)\cdots \varphi(x_n)\,
e^{\frac{i}{\hbar}S[\varphi]}\,e^{-\frac{i}{\hbar}\bar{S}[\bar{\varphi}]}\,
\left(\frac{S[\varphi]-\bar{S}[\bar{\varphi}]}{2i}\right)^j.
\label{eq:corr-expanded-qft}
\end{align}
As in Ref.~\cite{Giachello_Harmonic} (see Appendix A there),
one finds that for fixed $n$ the coefficients satisfy the large-$M$
behaviour
\begin{align}
c_j(M,n)\sim \frac{1}{M}\qquad \forall\,j\ge 1\,,
\label{eq:cj-asympt-qft}
\end{align}
while $c_0(M,n)\to 1$.
At this point it is convenient, as in the Lefschetz-thimble discussion
of Refs.~\cite{Witten:2010cx,Blau:2016vfc}, to treat $\varphi$ and
$\bar{\varphi}$ as independent variables integrated along independent
contours $\Gamma_\varphi$ and $\Gamma'_{\bar{\varphi}}$.  We can then
use the binomial theorem
\begin{align}
\left(\frac{S[\varphi]-\bar{S}[\bar{\varphi}]}{2i}\right)^j
=
\frac{1}{(2i)^j}\sum_{k=0}^{j}\binom{j}{k}\,
S[\varphi]^{\,j-k}\,\bar{S}[\bar{\varphi}]^{\,k}\,,
\end{align}
which allows us to factorize the holomorphic and anti-holomorphic
functional integrals in Eq.~\eqref{eq:corr-expanded-qft}:
\begin{align}
\Big\langle \varphi(x_1)\cdots \varphi(x_n)\Big\rangle_{M-n}
=&
\sum_{j=0}^{\infty}c_j(M,n)\frac{1}{(2i)^j}\sum_{k=0}^{j}\binom{j}{k}
\left[
\int_{\Gamma_\varphi}\mathcal{D}_M\varphi\;
\varphi(x_1)\cdots \varphi(x_n)\,
S[\varphi]^{\,j-k}\,e^{\frac{i}{\hbar}S[\varphi]}
\right]
\nonumber\\
&\left[
\int_{\Gamma'_{\bar{\varphi}}}\mathcal{D}_M\bar{\varphi}\;
\bar{S}[\bar{\varphi}]^{\,k}\,e^{-\frac{i}{\hbar}\bar{S}[\bar{\varphi}]}
\right].
\label{eq:dcorr-summation-qft}
\end{align}

We now discuss the large-$M$ limit of Eq.~\eqref{eq:dcorr-summation-qft}.
The crucial point is that, in the continuum limit taken at fixed
space-time volume, the leading term $j=0$ reproduces the Feynman weight,
while the terms $j\ge1$ are suppressed by Eq.~\eqref{eq:cj-asympt-qft}
under standard assumptions on the behavior of action insertions.

In $0+1$ dimensions (quantum mechanics), it is natural to assume that
the holomorphic action $S[\varphi]$ remains finite as $M\to\infty$,
implying $c_j(M,n)S[\varphi]^j\to 0$ for all $j\ge1$ (see
Ref.~\cite{Giachello_Harmonic}). In a genuine field theory, one must
interpret this statement in a renormalized sense: we assume that the
insertions of powers of the {\it renormalized} holomorphic action do not
overcome the $1/M$ suppression of $c_j(M,n)$ in the continuum limit at
fixed physical volume, namely they do not generate extra divergences in a renormalizable theory. Indeed, while the bare
action typically contains UV-divergent contributions, the continuum
limit of correlation functions is formulated in terms of the renormalized
action and renormalized composite operator insertions.
Therefore, all the insertions of powers of the \emph{renormalized}
holomorphic action produced by the expansion in
Eq.~\eqref{eq:dcorr-summation-qft} go smoothly to zero in the continuum
limit, as was also discussed in~\cite{Giachello_JHEP}. Concretely, in the continuum
limit taken at fixed physical space-time volume, insertions of the renormalized action into correlators remain finite, so that
\begin{equation}
  \frac{S[\varphi]}{M}\;\longrightarrow\;0\qquad \text{when}\qquad M\rightarrow\infty\,,
\end{equation}
and therefore the products $c_j(M,n)\,S[\varphi]^j$ are suppressed for
all $j\geq 1$ by the large-$M$ behavior of the coefficients
$c_j(M,n)$. Clearly, the assumption that $S[\varphi]$ remains finite in
the limit $M\rightarrow\infty$ would not equally apply to a
thermodynamic limit, where one expects the renormalized action to be
extensive $S[\varphi]\sim M$ \cite{Giachello_JHEP}.

Under the hypothesis that the theory is renormalizable, all terms
$j\ge1$ vanish in the continuum limit, and Eq.~\eqref{eq:dcorr-summation-qft}
simplifies to
\begin{align}
\Big\langle \varphi(x_1)\cdots \varphi(x_n)\Big\rangle_{M-n}
\cong
\left[
\int_{\Gamma_\varphi}\mathcal{D}_M\varphi\;
\varphi(x_1)\cdots \varphi(x_n)\,e^{\frac{i}{\hbar}S[\varphi]}
\right]
\left[
\int_{\Gamma'_{\bar{\varphi}}}\mathcal{D}_M\bar{\varphi}\;
e^{-\frac{i}{\hbar}\bar{S}[\bar{\varphi}]}
\right]
+O\!\left(\frac{1}{M}\right).
\label{eq:dressed-leading-qft}
\end{align}
Inserting Eq.~\eqref{eq:dressed-leading-qft} into the source expansion
Eq.~\eqref{eq:Omega-series-qft}, we also use the standard limit (valid
for fixed $n$)
\begin{equation}
\lim_{M\to\infty}
\left(\frac{1}{M}\right)^n\frac{\Gamma(M)}{\Gamma(M-n)}
=1\,,
\label{eq:coeffto1-qft}
\end{equation}
and obtain
\begin{align}
\Omega[\hbar M,J]
&\cong
\frac{1}{\Omega_\infty[\hbar,0]}
\sum_{n=0}^{\infty}\frac{1}{n!}
\left(\frac{i}{\hbar}\right)^n
\int d^{d+1}x_1\cdots d^{d+1}x_n\,
J(x_1)\cdots J(x_n)
\nonumber\\
&\hspace{1.6cm}\times
\left[
\int_{\Gamma_\varphi}\mathcal{D}\varphi\;
\varphi(x_1)\cdots \varphi(x_n)\,e^{\frac{i}{\hbar}S[\varphi]}
\right]
\left[
\int_{\Gamma'_{\bar{\varphi}}}\mathcal{D}\bar{\varphi}\;
e^{-\frac{i}{\hbar}\bar{S}[\bar{\varphi}]}
\right],
\label{eq:Omega-asympt-1-qft}
\end{align}
where $\mathcal{D}$ denotes the continuum functional measure. We now
recognize that the series in $n$ resums into the exponential of the
source term:
\begin{align}
\Omega[\hbar M,J]
&\cong
\frac{1}{\Omega_\infty[\hbar,0]}
\left[
\int_{\Gamma_\varphi}\mathcal{D}\varphi\;
\exp\!\left(\frac{i}{\hbar}S[\varphi]+\frac{i}{\hbar}\int d^{d+1}x\,J(x)\varphi(x)\right)
\right]
\left[
\int_{\Gamma'_{\bar{\varphi}}}\mathcal{D}\bar{\varphi}\;
e^{-\frac{i}{\hbar}\bar{S}[\bar{\varphi}]}
\right].
\label{eq:Omega-asympt-2-qft}
\end{align}
Finally, the anti-holomorphic functional integral factorizes both in the
numerator and in the denominator of the normalized definition
Eq.~\eqref{eq:micro-part-func-qft}, hence it cancels out. We thus obtain
the continuum-limit generating functional
\begin{align}
\lim_{M\to\infty}\Omega[\hbar M,J]
=
\Omega_\infty[\hbar,J]
=
\frac{
\int_{\Gamma_\varphi}\mathcal{D}\varphi\;
\exp\!\left(\frac{i}{\hbar}S[\varphi]+\frac{i}{\hbar}\int d^{d+1}x\,J(x)\varphi(x)\right)
}{
\int_{\Gamma_\varphi}\mathcal{D}\varphi\;
\exp\!\left(\frac{i}{\hbar}S[\varphi]\right)
}\,.
\label{eq:Omega-to-Z-qft-final}
\end{align}
By choosing $\Gamma_\varphi$ to implement the $i\epsilon$ prescription
(i.e. the standard Feynman integration cycle), Eq.~\eqref{eq:Omega-to-Z-qft-final}
is precisely the normalized Feynman generating functional
$\mathcal{Z}[\hbar,J]/\mathcal{Z}[\hbar,0]$. This proves
Eq.~\eqref{eq:equivalence-sq-fey-qft}, namely that the microcanonical
correlators generated by constrained symplectic quantization coincide,
in the continuum limit, with the Minkowskian correlators of the Feynman
path integral.

The argument just presented is crucial because it places on firm grounds
the correspondence between the microcanonical partition function
underlying CSQ and the Feynman path integral in $d+1$ dimensions. At the
same time, it is formal in the same sense as in Ref.~\cite{Giachello_Harmonic}:
it does not yet specify how the integration cycles $\Gamma_\varphi$ and
$\Gamma'_{\bar{\varphi}}$ should be selected constructively in the
microcanonical formulation. In constrained symplectic quantization, this
selection is realized dynamically by the underlying constrained
Hamiltonian flow in intrinsic time. We discuss this point in detail in
the following sections.

\section{Symplectic dynamics and the Dyson-Schwinger equations}
\label{sec:ds}
The dynamics of quantum fluctuations is generated by the Hamilton equations
\begin{align}
	\frac{\partial{\varphi}(x,\tau)}{\partial \tau} &= \frac{\delta \mathbb{H}_{\rm SQ}[\varphi,\bar{\varphi},\pi,\bar{\pi}]}{\delta\pi(x,\tau)} =\bar{\pi}(x,\tau) \label{eq:sq-Ham-eq-q} \\ 
	\frac{\partial\pi(x,\tau)}{\partial\tau} &= - \frac{\delta \mathbb{H}_{\rm SQ}[\varphi,\bar{\varphi},\pi,\bar{\pi}]}{\delta\varphi(x,\tau)} = i \frac{\delta S[\varphi]}{\delta\varphi(x,\tau)}, 
	\label{eq:sq-Ham-eq}
\end{align}
where the complex conjugate equations are understood.
The solutions of this system are further restricted by the conservation of the generalized action,
\begin{equation}\label{eq:constraint}
\mA_*  =\hbar M= \mathbb{H}_{\rm SQ}[\varphi_*,\bar{\varphi}_*,\pi_*,\bar{\pi}_*]
\end{equation}
with $\varphi_*$ and $\pi_*$, together with their complex conjugates, denoting solutions of Eq.~\eqref{eq:sq-Ham-eq} for generic initial conditions. The strategy to obtain the Schwinger-Dyson equation consists now in taking the average over the microcanonical measure in Eq.~\eqref{eq:micorcanonical-emasure} of both members in Eq.~\eqref{eq:sq-Ham-eq}:
\begin{align}
\label{eq:Ham-averaged}
\left\langle	\frac{\partial \pi(x,\tau)}{\partial \tau} \right\rangle_{\text{micro}} = i\left\langle	\frac{\delta S[\varphi]}{\delta \varphi(x,\tau)} \right\rangle_{\text{micro}}\,,
\end{align}
and then recalling that, due to the equivalence between dynamical and statistical average at the basis of the symplectic quantization approach, we can also write 
\begin{align}\label{eq:mean}
\left\langle	\frac{\partial \pi(x,\tau)}{\partial \tau} \right\rangle_{\text{micro}}=\lim_{ \tau\to \infty}\frac{1}{ \tau}\int_{0}^{ \tau}ds\, \frac{\partial \pi_*(x,s)}{\partial s}=\lim_{ \tau\to \infty}\frac{1}{ \tau}(\pi_*(x,\tau)-\pi_*(x,0)),
\end{align}
where the subscript asterisk for $\pi_*(x,\tau)$ denotes a solution of the Hamiltonian dynamics in the intrinsic time $\tau$. At this point we need to make a crucial observation: within the symplectic quantization approach physical solutions on the fixed-energy surface $\mathbb H_{\rm SQ}=\mA_*$ are bounded. This means, in particular, that $\pi_*(x,\tau)$ is a bounded function of time and we necessarily have
\begin{align}\label{eq:mean}
\lim_{ \tau\to \infty}\frac{\pi_*(x,\tau)}{ \tau} = 0,
\end{align}
from which, recalling the identity in Eq.~\eqref{eq:Ham-averaged}, we can conclude that 
\begin{equation}\label{eq:dys1}
\left\langle	\frac{\delta S[\varphi]}{\delta \varphi(x,\tau)} \right\rangle_{\text{micro}} = 0
\end{equation}
This is the SQ version of the "quantum equations of motion" for the multipoint correlation functions and, as we now show, the first member of the Dyson--Schwinger hierarchy. As for "quantum equations of motion", within this discussion we do not clearly refer to the flow with respect to the instrisic time $\tau$ but rather to the field-theoretic identities which allow to write the variation with respect to the Minkowskian time $x_0$ of a certain multipoint correlation function in terms of higher order correlation functions. The corresponding equation for a non-relativistic quantum particle in a bounded potential yields the Ehrenfest's theorem, as already shown in~\cite{Gradenigo_3}. Since in the previous section we proved that the microcanonical expectation values of holomorphic observables coincide with those obtained from the Feynman path-integral formulation of quantum field theory, Eq.~\eqref{eq:dys1} is precisely the first Dyson--Schwinger equation in the continuum limit at fixed $\mA_*=\hbar M$.

Let us consider now again Eq.~\eqref{eq:sq-Ham-eq}, multiply both members of it by a string of fields $\varphi_*(x_1,\tau)\ldots\varphi_*(x_n,\tau)$, and take the statistical average. By also exploiting the equivalence between statistical and dynamical averages, we get
\begin{align}\label{eq:mean2}
	&\left\langle	\frac{\partial \pi(x)}{\partial \tau} \varphi(x_1,\tau)\ldots\varphi(x_n,\tau)\right\rangle_{\text{micro}}\nonumber\\
	&=\lim_{ \tau\to \infty}\frac{1}{ \tau}\int_{0}^{ \tau}ds\, \frac{\partial \pi_*(x,s)}{\partial s}\varphi_*(x_1,s)\ldots\varphi_*(x_n,s)\nonumber\\
	&=\lim_{ \tau\to \infty}\frac{1}{ \tau}\left[\pi_*(x,s)\varphi_*(x_1,s)\ldots\varphi_*(x_n,s)\right]\big\rvert_{0}^\tau\nonumber\\
	&-\lim_{ \tau\to \infty}\frac{1}{ \tau}\sum_{i=1}^n\int_{0}^{ \tau}ds\, \pi_*(x,s)\bar{\pi}_*(x_i,s)\varphi_*(x_1,s)\ldots\varphi_*(x_{i-1},s)\varphi_*(x_{i+1},s)\ldots\varphi_*(x_n,s)\nonumber\\
	&=i\left\langle		\frac{\delta S[\varphi]}{\delta \varphi(x,\tau)}\varphi(x_1,\tau)\ldots\varphi(x_n,\tau)\right\rangle_{\text{micro}}
\end{align}
In the third line of Eq.~\eqref{eq:mean2} we integrated by parts and used the identity $\partial_\tau{\varphi}_*(x,s)=\bar{\pi}_*(x,s)$ obeyed by the solutions of the symplectic equations. The boundary term again vanishes by boundedness of the constrained flow:
\begin{equation}
\lim_{ \tau\to \infty}\frac{1}{ \tau}\left[\pi_*(x,s)\varphi_*(x_1,s)\ldots\varphi_*(x_n,s)\right]\big\rvert_{0}^\tau =0
\end{equation}
The term in the fourth line of Eq.~\eqref{eq:mean2} can instead be written as an ensemble average:
\begin{align}
&\lim_{ \tau\to \infty}\frac{1}{ \tau}\sum_{i=1}^n\int_{0}^{ \tau}ds\, \pi_*(x,s)\bar{\pi}_*(x_i,s)\varphi_*(x_1,s)\ldots\varphi_*(x_{i-1},s)\varphi_*(x_{i+1},s)\ldots\varphi_*(x_n,s)\nonumber\\
&=\frac{1}{\Omega[\mA]}~\sum_{i=1}^n\,\int \mD\varphi\mD\bar \varphi\mD\pi\mD\bar\pi~\delta\left(\mA-\mathbb{H}_{\rm SQ}[\varphi,\bar{\varphi},\pi,\bar{\pi}]\right)\pi(x)\bar{\pi}(x_i)\varphi(x_1)\ldots\varphi(x_{i-1})\varphi(x_{i+1})\ldots\varphi(x_n)\,.
\end{align}
At this stage we use the fact that $\mathbb{H}_{\rm SQ}[\varphi,\bar{\varphi},\pi,\bar{\pi}]$ depends on the momenta through the non-propagating combination $\pi(x)\bar{\pi}(x)$. Under $\pi(x)\to-\pi(x)$, the measure and the constraint remain invariant. It follows that the only non-vanishing contributions to the expectation value arise when $x=x_i$. Indeed, for $x\neq x_i$ the integrand is odd in $\pi(x)$, whereas the measure and the constraint are even in $\pi$, so the $\pi(x)$ integral vanishes. Only when $x=x_i$ does the factor become $\pi(x)\bar\pi(x)$ and survive. In this case the expectation value of $\pi(x)\bar{\pi}(x)$ reduces to a constant, and in Appendix~\ref{app:alpha} we prove that, upon fixing $\mA_* = \hbar M$, one can write in the continuum limit:
\begin{align}
\left\langle	\frac{\partial \pi(x)}{\partial \tau} \varphi(x_1,\tau)\ldots\varphi(x_n,\tau)\right\rangle_{\text{micro}} =-\hbar \sum_{i=1}^n\delta(x-x_i)\left\langle\varphi(x_1)\ldots\varphi(x_{i-1})\varphi(x_{i+1})\ldots\varphi(x_n)\right\rangle_{\text{micro}}\,,
\end{align}
Combining this identity with the last line of Eq.~\eqref{eq:mean2}, we obtain
\begin{align}
\label{eq:general-Dyson}
\left\langle		\frac{\delta S[\varphi]}{\delta \varphi(x)}\varphi(x_1)\ldots\varphi(x_n)\right\rangle_{\text{micro}}=i\hbar \sum_{i=1}^n\delta(x-x_i)\left\langle\varphi(x_1)\ldots\varphi(x_{i-1})\varphi(x_{i+1})\ldots\varphi(x_n)\right\rangle_{\text{micro}}\,.
\end{align}
If we finally consider the identity
\begin{align}
\Big\langle ~\# ~\Big\rangle_{\text{micro}}
=
\Big\langle ~\# ~\Big\rangle_{\text{Feynman}}\,,
\end{align}
between microcanonical averages and averages taken with respect the Feynman path integral, which we proved in Sec.~\ref{sec:equivalence}, Eq.~\eqref{eq:general-Dyson} becomes the standard Dyson--Schwinger identity for multipoint correlation functions,
\begin{align}
\left\langle
\frac{\delta S[\varphi]}{\delta \varphi(x)}
\varphi(x_1)\ldots\varphi(x_n)
\right\rangle_{\text{Feynman}}
=
i\hbar \sum_{i=1}^n\delta(x-x_i)
\left\langle
\varphi(x_1)\ldots\varphi(x_{i-1})
\varphi(x_{i+1})\ldots\varphi(x_n)
\right\rangle_{\text{Feynman}}\,.
\label{eq:general-Dyson-Feynman}
\end{align}
For the free scalar field,
\begin{align}
S_{\rm free}[\varphi]
=
\int d^{d+1}x\,
\left[
\frac{1}{2}\,\partial_\mu\varphi\,\partial^\mu\varphi
-\frac{1}{2}m^2\varphi^2
\right],
\end{align}
and therefore
\begin{align}
\frac{\delta S_{\rm free}[\varphi]}{\delta \varphi(x)}
=
-\left(\partial_{x_0}^2-\nabla^2+m^2\right)\varphi(x)\,.
\end{align}
Hence the Dyson--Schwinger hierarchy can be written explicitly as
\begin{align}
&\left\langle
\left(\partial_{x_0}^2-\nabla^2+m^2\right)\varphi(x)
\varphi(x_1)\ldots\varphi(x_n)
\right\rangle_{\text{Feynman}}
=\nonumber\\
&-i\hbar \sum_{i=1}^n\delta(x-x_i)
\left\langle
\varphi(x_1)\ldots\varphi(x_{i-1})
\varphi(x_{i+1})\ldots\varphi(x_n)
\right\rangle_{\text{Feynman}}\,.
\label{eq:free-Dyson-hierarchy}
\end{align}
For $n=1$, one obtains the propagator equation
\begin{align}
\left\langle
\left(\partial_{x_0}^2-\nabla^2+m^2\right)\varphi(x)
\varphi(x_1)
\right\rangle_{\text{Feynman}}
=
-i\hbar\,\delta(x-x_1)\,.
\label{eq:free-Dyson-n1}
\end{align}
For $n=2$, one finds
\begin{align}
&\left\langle
\left(\partial_{x_0}^2-\nabla^2+m^2\right)\varphi(x)
\varphi(x_1)\varphi(x_2)
\right\rangle_{\text{Feynman}}
\nonumber\\
&\qquad =
-i\hbar\,\delta(x-x_1)
\left\langle \varphi(x_2)\right\rangle_{\text{Feynman}}
-i\hbar\,\delta(x-x_2)
\left\langle \varphi(x_1)\right\rangle_{\text{Feynman}}\,.
\label{eq:free-Dyson-n2}
\end{align}
For $n=3$, one obtains
\begin{align}
&\left\langle
\left(\partial_{x_0}^2-\nabla^2+m^2\right)\varphi(x)
\varphi(x_1)\varphi(x_2)\varphi(x_3)
\right\rangle_{\text{Feynman}}
\nonumber\\
&\qquad =
-i\hbar\,\delta(x-x_1)
\left\langle \varphi(x_2)\varphi(x_3)\right\rangle_{\text{Feynman}}
-i\hbar\,\delta(x-x_2)
\left\langle \varphi(x_1)\varphi(x_3)\right\rangle_{\text{Feynman}}
\nonumber\\
&\qquad\quad
-i\hbar\,\delta(x-x_3)
\left\langle \varphi(x_1)\varphi(x_2)\right\rangle_{\text{Feynman}}\,.
\label{eq:free-Dyson-n3}
\end{align}
The first of these equations states that the two-point function is the Green function of the Minkowskian Klein--Gordon operator, with the normalization fixed by the contact term. The higher equations show explicitly how the same contact structure propagates through the full hierarchy of multipoint correlators. We have therefore shown that the intrinsic-time Hamiltonian flow of constrained symplectic quantization encodes the full hierarchy of quantum equations satisfied by the correlators.
%
\section{Scalar Field: constrained equations of motion}
\label{sec:SF-dynamics}
In the two previous sections, Sec.~\ref{sec:equivalence} and Sec.~\ref{sec:ds} have shown in a completely general way that for a generic interacting scalar field theory, provided it is renormalizable, the constrained SQ microcanonical functional is equivalent to the Feynman generating functional and that, in particular, the precise form of the Dyson-Schwinger identities between multipoint correlation functions can be derived. Now, in order to prepare the ground for numerical tests, we focus on a particular case, the one of the free Lorenz-invariant real scalar field. As will be clear in the forthcoming discussion, when the time comes to move from general proofs to numerical tests, the details of the model matter. In particular, the choice of constraints which guaratantee the stability of the dynamics generated by $\mathbb{H}_{\textrm{SQ}}$ must be done with care and turns out to be more or less simple depending on the intricacy of the interactions and the symmetries of the model. We now derive the corresponding constrained Hamiltonian dynamics for a free real scalar field in $d+1$ Minkowski dimensions. The physical coordinate time $x_0$ is treated as one of the space-time coordinates of the field configuration, while the actual sampling evolution is the intrinsic time $\tau$. Real-time observables are therefore extracted from correlations along the $x_0$ direction, whereas the quantum fluctuations are generated by the constrained Hamiltonian flow in $\tau$.

Let $x=(x_0,\mathbf{x})$ with $\mathbf{x}\in\mathbb{R}^d$, and consider the free massive scalar field with action
\begin{align}
  S[\varphi] &= \int_{\mathcal{V}} d^{d+1}x \left[\frac{1}{2}\,\partial_\mu \varphi(x)\,\partial^\mu \varphi(x) - \frac{1}{2}\,m^2\,\varphi^2(x)\right]\nonumber\\
  &= \int_{\mathcal{V}} d^{d+1}x \left[\frac{1}{2}\,(\partial_{x_0}\varphi)^2 - \frac{1}{2}\,(\nabla\varphi)^2 - \frac{1}{2}\,m^2\,\varphi^2\right],
\label{eq:sf-action}
\end{align}
where $\mathcal{V}$ denotes the spacetime domain on which we impose boundary conditions, $\nabla$ is the gradient in the $d$ spatial directions, and the metric signature is $(+,-,\ldots,-)$. Recall now the expression for the generalized Hamiltonian (Eq. \eqref{eq:csq-hamiltonian}):
\begin{equation}
  \mathbb{H}_{\text{SQ}}[\pi,\bar{\pi},\varphi,\bar{\varphi}]
  \;=\;
  \int_{\mathbb{R}^{d+1}}\! d^{d+1}x\;\bar{\pi}(x,\tau)\,\pi(x,\tau)
  \;+\;2\,\Im S[\varphi,\bar{\varphi}],
  \qquad
  \Im S[\varphi,\bar{\varphi}]\equiv \frac{S[\varphi]-\bar S[\bar{\varphi}]}{2i},
  \label{eq:sf-hamiltonian}
\end{equation}
where $\Im S[\varphi,\bar{\varphi}]$ is defined solely by the holomorphic/anti-holomorphic pair (and there is no mixed action $S[\varphi,\bar{\varphi}]$ in the theory). For the free scalar action~\eqref{eq:sf-action} one may write $2\Im S$ explicitly as
\begin{align}
  2\,\Im S[\varphi,\bar{\varphi}]
  &\;=\;
  \frac{1}{i}\Big(S[\varphi]-\bar S[\bar{\varphi}]\Big)\\\nonumber
  &\;=\;
  \frac{1}{2i}\int_{\mathbb{R}^{d+1}}\! d^{d+1}x\;
  \Big[
    (\partial_{x_0}\varphi)^2-(\nabla\varphi)^2-m^2\varphi^2
    -(\partial_{x_0}\bar{\varphi})^2+(\nabla\bar{\varphi})^2+m^2\bar{\varphi}^{\,2}
  \Big],
  \label{eq:ImS-explicit-phibarphi}
\end{align}
so that
\begin{equation}
\begin{aligned}
  \mathbb{H}[\pi,\bar{\pi},\varphi,\bar{\varphi}]
  &= \int_{\mathbb{R}^{d+1}}\! d^{d+1}x\;\bar{\pi}(x,\tau)\,\pi(x,\tau) \\
  &\quad + \frac{1}{2i}\int_{\mathbb{R}^{d+1}}\! d^{d+1}x\Big[(\partial_{x_0}\varphi)^2-(\nabla\varphi)^2-m^2\varphi^2-(\partial_{x_0}\bar{\varphi})^2+(\nabla\bar{\varphi})^2+m^2\bar{\varphi}^{\,2}\Big].
\end{aligned}
\label{eq:H-explicit-phibarphi}
\end{equation}
Equivalently, integrating by parts on $\mathbb{R}^{d+1}$ (and discarding boundary terms at infinity),
\begin{align}
  S[\varphi]
  \;=\;
  -\frac12\int_{\mathbb{R}^{d+1}}\! d^{d+1}x\;
  \varphi\,\big(\partial_{x_0}^2-\nabla^2+m^2\big)\varphi,
      \\\nonumber
  \bar S[\bar{\varphi}]
  \;=\;
  -\frac12\int_{\mathbb{R}^{d+1}}\! d^{d+1}x\;
  \bar{\varphi}\,\big(\partial_{x_0}^2-\nabla^2+m^2\big)\bar{\varphi},
\end{align}
and we get
\begin{equation}
  2\,\Im S[\varphi,\bar{\varphi}]
  \;=\;
  -\frac{1}{2i}\int_{\mathbb{R}^{d+1}}\! d^{d+1}x\;
  \Big[
    \varphi\,\big(\partial_{x_0}^2-\nabla^2+m^2\big)\varphi
    -\bar{\varphi}\,\big(\partial_{x_0}^2-\nabla^2+m^2\big)\bar{\varphi}
  \Big].
  \label{eq:ImS-by-parts-phibarphi}
\end{equation}
The Hamilton equations along the intrinsic time are 
\begin{equation}
\begin{aligned}
  \partial_\tau \varphi(x,\tau) &= \frac{\delta\mathbb{H}}{\delta \pi(x,\tau)}=\bar{\pi}(x,\tau),\\
  \partial_\tau \bar{\varphi}(x,\tau) &= \frac{\delta\mathbb{H}}{\delta \bar{\pi}(x,\tau)}=\pi(x,\tau), \\
  \partial_\tau \pi(x,\tau) &= -\frac{\delta\mathbb{H}}{\delta \varphi(x,\tau)}
  = i\,\frac{\delta S[\varphi]}{\delta \varphi(x,\tau)}, \\
  \partial_\tau \bar{\pi}(x,\tau) &= -\frac{\delta\mathbb{H}}{\delta \bar{\varphi}(x,\tau)}
  = -i\,\frac{\delta \bar{S}[\bar{\varphi}]}{\delta \bar{\varphi}(x,\tau)},
\end{aligned}
\label{eq:sf-hamilton-eqs}
\end{equation}
which combine into the mixed holomorphic/anti-holomorphic equations of motion
\begin{equation}
  -i\,\frac{d^2}{d\tau^2}\bar{\varphi}(x,\tau)
  = -\left(\partial_{x_0}^2-\nabla^2+m^2\right)\varphi(x,\tau),
\label{eq:sf-complex-eom}
\end{equation}
together with its complex conjugate equation. At this point, the key technical and conceptual issue is the following: generic solutions of Eq.~\eqref{eq:sf-complex-eom} are unbounded in $\tau$ unless one selects an appropriate stable manifold in the complexified phase space, and this selection parallels (and, in fact, will be identified with) the choice of convergent integration contours in the corresponding microcanonical functional integral. To make the stability structure explicit, we diagonalize the dynamics by expanding in eigenmodes of the quadratic kernel, with the precise mode basis determined by the boundary conditions imposed on~$\mathcal{V}$. Considering for now periodic boundary conditions on a $(d+1)$-dimensional box $\mathcal{V}=[0,T]\times\prod_{i=1}^d[0,L_i]$ we write
\begin{equation}
  \varphi(x,\tau)=\frac{1}{\sqrt{T\,\prod_{i=1}^{d}L_i}}\sum_{k}\,e^{i k\cdot x}\,\varphi(k,\tau), \qquad
  k\cdot x \equiv k_0 x_0+\mathbf{k}\cdot\mathbf{x}, \qquad
  k_\mu=\left(\frac{2\pi n_0}{T},\frac{2\pi n_1}{L_1},\ldots,\frac{2\pi n_d}{L_d}\right),
\label{eq:sf-fourier}
\end{equation}
with integers $n_\mu\in\mathbb{Z}$ and the obvious generalization to a finite truncation in numerical implementations. The diagonalization procedure is completely analogous for any other choice of boundary conditions: one expands the field on a complete orthonormal basis of eigenfunctions compatible with the prescribed boundaries, which leads again to a discrete set of mode labels and to the same mode-by-mode decoupling once $k_\mu$ is understood as the corresponding eigenvalue set. In all cases one finds for each mode the decoupled equation
\begin{equation}
  \frac{d^2}{d\tau^2}\,\bar{\varphi}(k,\tau) + i\,\omega^2(k)\,\varphi(k,\tau)=0, \qquad
  \omega^2(k)\equiv -k_0^2+\mathbf{k}^2+m^2,
\label{eq:sf-mode-eom}
\end{equation}
where $\omega^2(k)$ is the Minkowskian quadratic form appearing in the free action and its sign depends on the mode. Writing $\varphi(k,\tau)=\varphi_R(k,\tau)+i\,\varphi_I(k,\tau)$, Eq.~\eqref{eq:sf-mode-eom} is equivalent to the coupled real system
\begin{align}
  \ddot{\varphi}_R(k,\tau) - \omega^2(k)\,\varphi_I(k,\tau) &= 0 \nonumber\\
  \ddot{\varphi}_I(k,\tau) - \omega^2(k)\,\varphi_R(k,\tau) &= 0,
\label{eq:sf-coupled-real-imag}
\end{align}
which immediately shows that unless $\varphi_R$ and $\varphi_I$ are tied together by a linear constraint, the flow explores directions that lead to exponential growth in~$\tau$ for part of the spectrum. The natural choice that keeps each mode on a bounded manifold is the direct scalar-field analogue of what we found for the Harmonic oscillator~\cite{Giachello_Harmonic}: for each mode $k$ we impose
\begin{align}
  \omega^2(k)>0~~~&\Longrightarrow~~~\varphi_I(k,\tau)=-\varphi_R(k,\tau)\quad\&\quad \pi_I(k,\tau)=-\pi_R(k,\tau), \nonumber\\
  \omega^2(k)<0~~~&\Longrightarrow~~~\varphi_I(k,\tau)=+\varphi_R(k,\tau)\quad\&\quad \pi_I(k,\tau)=+\pi_R(k,\tau),
\label{eq:sf-stable-surface0}
\end{align}
or, in compact form,
\begin{equation}
  \varphi_I(k,\tau)=-\sgn[\omega^2(k)]\,\varphi_R(k,\tau)\qquad\pi_I(k,\tau)=-\sgn[\omega^2(k)]\,\pi_R(k,\tau),
\label{eq:sf-stable-surface}
\end{equation}
which reduces Eq.~\eqref{eq:sf-coupled-real-imag} to the bounded oscillatory motion
\begin{align}
  \ddot{\varphi}_R(k,\tau)+|\omega^2(k)|\,\varphi_R(k,\tau)=0, \qquad
  \ddot{\varphi}_I(k,\tau)+|\omega^2(k)|\,\varphi_I(k,\tau)=0.
\label{eq:sf-stable-motion}
\end{align}
As in~\cite{Giachello_Harmonic}, the relationship between the real and imaginary parts of the field is not an accidental degeneracy: it expresses the fact that the constrained dynamics lives on a one-dimensional manifold per mode, so that the analytic prolongation to $\mathbb{C}$ does not double the physical degrees of freedom. In particular, once $\varphi_R(k,\tau)$ is determined, $\varphi_I(k,\tau)$ is fixed by Eq.~\eqref{eq:sf-stable-surface}, and the same holds for the momenta. Hence the constrained symplectic dynamics can be simulated by evolving only a single real field component with mode-dependent sign constraints. \par
The remaining crucial step is to connect the stable manifold selected by Eq.~\eqref{eq:sf-stable-surface} with the complex integration contours $\boldsymbol{\Gamma}(x)$ that appear in the microcanonical functional integral and that, upon integrating out the momenta, reproduce the usual Feynman measure. Along the constrained Hamiltonian flow the instantaneous value of local quantities such as $\varphi^n(x,\tau)$ or bilocals such as $\varphi(x,\tau)\varphi(y,\tau)$ is generically complex, because the fields are complex on the sampling manifold. Nevertheless, physical correlators are recovered because the microcanonical average is evaluated on a nontrivial contour $\boldsymbol{\Gamma}$ in the complexified field space that is precisely the contour compatible with the constraints. For a generic operator insertion $\mathcal{O}[\varphi]$ and boundary condition choice the symplectic-quantization prescription is to compute the $\tau$-time average and to assume the equivalence between the long-$\tau$ dynamical average and the corresponding microcanonical functional integral:
\begin{align}
  \lim_{\Delta\tau\to\infty}\frac{1}{\Delta\tau}\int_{\tau_0}^{\tau_0+\Delta\tau}\!d\tau\;\mathcal{O}[\varphi(\cdot,\tau)]
  \;\cong\;
  \frac{1}{\Omega_\infty[\hbar]}
  \int_{\boldsymbol{\Gamma}_{\mathsf{BC}}}
  \mathcal{D}\bar{\pi}\,\mathcal{D}\pi\,\mathcal{D}\bar{\varphi}\,\mathcal{D}\varphi\;
  \mathcal{O}[\varphi]\;
  \delta\!\left(\mathcal{A}-\mathbb{H}[\varphi,\bar{\varphi},\pi,\bar{\pi}]\right),
\label{eq:sf-ergodicity-sq}
\end{align}
where $\boldsymbol{\Gamma}_{\mathsf{BC}}$ denotes the contour in complexified phase space compatible with the boundary conditions and with the dynamical constraints. The matching with standard QFT correlators is then achieved by the same contour-deformation argument: in the usual Feynman representation one integrates over real field configurations at each spacetime point, whereas in symplectic quantization one integrates over a deformed contour, and expectation values are invariant under such deformations provided no singularities are crossed and the integral remains convergent.\par
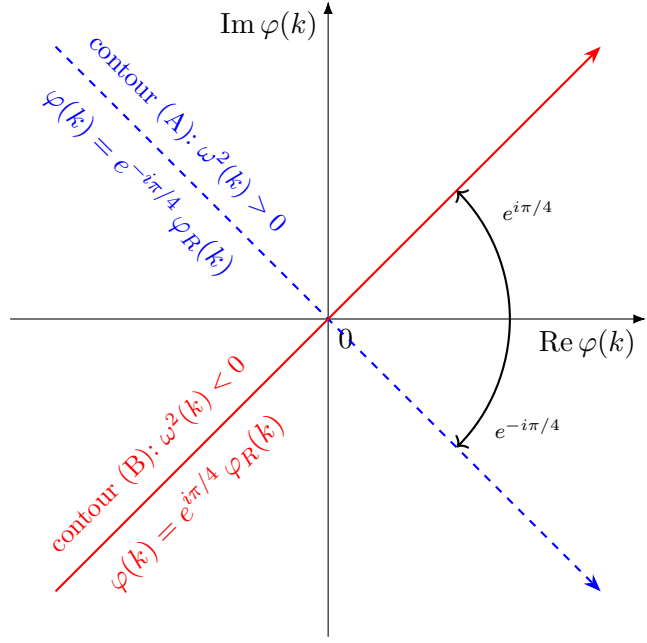
\begin{figure}[H]
\centering
\begin{tikzpicture}[scale=1.2]
    \draw[-{Latex}] (-3.5,0) -- (3.5,0) node[below left] {$\operatorname{Re}\varphi(k)$};
    \draw[-{Latex}] (0,-3.5) -- (0,3.5) node[below left] {$\operatorname{Im}\varphi(k)$};
    \node at (0,0) [below right] {$0$};
    
    \draw[blue, thick, dashed, -{Stealth}] (-3, 3) -- (3, -3);
    \node[blue, rotate=-45, above, xshift=-3.05cm, yshift=0.1cm] {\small contour (A): $\omega^2(k)>0$};
    \node[blue, rotate=-45, below, xshift=-3.05cm, yshift=-0.1cm] {$\varphi(k)=e^{-i\pi/4}\,\varphi_R(k)$};

    \draw[red, thick, -{Stealth}] (-3, -3) -- (3, 3);
    \node[red, rotate=45, above, xshift=-2.95cm, yshift=0.1cm] {\small contour (B): $\omega^2(k)<0$};
    \node[red, rotate=45, below, xshift=-2.95cm, yshift=-0.1cm] {$\varphi(k)=e^{i\pi/4}\,\varphi_R(k)$};

    \draw[->, thick] (2,0) arc[start angle=0, end angle=45, radius=2];
    \node at (2.2,1.2) {\scriptsize $e^{i\pi/4}$};

    \draw[->, thick] (2,0) arc[start angle=0, end angle=-45, radius=2];
    \node at (2.2,-1.2) {\scriptsize $e^{-i\pi/4}$};
\end{tikzpicture}
\caption{Integration contours in the complex $\varphi(k)$ plane. Each mode is rotated so that the holomorphic weight becomes real and exponentially damped. Contour (A) corresponds to $\omega^2(k)>0$ and is rotated by $-\pi/4$; contour (B) corresponds to $\omega^2(k)<0$ and is rotated by $+\pi/4$. This is equivalent to imposing the linear relations among Fourier components defining the stable manifold of the symplectic dynamics.}
\label{fig:sf-contours}
\end{figure}

In practice, imposing Eq.~\eqref{eq:sf-stable-surface0} in mode space corresponds to a mode-dependent rotation of the integration contour in the complex $\varphi(k)$ plane, completely analogous to the harmonic-oscillator case~\cite{Giachello_Harmonic,Giachello_lattice2025}. For each momentum mode $k$ the constraint $\varphi_I(k)=-\varphi_R(k)$ for $\omega^2(k)>0$ and $\varphi_I(k)=+\varphi_R(k)$ for $\omega^2(k)<0$ is equivalent to rotating the contour by an angle $\pi/4$ clockwise when $\omega^2(k)>0$ and counterclockwise when $\omega^2(k)<0$, so that the holomorphic weight becomes a convergent Gaussian after the analytic prolongation $\varphi\mapsto\varphi_R+i\varphi_I$ is restricted to the stable manifold. To make the correspondence explicit, let us adopt periodic boundary conditions, so that the functional is expressed as a trace. Denoting the spacetime volume by $V_{d+1}\equiv T\prod_{i=1}^{d}L_i$, we write
\begin{align}
  \Tr\!\left[e^{-iT\hat{H}/\hbar}\right]
  \;=\;
  \int_{\substack{\varphi(x_0+T,\mathbf{x})=\varphi(x_0,\mathbf{x})\\
                  \varphi(x_0,\mathbf{x}+\mathbf{L})=\varphi(x_0,\mathbf{x})}}
  \!\!\mathcal{D}\varphi~e^{\frac{i}{\hbar}S[\varphi]}
  \;=\;
  \int_{\text{P.B.}}\!\mathcal{D}\varphi~e^{\frac{i}{\hbar}S[\varphi]}\,,
  \label{eq:sf-trace-periodic}
\end{align}
Here and in the following, $\int_{\text{P.B.}}\mathcal{D}\varphi$ denotes the functional integral over field configurations satisfying periodic boundary conditions on $\mathcal V=[0,T]\times\mathbb{T}^d$. In the standard Feynman representation the integration is performed over the real axis at each space-time point, namely $\int\mathcal{D}\varphi=\prod_x\int_{-\infty}^{\infty}d\varphi(x)$. In the CSQ microcanonical formulation, instead, the integration is performed over a contour $\boldsymbol{\Gamma}(x)$ in the complexified field space selected by the stability constraint. For the free theory the integrand is holomorphic and the contour can be deformed without crossing singularities. We may therefore write
\begin{align}
  \langle \hat{\varphi}^2(x)\rangle_{\text{P.B.}} &= \frac{\int_{-\infty}^{\infty}\prod_{x} d\varphi(x)\;\varphi^2(x)\;e^{\frac{i}{\hbar}S[\varphi]}}
       {\int_{-\infty}^{\infty}\prod_{x} d\varphi(x)\;e^{\frac{i}{\hbar}S[\varphi]}}
  \label{eq:sf-equivalence-1}\\
  &= \frac{\int \prod_{\gamma_{\varphi(x)}} d\varphi(x)\;\varphi^2(x)\;e^{\frac{i}{\hbar}S[\varphi]}} {\int \prod_{\gamma_{\varphi(x)}} d\varphi(x)\;e^{\frac{i}{\hbar}S[\varphi]}}
  \label{eq:sf-equivalence-2}\\
  &= \frac{\int \prod_{\gamma_{\varphi(x)}} d\varphi_R(x)\,d\varphi_I(x)\;[\varphi_R(x)+i\varphi_I(x)]^2\;e^{\frac{i}{\hbar}S[\varphi_R+i\varphi_I]}}{\int \prod_{\gamma_{\varphi(x)}} d\varphi_R(x)\,d\varphi_I(x)\;e^{\frac{i}{\hbar}S[\varphi_R+i\varphi_I]}}\,,
  \label{eq:sf-equivalence-3}
\end{align}
where $\gamma_{\varphi(x)}$ denotes the contour in the complex plane selected at each point $x$ by the stability constraint, and the last line makes explicit the analytic prolongation $\varphi=\varphi_R+i\varphi_I$. Using the normalized mode expansion of Eq.~\eqref{eq:sf-fourier}, and integrating by parts (boundary terms vanish by periodicity), the free action becomes
\begin{align}
  S[\varphi] &= -\frac{1}{2}\int_{\mathcal{V}} d^{d+1}x\;\varphi(x)\left(\partial_{x_0}^2-\nabla^2+m^2\right)\varphi(x) = \frac{1}{2}\sum_{k}\omega^2(k)\,\varphi(k)\varphi(-k),
  \label{eq:sf-action-diagonal}
\end{align}
with $\omega^2(k)\equiv -k_0^2+\mathbf{k}^2+m^2$. Since $\varphi(x)\in\mathbb{C}$ along the constrained symplectic flow, one cannot identify $\varphi(-k)$ with $\varphi^{*}(k)$; rather, one writes $\varphi(k)=\varphi_R(k)+i\varphi_I(k)$ and imposes the constraints directly in mode space. Proceeding exactly as in the harmonic-oscillator case~\cite{Giachello_Harmonic}, we rewrite the diagonal action in terms of real and imaginary parts and then restrict the functional integration to the stable manifold determined by Eq.~\eqref{eq:sf-stable-surface0}. The result is that each mode is integrated along a contour rotated by $\pm\pi/4$, and the holomorphic weight becomes a convergent Gaussian:
\begin{align}
  \exp\!\left(\frac{i}{\hbar}S[\varphi]\right)
  =
  \exp\!\left(
  -\frac{1}{\hbar}\sum_{\{k\,|\,\omega^2(k)\ge 0\}}\omega^2(k)\,|\varphi_R(k)|^2
  +\frac{1}{\hbar}\sum_{\{k\,|\,\omega^2(k)< 0\}}\omega^2(k)\,|\varphi_R(k)|^2
  \right),
  \label{eq:sf-factor-measure-k}
\end{align}
where we used that $\varphi_R(x)$ and $\varphi_I(x)$ are real functions, so that $\varphi_R(-k)=\varphi_R^{*}(k)$ and $\varphi_I(-k)=\varphi_I^{*}(k)$, and we then applied the constraint $\varphi_I(k)=-\varphi_R(k)$ for $\omega^2(k)>0$ and $\varphi_I(k)=+\varphi_R(k)$ for $\omega^2(k)<0$. Equivalently, the same restriction is described as a rotation of the contour for each Fourier component in the complex $\varphi(k)$ plane, as shown in Fig.~\ref{fig:sf-contours}. 
With the contour correspondence made explicit, we can now state the operational identification between standard quantum-field-theory correlators and $\tau$-averages along the constrained symplectic flow. In complete analogy with the harmonic oscillator, we consider the two-point function in mode space,
\begin{align}
  \left\langle \varphi(k)\varphi(-k)\right\rangle_{\text{P.B.}}
  &=
  \frac{\int_{\boldsymbol{\Gamma}_{\text{P.B.}}}\mathcal{D}\varphi~\varphi(k)\varphi(-k)~e^{\frac{i}{\hbar}S[\varphi]}}
       {\int_{\boldsymbol{\Gamma}_{\text{P.B.}}}\mathcal{D}\varphi~e^{\frac{i}{\hbar}S[\varphi]}}
  =
  \lim_{\Delta\tau\to\infty}\frac{1}{\Delta\tau}\int_{\tau_0}^{\tau_0+\Delta\tau}\!d\tau~\varphi(k,\tau)\varphi(-k,\tau),
  \label{eq:sf-two-point-path-fourier}
\end{align}
and we then decompose the kernel into real and imaginary parts as
\begin{align}
  \left\langle \varphi(k)\varphi(-k)\right\rangle_{\text{P.B.}}
  &=
  \left\langle \varphi_R(k)\varphi_R(-k)\right\rangle
  -\left\langle \varphi_I(k)\varphi_I(-k)\right\rangle
  + i\left[\left\langle \varphi_R(k)\varphi_I(-k)\right\rangle
  +\left\langle \varphi_I(k)\varphi_R(-k)\right\rangle\right],
  \label{eq:sf-two-point-complete}
\end{align}
where the averages on the right-hand side are understood either as functional averages on $\boldsymbol{\Gamma}_{\text{P.B.}}$ or, equivalently, as long-$\tau$ averages along the constrained dynamics. Imposing the constraints \eqref{eq:sf-stable-surface0} in the Gaussian functional integral yields the same simplification pattern as in the harmonic oscillator: the correlator becomes a pure multiple of the real-part correlator sampled by the constrained flow, with the sign determined by $\omega^2(k)$. In particular, one finds
\begin{align}
  \omega^2(k)<0 ~~~&\Longrightarrow~~~\left\langle \varphi(k)\varphi(-k)\right\rangle_{\text{P.B.}}=\frac{2i}{\omega^2(k)}
  =-2i\,\left\langle \varphi_R(k)\varphi_R(-k)\right\rangle, \nonumber\\
  \omega^2(k)>0 ~~~&\Longrightarrow~~~\left\langle \varphi(k)\varphi(-k)\right\rangle_{\text{P.B.}}=\frac{2i}{\omega^2(k)}
  =+2i\,\left\langle \varphi_R(k)\varphi_R(-k)\right\rangle,
  \label{eq:sf-corresp-qft-sq}
\end{align}
and the real-part correlator on the right-hand side is precisely the quantity directly accessible in symplectic quantization as a dynamical average,
\begin{align}
  \left\langle \varphi_R(k)\varphi_R(-k)\right\rangle_{\Delta\tau}
  =
  \lim_{\Delta\tau\to\infty}\frac{1}{\Delta\tau}\int_{\tau_0}^{\tau_0+\Delta\tau}\!d\tau~\varphi_R(k,\tau)\varphi_R(-k,\tau).
  \label{eq:sf-ensemble-dynamical-av}
\end{align}
where the fields on the right-hand side of Eq.~\eqref{eq:sf-ensemble-dynamical-av} denote the solutions of the symplectic-quantization Hamiltonian dynamics. The corresponding position-space correlator $\langle\varphi(x)\varphi(y)\rangle$ follows by inverse mode transform, and the same strategy applies to other observables. Finally, other types of boundary conditions are incorporated in an analogous way: one enforces the chosen boundary constraints \emph{for all} $\tau$, evolves the constrained Hamiltonian flow in $\tau$ in the bulk, and computes long-$\tau$ averages of the desired insertions. The resulting quantities reproduce the corresponding path-integral correlators with the same boundary conditions once the induced contour $\boldsymbol{\Gamma}_{\mathsf{BC}}$ is taken into account. This completes the contour/stable-manifold correspondence for the scalar field and sets the stage for the explicit dictionary between standard quantum-field-theory observables and the quantities directly sampled by symplectic-quantization dynamics.

%
%
%
%
\section{Constrained symplectic dynamics: numerical algorithm}
\label{sec:numerical-algorithm}
In this section we describe the numerical integration of the constrained symplectic dynamics for the scalar field. The discussion has two parts. First, we discretize the $\tau$-Hamiltonian flow generated by a separable Hamiltonian on a space-time lattice. Second, we explain how the stability constraints are enforced in practice so as to keep the evolution on the stable manifold and suppress the exponentially growing directions produced by finite-step errors. As in the harmonic-oscillator case~\cite{Giachello_Harmonic}, this constraint step does not modify the target theory since it only ensures that the numerical evolution remains stable.\par
To keep this section self contained, we start by recalling the structure of the separable Hamiltonian driving the $\tau$-evolution,
\begin{align}
\mathbb{H}_{\text{SQ}}[\varphi,\bar\varphi,\pi,\bar\pi]=\mathbb{K}[\pi,\bar\pi]+\mathbb{V}[\varphi,\bar\varphi],
\end{align}
with kinetic term
\begin{align}
\mathbb{K}[\pi,\bar\pi]=\int_{\mathcal V} d^2x~\pi(x,\tau)\,\bar\pi(x,\tau),
\end{align}
and potential term given by the imaginary part of the action,
\begin{align}
\mathbb{V}[\varphi,\bar\varphi]=2\,\mathrm{Im}\,S[\varphi,\bar\varphi]=\frac{S[\varphi]-\bar S[\bar \varphi]}{i}.
\end{align}
The Hamiltonian equations generated by $\mathbb{H}$ take the universal form
\begin{align}
\partial_\tau\varphi(x,\tau)&=\bar\pi(x,\tau), \nonumber\\
\partial_\tau{\bar\varphi}(x,\tau)&=\pi(x,\tau), \nonumber\\
\partial_\tau\pi(x,\tau)&=-\,i\,\frac{\delta S}{\delta \varphi(x)}\Bigg|_{\varphi(\cdot,\tau)}, \nonumber\\
\partial_\tau{\bar\pi}(x,\tau)&=+\,i\,\frac{\delta \overline{S}}{\delta \bar\varphi(x)}\Bigg|_{\bar\varphi(\cdot,\tau)}.
\label{eq:tau-ham-eq}
\end{align}
For notational simplicity, and because this is the case used in all numerical experiments presented below, in this section we specialize to $d=1$, i.e. to a $1+1$ dimensional spacetime domain $\mathcal V=[0,T]\times[0,L]$. The generalization to $d>1$ is straightforward and amounts to replacing the single spatial coordinate by $\mathbf{x}\in\mathbb{R}^d$, the integral $d^2x$ by $d^{d+1}x$, and the one-dimensional spatial Laplacian by $\nabla^2$. Denoting $x\equiv(x_0,x_1)$, periodicity means
\begin{align}
\varphi(x_0+T,x_1)=\varphi(x_0,x_1),\qquad 
\varphi(x_0,x_1+L)=\varphi(x_0,x_1),
\end{align}
(and analogously for $\bar\varphi$), and therefore also
\begin{align}
\partial_\mu\varphi(x_0+T,x_1)=\partial_\mu\varphi(x_0,x_1),\qquad 
\partial_\mu\varphi(x_0,x_1+L)=\partial_\mu\varphi(x_0,x_1),\qquad \mu=0,1.
\end{align}
Starting from the standard quadratic action in first-derivative form,
\begin{align}
S_0[\varphi]
=
\frac{1}{2}\int_{0}^{T}\!dx_0\int_{0}^{L}\!dx_1\,
\left[(\partial_{x_0}\varphi)^2-(\partial_{x_1}\varphi)^2-m^2\varphi^2\right],
\label{eq:bulk-action-minkowski}
\end{align}
one integration by parts in each direction yields
\begin{align}
\int_{\mathcal V}\! d^2x\,(\partial_\mu\varphi)^2
=
-\int_{\mathcal V}\! d^2x\,\varphi\,\partial_\mu^2\varphi
\;+\;
\int_{\partial\mathcal V}\! ds_\mu~\varphi\,\partial_\mu\varphi,
\qquad \mu=0,1,
\end{align}
so that the only extra contributions are the boundary terms on $\partial\mathcal V$. With periodicity, the contributions from opposite sides cancel identically. Explicitly,
\begin{align}
\int_{0}^{L}\!dx_1\;\Big[\varphi\,\partial_{x_0}\varphi\Big]_{0}^{T}=0,
\qquad
\int_{0}^{T}\!dx_0\;\Big[\varphi\,\partial_{x_1}\varphi\Big]_{0}^{L}=0,
\end{align}
since $\varphi(T,x_1)=\varphi(0,x_1)$, $\partial_{x_0}\varphi(T,x_1)=\partial_{x_0}\varphi(0,x_1)$ and similarly at $x_1=0,L$. For other boundary prescriptions the same partial-integration step produces the same boundary functional, but it is eliminated in different ways. For Dirichlet boundaries, one requires
\begin{align}
\varphi\big|_{\partial\mathcal V}=0
\qquad
\text{(or, more generally, $\varphi\big|_{\partial\mathcal V}=\varphi_{\rm bnd}$ fixed for all $\tau$)},
\end{align}
so that the boundary term vanishes identically or reduces to a $\tau$-independent constant. For Neumann boundaries, one instead imposes vanishing normal derivative,
\begin{align}
\partial_n\varphi\big|_{\partial\mathcal V}=0,
\end{align}
which again sets the boundary contribution to zero. In the mixed setting, one combines the two: the field is fixed on the chosen Dirichlet components of $\partial\mathcal V$, while the remaining components are left free in the Neumann sense (implemented by enforcing $\partial_n\varphi=0$ at the boundary).\par
Substituting back into Eq.~\eqref{eq:bulk-action-minkowski} one obtains the equivalent second-derivative representation
\begin{align}
S_0[\varphi]
=
\frac{1}{2}\int_{\mathcal V}\! d^2x\;
\varphi(x)\,\Big[-\partial_{x_0}^2+\partial_{x_1}^2-m^2\Big]\varphi(x),
\end{align}
From Eq.~\eqref{eq:tau-ham-eq}, by separating real and imaginary parts,
\begin{align}
\varphi(x,\tau)=\varphi_R(x,\tau)+i\,\varphi_I(x,\tau),\qquad
\pi(x,\tau)=\pi_R(x,\tau)+i\,\pi_I(x,\tau),
\end{align}
and analogously for $\bar\varphi,\bar\pi$, one obtains in coordinate space the coupled system
\begin{align}
\ddot{\varphi}_R(x_0,x_1;\tau)
&=
\left(\frac{\partial^2}{\partial x_0^2}-\frac{\partial^2}{\partial x_1^2}+m^2\right)\varphi_I(x_0,x_1;\tau), \nonumber\\
\ddot{\varphi}_I(x_0,x_1;\tau)
&=
\left(\frac{\partial^2}{\partial x_0^2}-\frac{\partial^2}{\partial x_1^2}+m^2\right)\varphi_R(x_0,x_1;\tau).
\label{eq:coupled-real-imag-space-sf}
\end{align}
where $x=(x_0,x_1)$ and the differential operator acts on the $x$-dependence of the fields. In Fourier space, the same equations take the mode-decoupled form
\begin{align}
\ddot{\varphi}_R(k,\tau) &= \omega^2(k)\,\varphi_I(k,\tau), \nonumber\\
\ddot{\varphi}_I(k,\tau) &= \,\omega^2(k)\,\varphi_R(k,\tau),
\label{eq:coupled-real-imag-Fourier-sf}
\end{align}
with dispersion relation
\begin{align}
\omega^2(k)=-k_0^2+k_1^2+m^2,
\label{eq:dispersion-sf}
\end{align}
where $k=(k_0,k_1)$ denotes the appropriate lattice/continuum momenta associated to the chosen boundary conditions (for the explicit hybrid mode decomposition we refer to Apps.~\ref{app:mixed-lattice-implementation} and \ref{app:fixedfixed-terminal-implementation}).\par
As in the harmonic-oscillator case~\cite{Giachello_Harmonic}, the constraint selecting the stable manifold is most transparent mode by mode: for each $k$ the sign of $\omega^2(k)$ determines whether the coupled pair of Eq.~\eqref{eq:coupled-real-imag-Fourier-sf} corresponds to an elliptic (oscillatory) or hyperbolic (exponentially unstable) sector, and the stable manifold amounts to enforcing a fixed relative phase between $\varphi_R(k,\tau)$ and $\varphi_I(k,\tau)$, following the stable manifold of Eq.~\eqref{eq:sf-stable-surface0}. In practice, however, we aim at a numerical scheme that remains local in coordinate space (especially once interactions are included), and therefore we cannot enforce the mode-by-mode constraint by solving directly Eq.~\eqref{eq:coupled-real-imag-Fourier-sf}. The convenient strategy is to decompose the real part of the field into two components supported on complementary spectral regions, according to the sign of $\omega^2(k)$:
\begin{align}
\varphi_R(x,\tau)=\varphi_R^{O}(x,\tau)+\varphi_R^{E}(x,\tau),
\end{align}
with
\begin{align}
\varphi_R^{O}(x,\tau) &= \sum_{\{k~|~\omega^2(k)>0\}} e^{ik\cdot x}\,\varphi_R(k,\tau), \label{eq:phiO-def}\\
\varphi_R^{E}(x,\tau) &= \sum_{\{k~|~\omega^2(k)<0\}} e^{ik\cdot x}\,\varphi_R(k,\tau), \label{eq:phiE-def}
\end{align}
where the sums are understood as the appropriate discrete sums/integrals dictated by the boundary conditions. For the mixed temporal/spatial setting we refer to Apps.~\ref{app:mixed-lattice-implementation} and \ref{app:fixedfixed-terminal-implementation}, while for fully periodic boundaries $k_0=2\pi n_0/T$, $k_1=2\pi n_1/L$. Once the stable-manifold constraint is imposed in momentum space, the $\tau$-dynamics can be expressed entirely in terms of $\varphi_R^{O}$ and $\varphi_R^{E}$, yielding two \emph{decoupled} coordinate-space equations with opposite sign:
\begin{align}
\frac{d^2}{d\tau^2}\varphi_R^{O}(x,\tau)
&=
-\left(\partial_{x_0}^2-\partial_{x_1}^2+m^2\right)\varphi_R^{O}(x,\tau), \nonumber\\
\frac{d^2}{d\tau^2}\varphi_R^{E}(x,\tau)
&=
+\left(\partial_{x_0}^2-\partial_{x_1}^2+m^2\right)\varphi_R^{E}(x,\tau).
\label{eq:even-odd-dynamics-sf}
\end{align}
Formally, both equations are stable on their respective spectral supports: $\varphi_R^{O}$ is built only out of modes with $\omega^2>0$ and therefore evolves as a collection of oscillators, while $\varphi_R^{E}$ is built only out of modes with $\omega^2<0$ and evolves with the sign flipped so as to remove exponential growth. The crucial numerical issue is that a finite-step discretization of Eq.~\eqref{eq:even-odd-dynamics-sf} does not preserve the spectral support exactly. Even if $\varphi_R^{O}$ is initialized as a superposition of $\omega^2(k)>0$ modes only, during the $\tau$-integration small components with $\omega^2(k)<0$ are generated by roundoff and truncation errors: once present, these components grow exponentially and eventually dominate. The same phenomenon occurs for $\varphi_R^{E}$ with leakage into the opposite sector. We cure this instability by repeatedly projecting back onto the intended spectral subspaces.\par
On the lattice we discretize $\mathcal V$ on a $N_0\times N_1$ lattice,
\begin{align}
x_0^{(\ell_0)}=\ell_0 a,\qquad \ell_0=0,\dots,N_0-1,\qquad
x_1^{(\ell_1)}=\ell_1 a,\qquad \ell_1=0,\dots,N_1-1,
\end{align}
and, for periodic boundaries, we use the standard discrete Fourier representation
\begin{align}
\varphi_R(\ell_0,\ell_1;\tau)=\frac{1}{\sqrt{N_0N_1}}
\sum_{n_0=-N_0/2}^{N_0/2-1}\sum_{n_1=-N_1/2}^{N_1/2-1}
e^{i(k_0^{(n_0)}x_0^{(\ell_0)}+k_1^{(n_1)}x_1^{(\ell_1)})}\,\hat\varphi_R(n_0,n_1;\tau),
\end{align}
with lattice momenta $k_0^{(n_0)}=2\pi n_0/T$, $k_1^{(n_1)}=2\pi n_1/L$. The lattice dispersion entering the mode classification is
\begin{align}
\omega^2(n_0,n_1)
=
-\left(\frac{2}{a}\sin\frac{k_0^{(n_0)}a}{2}\right)^2
+
\left(\frac{2}{a}\sin\frac{k_1^{(n_1)}a}{2}\right)^2
+m^2,
\label{eq:lattice-dispersion-sf}
\end{align}
and we define the lattice even/odd components by splitting the Fourier sum according to the sign of $\omega^2(n_0,n_1)$, in direct analogy with Eqs.~\eqref{eq:phiO-def}, \eqref{eq:phiE-def}. The coordinate-space eqs.~\eqref{eq:even-odd-dynamics-sf} are then discretized with the standard nearest-neighbour Laplacian in the $(x_0,x_1)$ directions and integrated in $\tau$ with a leapfrog symplectic scheme. To keep the evolution on the stable manifold, we implement the following projection protocol:
\begin{enumerate}
\item[{\bf 1)}] At $\tau=0$ compute $\hat\varphi_R(n_0,n_1;0)$ and set to zero all Fourier components that do not belong to the target subspace:
\begin{align}
\hat\varphi_R^{O}(n_0,n_1;0)=0 \quad \forall~(n_0,n_1)~|~\omega^2(n_0,n_1)<0,\nonumber\\
\hat\varphi_R^{E}(n_0,n_1;0)=0 \quad \forall~(n_0,n_1)~|~\omega^2(n_0,n_1)>0.
\end{align}
\item[{\bf 2)}] Transform back to coordinate space and perform one leapfrog step for the lattice fields $\varphi_R^{O}(\ell_0,\ell_1;\tau)$ and $\varphi_R^{E}(\ell_0,\ell_1;\tau)$.
\item[{\bf 3)}] After the update, transform again to Fourier space and re-apply the projection,
\begin{align}
\hat\varphi_R^{O}(n_0,n_1;\tau+\delta\tau)=0 \quad \forall~(n_0,n_1)~|~\omega^2(n_0,n_1)<0,\nonumber\\
\hat\varphi_R^{E}(n_0,n_1;\tau+\delta\tau)=0 \quad \forall~(n_0,n_1)~|~\omega^2(n_0,n_1)>0,
\end{align}
and iterate from {\bf 2)}.
\end{enumerate}
Finally, once the stable-manifold constraint is enforced mode by mode, the imaginary part is reconstructed uniquely from $\varphi_R(n,\tau)$ (and similarly for the conjugate field), so that evolving $\varphi_R^{O}$ and $\varphi_R^{E}$ is sufficient to control the full complex field $\varphi=\varphi_R+i\varphi_I$ throughout the simulation.\par
The leapfrog algorithm keeps track of both $\varphi(\ell_0,\ell_1;\tau)$ and
$\pi(\ell_0,\ell_1;\tau)$, so that at any intrinsic time $\tau$ we can monitor the value of the generalized Hamiltonian
\begin{align}
\mathbb{H}_\tau[\pi,\bar\pi,\varphi,\bar\varphi]
=
\mathbb{K}_\tau[\pi,\bar\pi]+\mathbb{V}_\tau[\varphi,\bar\varphi].
\end{align}
For the free scalar field discretized on a $N_0\times N_1$ lattice, the kinetic part reads
\begin{align}
\mathbb{K}_\tau[\pi,\bar\pi]
=
\frac{1}{2}\sum_{\ell_0,\ell_1}\Big[\pi_R^2(\ell_0,\ell_1;\tau)+\pi_I^2(\ell_0,\ell_1;\tau)\Big],
\end{align}
while the potential term is the lattice version of $2\,\mathrm{Im}\,S_0$ and can be written in compact nearest-neighbour form as
\begin{align}
\mathbb{V}_\tau[\varphi,\bar\varphi]
&=
-\frac{1}{2}\sum_{\ell_0,\ell_1}\Big[
\varphi_R(\ell_0,\ell_1;\tau)\,\Delta\,\varphi_I(\ell_0,\ell_1;\tau)
+
\varphi_I(\ell_0,\ell_1;\tau)\,\Delta\,\varphi_R(\ell_0,\ell_1;\tau)
\nonumber\\
&\hspace{3.2cm}
+
m^2\,\varphi_R(\ell_0,\ell_1;\tau)\,\varphi_I(\ell_0,\ell_1;\tau)
\Big],
\label{eq:discrete-Hamiltonian-sf}
\end{align}
where $\Delta$ denotes the lattice discretization of the Minkowskian operator $(\partial_{x_0}^2-\partial_{x_1}^2)$ acting on the coordinate dependence,
\begin{align}
\Delta\,\psi(\ell_0,\ell_1)
=
\frac{\psi(\ell_0+1,\ell_1)-2\psi(\ell_0,\ell_1)+\psi(\ell_0-1,\ell_1)}{a^2}
-
\frac{\psi(\ell_0,\ell_1+1)-2\psi(\ell_0,\ell_1)+\psi(\ell_0,\ell_1-1)}{a^2},
\end{align}
with the understanding that the boundary prescription is implemented in the evaluation of the nearest neighbours.\par
By monitoring $\mathbb{H}_\tau$ along the trajectory we test the conservation of the total generalized action
$E=\mathbb{H}_\tau[\pi,\bar\pi,\varphi,\bar\varphi]$, which provides the primary diagnostic for the correct functioning of the symplectic integrator in presence of repeated spectral projections. In addition, in the regime where the long-$\tau$ evolution relaxes to a stationary state, we observe equipartition between kinetic and potential contributions,
\begin{align}
\Big\langle \mathbb{K}_\tau[\pi,\bar\pi]\Big\rangle
=
\Big\langle \mathbb{V}_\tau[\varphi,\bar\varphi]\Big\rangle,
\end{align}
where $\langle\cdot\rangle$ denotes an average over $\tau$-fluctuations in the stationary window. 
\begin{figure}[H]
\centering
	\includegraphics[width=0.75\linewidth]{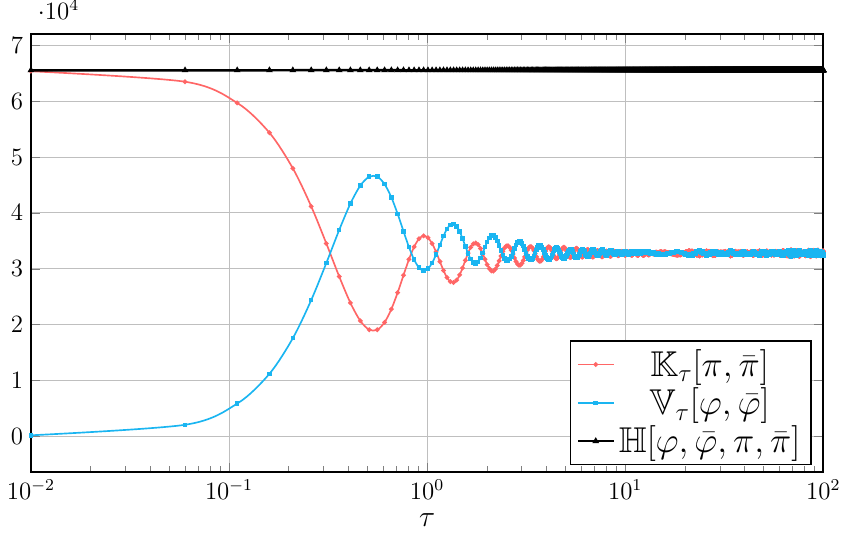}
	\caption{Time evolution of the kinetic ($E_{kin}=\mathbb{K}_\tau[\pi,\bar{\pi}]$) and potential ($E_{pot}=\mathbb{V}_\tau[\varphi,\bar\varphi]$) energy for a scalar field with $m=0.6$, $a=0.5$, with integration parameters $d\tau=0.01$, and total simulation time $\Delta \tau=100$. The system size is $N_0\times N_1=256\times 256=65536$, and we use periodic boundaries.}
	\label{fig:fig1-energy_conservation}
\end{figure}
%
%
%
%
\section{Constrained symplectic dynamics: numerical results}
\label{sec:numericalresults}
In this section we test the constrained symplectic algorithm against exact results for the free scalar field. The aim is twofold. First, we verify that the constrained $\tau$-flow is implemented correctly at the numerical level. Second, we show that the method gives direct access to genuinely Minkowskian observables, including real-time correlators and contact terms, within a stable deterministic evolution. In Sec.~\ref{subsec:two-point-periodic} we consider the fully periodic setup, where the functional integral computes a trace, and compare the measured two-point function with the exact lattice free-field prediction in both momentum and coordinate space. In Sec.~\ref{sec:lat-commutator} we test the equal-time canonical structure by extracting $\hbar$ from the commutator trace. In Sec.~\ref{sec:lat-dyson} we verify the Dyson--Schwinger identities derived in Sec.~\ref{sec:ds}, including the expected contact term at coincident points. Finally, in Sec.~\ref{sec:initial-value-dynamics} we turn to an initial-value formulation and study correlators anchored at a fixed preparation slice, which allow us to extract mode frequencies and visualize causal propagation from a localized initial pulse.
%
%
%
%
\subsection{Two-point correlation function (periodic boundaries)}
\label{subsec:two-point-periodic}
We first consider for simplicity the free scalar field on a Minkowski rectangular domain $\mathcal V=[0,T]\times[0,L]$ with \emph{periodic} boundary conditions in both directions,
\begin{equation}
  \varphi(x_0+T,x_1)=\varphi(x_0,x_1),\qquad
  \varphi(x_0,x_1+L)=\varphi(x_0,x_1),
  \label{eq:per-bc-field}
\end{equation}
and we study the time-ordered two-point function
\begin{equation}
  C(x_0,x_1)\;\equiv\;\big\langle \mathrm T\{\hat\varphi(x_0,x_1)\hat\varphi(0,0)\}\big\rangle_{\mathrm{P.B.}} .
  \label{eq:C-def-per}
\end{equation}
Periodicity in coordinate time turns the real-time functional integral into a trace. Denoting by $\hat H$ the field Hamiltonian and by $\hat U(T)\equiv e^{-i\hat H T/\hbar}$ the time-evolution operator, the periodic partition function and correlator are
\begin{equation}
  Z_{\mathrm{P.B.}}(T)\;=\;\int_{\mathrm{P.B.}}\!\mathcal D\varphi\;e^{\,\frac{i}{\hbar}S[\varphi]}
  \;=\;\Tr\!\big(e^{-\frac{i}{\hbar}\hat H T}\big),
  \label{eq:Z-trace-per}
\end{equation}
\begin{equation}
  C(x_0,x_1)
  \;=\;
  \frac{1}{Z_{\mathrm{P.B.}}(T)}\,
  \Tr\!\Big(e^{-i\hat H T/\hbar}\,\mathrm T\{\hat\varphi(x_0,x_1)\hat\varphi(0,0)\}\Big),
  \label{eq:C-trace-per}
\end{equation}
with the subscript $\mathrm{P.B.}$ indicating periodic boundary conditions. The trace representation makes explicit that the spectrum contributing to $C$ is discrete on the torus: inserting a complete set of energy--momentum eigenstates yields a sum over oscillatory harmonics whose frequencies are energy differences. The \emph{mass gap} is identified with the lowest one-particle energy at zero spatial momentum, i.e.\ the lightest pole at $k_1=0$, and it controls both the dominant temporal oscillation and the spatial correlation length.\par
In the numerical implementation we discretize $\mathcal V$ on an $N_0\times N_1$ lattice with spacing $a$,
\begin{align}
  &x_0^{(\ell_0)}=\ell_0 a,\qquad \ell_0=0,\dots,N_0-1,\\\nonumber
  &x_1^{(\ell_1)}=\ell_1 a,\qquad \ell_1=0,\dots,N_1-1,\\\nonumber
  & T=N_0 a,\quad L=N_1 a,
  \label{eq:lattice-geom-per}
\end{align}
and enforce periodicity $\varphi(\ell_0+N_0,\ell_1)=\varphi(\ell_0,\ell_1)$, $\varphi(\ell_0,\ell_1+N_1)=\varphi(\ell_0,\ell_1)$. It is then convenient to work in discrete Fourier space. We define
\begin{equation}
  \tilde\varphi(n_0,n_1)
  \;\equiv\;
  \frac{1}{\sqrt{N_0N_1}}
  \sum_{\ell_0=0}^{N_0-1}\sum_{\ell_1=0}^{N_1-1}
  e^{-i\left(k_0^{(n_0)}x_0^{(\ell_0)}+k_1^{(n_1)}x_1^{(\ell_1)}\right)}
  \,\varphi(\ell_0,\ell_1),
  \label{eq:FT-def-2d}
\end{equation}
with inverse
\begin{equation}
  \varphi(\ell_0,\ell_1)
  \;=\;
  \frac{1}{\sqrt{N_0N_1}}
  \sum_{n_0=-N_0/2}^{N_0/2-1}\sum_{n_1=-N_1/2}^{N_1/2-1}
  e^{\,i\left(k_0^{(n_0)}x_0^{(\ell_0)}+k_1^{(n_1)}x_1^{(\ell_1)}\right)}
  \,\tilde\varphi(n_0,n_1),
  \label{eq:invFT-def-2d}
\end{equation}
where the discrete Minkowskian momenta are
\begin{align}
  k_0^{(n_0)}=\frac{2\pi n_0}{T},\qquad
  n_0=-\frac{N_0}{2},\dots,\frac{N_0}{2}-1,\\\nonumber	
  k_1^{(n_1)}=\frac{2\pi n_1}{L},\qquad
  n_1=-\frac{N_1}{2},\dots,\frac{N_1}{2}-1.
  \label{eq:kn0-kn1}
\end{align}
For the free theory the momentum-space correlator is diagonal,
\begin{equation}
  \Big\langle \tilde\varphi(n_0,n_1)\,\tilde\varphi(-n_0,-n_1)\Big\rangle_{\mathrm{P.B.}}
  \;=\;\widetilde C_{\mathrm{P.B.}}\!\left(k_0^{(n_0)},k_1^{(n_1)}\right),
  \label{eq:diag-prop}
\end{equation}
and, for the standard nearest-neighbor lattice discretization of the quadratic kernel, its analytic form is
\begin{equation}
  \widetilde C_{\mathrm{P.B.}}\!\left(k^{(n_0)}_0,k^{(n_1)}_1\right)
  \;=\;
  \frac{i}{\frac{4}{a^2}\sin^2\!\left(\frac{k_0^{(n_0)} a}{2}\right)
         -\frac{4}{a^2}\sin^2\!\left(\frac{k_1^{(n_1)} a}{2}\right)
         -m^2}\,.
  \label{eq:lat-prop-mink-2d}
\end{equation}
In particular, the mass gap is the location of the lightest pole at zero
spatial momentum,
\begin{equation}
  k_1=0:\qquad
  \widetilde C_{\mathrm{P.B.}}(k_0,0)\;=\;\frac{i}{\frac{4}{a^2}\sin^2\!\left(\frac{k_0^{(n_0)} a}{2}\right)-m^2},
  \label{eq:lat-prop-k10}
\end{equation}
which in the continuum limit reduces to the familiar pole at $k_0=\pm m$.\par
\begin{figure}[H]
  \centering
  \includegraphics[width=1\linewidth]{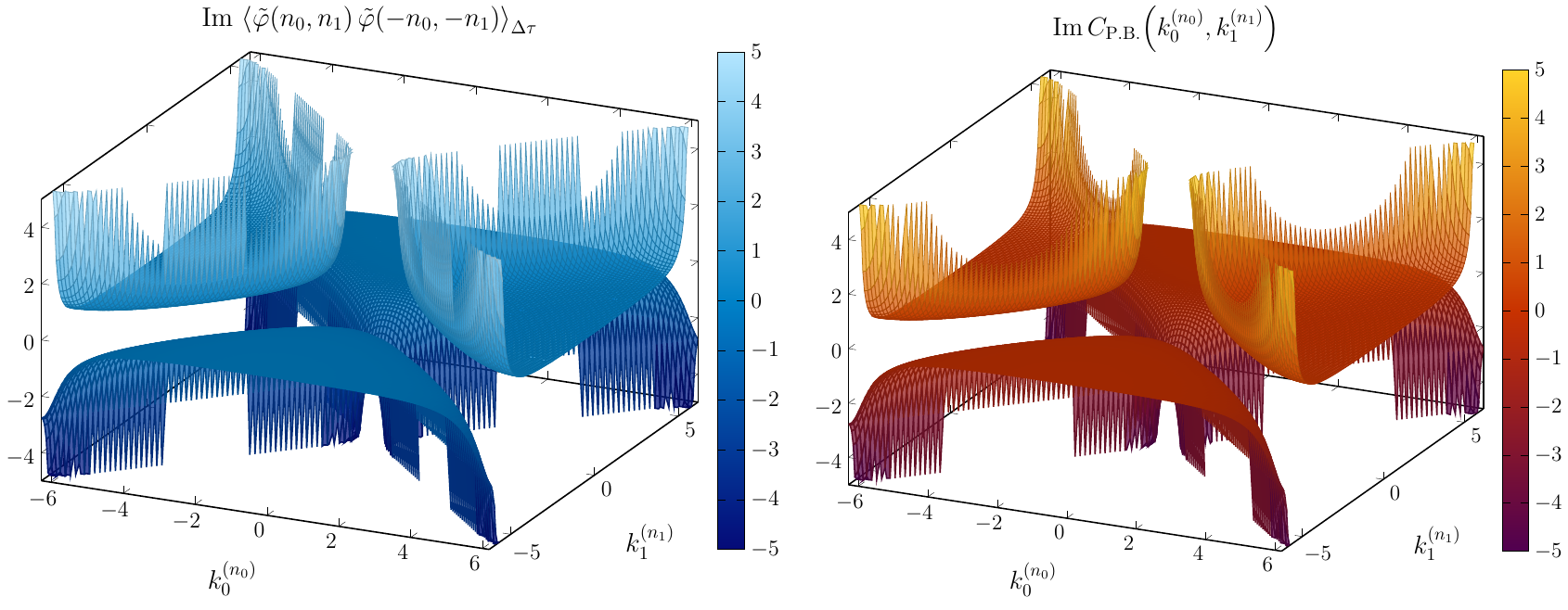}
  \caption{
    Momentum-space propagator with periodic boundary conditions.
    \textbf{Right:} analytic prediction for the free scalar propagator
    $\Im\,\widetilde C_{\mathrm{P.B.}}\!\left(k_0^{(n_0)},k_1^{(n_1)}\right)$,
    cf.~Eq.~\eqref{eq:lat-prop-mink-2d}.
    \textbf{Left:} numerical estimate from symplectic-quantization dynamics,
    $\Im\,\langle \tilde\varphi(n_0,n_1)\tilde\varphi(-n_0,-n_1)\rangle_{\Delta\tau}$,
    showing perfect agreement with the theoretical prediction. Simulation parameters: mass $m=0.6$, lattice size $a=0.5$ on a $N_0\times N_1=256\times256=65536$ square lattice.
  }
  \label{fig:propagator-momentum-periodic}
\end{figure}

In symplectic-quantization simulations we estimate the correlator as a \emph{dynamical} intrinsic-time average over a finite window $\Delta\tau$, which we denote by $\langle\cdots\rangle_{\Delta\tau}$, with convergence $\langle\cdots\rangle_{\Delta\tau}\to\langle\cdots\rangle_{\mathrm{P.B.}}$ in the limit $\Delta\tau\to\infty$. Writing each Fourier mode as
\begin{equation}
  \tilde\varphi(n_0,n_1)\;\equiv\;\tilde\varphi_R(n_0,n_1)+i\,\tilde\varphi_I(n_0,n_1),
  \qquad \tilde\varphi_{R,I}(n_0,n_1)\in\mathbb R,
  \label{eq:phiRphiI}
\end{equation}
we have
\begin{align}
  \Big\langle \tilde\varphi(n_0,n_1)\,\tilde\varphi(-n_0,-n_1)\Big\rangle_{\Delta\tau}
  =&
  \Big\langle
    \tilde\varphi_R(n_0,n_1)\tilde\varphi_R(-n_0,-n_1)
   -\tilde\varphi_I(n_0,n_1)\tilde\varphi_I(-n_0,-n_1)
  \Big\rangle_{\Delta\tau}
  \nonumber\\
  &
  +\,i\Big\langle
    \tilde\varphi_R(n_0,n_1)\tilde\varphi_I(-n_0,-n_1)
   +\tilde\varphi_I(n_0,n_1)\tilde\varphi_R(-n_0,-n_1)
  \Big\rangle_{\Delta\tau}.
  \label{eq:modecorr-RI}
\end{align}
Therefore, in the large-$\Delta\tau$ limit, to recover the correct theoretical expectation value of Eq.~\eqref{eq:lat-prop-mink-2d}, we expect the real part to be suppressed while the imaginary part converges to the analytic prediction \eqref{eq:lat-prop-mink-2d},
\begin{align}
  \Re\,\Big\langle \tilde\varphi(n_0,n_1)\,\tilde\varphi(-n_0,-n_1)\Big\rangle_{\Delta\tau}
  &\xrightarrow[\Delta\tau\to\infty]{}0,
  \label{eq:re-vanish}\\
  \Im\,\Big\langle \tilde\varphi(n_0,n_1)\,\tilde\varphi(-n_0,-n_1)\Big\rangle_{\Delta\tau}
  &\xrightarrow[\Delta\tau\to\infty]{}
  \frac{1}{\frac{4}{a^2}\sin^2\!\left(\frac{k_0^{(n_0)} a}{2}\right)
         -\frac{4}{a^2}\sin^2\!\left(\frac{k_1^{(n_1)} a}{2}\right)
         -m^2}\,.
  \label{eq:im-conv-prop}
\end{align}
This is precisely what happens in the numerical simulations. Figure~\ref{fig:propagator-momentum-periodic} shows excellent agreement between the analytic prediction and the numerical CSQ measurement. The pole structure in $(k_0,k_1)$ is directly visible and the gap is read off from the lightest singularity at $k_1=0$.
Once $\widetilde C(k_0^{(n_0)},k_1^{(n_1)})$ is obtained, the full coordinate-space correlator is obtained by the inverse discrete transform,
\begin{align}
  C(\ell_0,\ell_1)
  &\;\equiv\;\big\langle \varphi(\ell_0,\ell_1)\varphi(0,0)\big\rangle_{\mathrm{P.B.}}\\\nonumber
  &\;=\;
  \frac{1}{N_0N_1}
  \sum_{n_0,n_1}
  e^{\,i\left(k_0^{(n_0)}x_0^{(\ell_0)}+k_1^{(n_1)}x_1^{(\ell_1)}\right)}
  \,\Big\langle \tilde\varphi(n_0,n_1)\tilde\varphi(-n_0,-n_1)\Big\rangle_{\mathrm{P.B.}}.
  \label{eq:C-invFT-lattice}
\end{align}
In practice, however, directly analyzing the resulting $C(\ell_0,\ell_1)$ on the torus is often unwieldy because periodic images in both directions compete and the notion of a single ``long-distance'' regime is obscured. For this reason we emphasize instead two \emph{marginalized} correlators, obtained by summing over one coordinate. Operationally, these sums implement discrete orthogonality relations and therefore act as Kronecker/Dirac projectors selecting either the $k_1=0$ (zero spatial momentum) or the $k_0=0$ (zero frequency) slice of the same momentum-space propagator.\par
The spatially marginalized correlator projects onto \emph{zero spatial momentum} and isolates the gap as a temporal harmonic:
\begin{equation}
  C_T(\ell_0)
  \;\equiv\;
  \frac{1}{N_1}\sum_{\ell_1=0}^{N_1-1} C(\ell_0,\ell_1)
  \;=\;
  \frac{1}{N_0}\sum_{n_0=-N_0/2}^{N_0/2-1}
  e^{\,i k_0^{(n_0)} x_0^{(\ell_0)}}\;
  \Big\langle \tilde\varphi(n_0,0)\tilde\varphi(-n_0,0)\Big\rangle_{\mathrm{P.B.}}.
  \label{eq:CT-proj-2d}
\end{equation}
Indeed, inserting the inverse Fourier representation of $C(\ell_0,\ell_1)$ and using the discrete identity
\begin{equation}
  \frac{1}{N_1}\sum_{\ell_1=0}^{N_1-1}e^{\,i k_1^{(n_1)}x_1^{(\ell_1)}}
  \;=\;\delta_{n_1,0},
  \qquad
  n_1=-\frac{N_1}{2},\dots,\frac{N_1}{2}-1,
  \label{eq:delta-k1}
\end{equation}
one sees that the spatial sum is a lattice delta-function selecting $n_1=0$, i.e.\ $k_1=0$. Therefore $C_T$ is nothing but the inverse transform of the $k_1=0$ slice of the momentum-space propagator. In the free theory this slice has its lightest pole at $k_0=\pm m$, so $C_T(\ell_0)$ exhibits a clean oscillatory signal whose fundamental frequency is the mass gap.
\begin{figure}[H]
  \centering
  \includegraphics[width=0.8\linewidth]{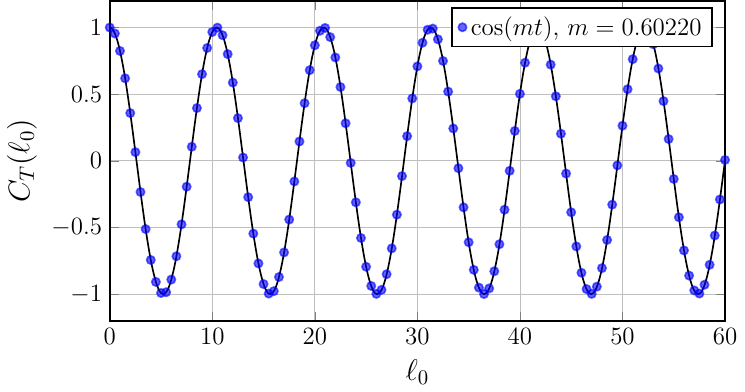}
\caption{
Zero-spatial-momentum correlator used to extract the mass gap. 
The plotted quantity is $C_T(\ell_0)$, obtained by averaging the two-point function over the spatial coordinate, this projects the propagator onto the $k_1=0$ sector. 
The data oscillate in physical time, and the fitted oscillation frequency gives the propagator pole, i.e. the mass gap $m$. 
Simulation parameters: $m=0.6$, $a=0.5$, and $N_0\times N_1=256\times256$.
}
  \label{fig:CT-marginal-periodic}
\end{figure}
Similarly, the temporal marginalization projects onto \emph{zero frequency} and isolates the spatial decay length:
\begin{equation}
  C_L(\ell_1)
  \;\equiv\;
  \frac{1}{N_0}\sum_{\ell_0=0}^{N_0-1} C(\ell_0,\ell_1)
  \;=\;
  \frac{1}{N_1}\sum_{n_1=-N_1/2}^{N_1/2-1}
  e^{\,i k_1^{(n_1)} x_1^{(\ell_1)}}\;
  \Big\langle \tilde\varphi(0,n_1)\tilde\varphi(0,-n_1)\Big\rangle_{\mathrm{P.B.}}.
  \label{eq:CL-proj-2d}
\end{equation}
Here the average over $\ell_0$ enforces $n_0=0$ by the discrete orthogonality relation
\begin{equation}
  \frac{1}{N_0}\sum_{\ell_0=0}^{N_0-1}e^{\,i k_0^{(n_0)}x_0^{(\ell_0)}}
  \;=\;\delta_{n_0,0},
  \qquad
  n_0=-\frac{N_0}{2},\dots,\frac{N_0}{2}-1,
  \label{eq:delta-k0}
\end{equation}
so $C_L$ is the inverse transform of the $k_0=0$ slice of the propagator. For the free scalar one has $\widetilde C(0,k_1)=i/(-k_1^2-m^2)$, and the corresponding coordinate profile is governed by the same scale $m$, producing an exponential decay in $|x_1|$.
\begin{figure}[H]
  \centering
  \includegraphics[width=0.8\linewidth]{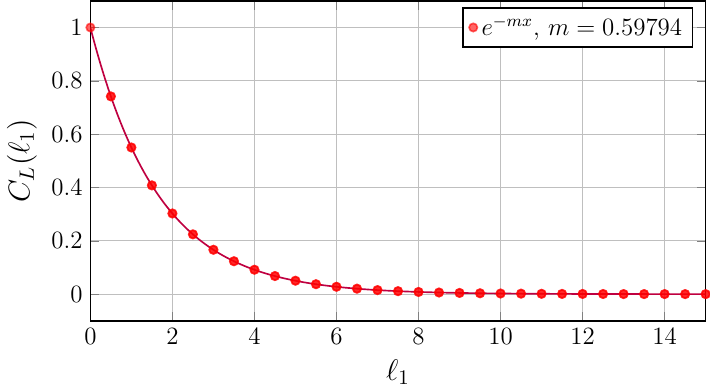}
  \caption{
Zero-frequency correlator used to measure the same mass scale from spatial decay. 
The plotted quantity is $C_L(\ell_1)$, obtained by averaging the two-point function over the time coordinate, this projects the propagator onto the $k_0=0$ sector. 
The resulting spatial profile decays exponentially, and the decay rate is controlled by the mass gap $m$. 
Simulation parameters: $m=0.6$, $a=0.5$, and $N_0\times N_1=256\times256$.
}
  \label{fig:CL-marginal-periodic}
\end{figure}
In summary, periodic boundaries make the full two-dimensional correlator $C_{\mathrm{P.B.}}(\ell_0,\ell_1)$ difficult to interpret directly, since temporal and spatial images compete on the torus. The momentum-space propagator, instead, exposes the gap unambiguously through its lightest pole at $k_1=0$. The two marginalizations of Eqs.~\eqref{eq:CT-proj-2d}--\eqref{eq:CL-proj-2d} are the coordinate-space counterparts of this statement: the lattice sums act as discrete delta functions, projecting the data onto the $k_1=0$ and $k_0=0$ slices of $\widetilde C_{\mathrm{P.B.}}(k_0,k_1)$. As a result, $C_T(\ell_0)$ displays a clean single-frequency oscillation and $C_L(\ell_1)$ a clean exponential decay, and in both cases a one-parameter fit returns the same input value of the mass gap $m$.

%
%
%
%
\subsection{Uncertainty principle (periodic boundaries)}
\label{sec:lat-commutator}
A fundamental consistency check for any deterministic fictitious dynamics that aims at reproducing Minkowskian quantum field theory is the ability to recover the canonical equal-time commutation relations and, hence, the uncertainty principle. For a real scalar field the canonical algebra reads
\begin{equation}
  \big[\hat\varphi(x_0,x_1),\hat p(x_0,y_1)\big]
  \;=\; i\hbar\,\delta(x_1-y_1),
  \label{eq:canonical-comm-rel-field}
\end{equation}
where $\hat p(x_0,x_1)\equiv \partial_{x_0}\hat\varphi(x_0,x_1)$ is the momentum canonically conjugate to $\hat\varphi$ with respect to the physical time $x_0$. On a spatial lattice with periodic boundary conditions, this becomes a Kronecker delta,
\begin{equation}
  \big[\hat\varphi(x_0,\ell_1),\hat p(x_0,\ell_1')\big]
  \;=\; i\hbar\,\frac{1}{a}\,\delta_{\ell_1,\ell_1'}\qquad
  (\text{periodic in }\ell_1),
  \label{eq:canonical-comm-rel-lat}
\end{equation}
with $a$ the lattice spacing. In the following we show how the symplectic-quantization dynamics reproduces Eq.~\eqref{eq:canonical-comm-rel-lat} in the form of a parameter-free numerical determination of $\hbar$.\par
With periodic boundary conditions in coordinate time the real-time path integral computes a trace, and we consider the normalized, spatially averaged commutator trace
\begin{equation}
  c(\hbar)=\frac{1}{N_1 Z_{\rm P.B.}}\sum_{\ell_1}\Tr\!\left(e^{-i\hat H T/\hbar}\,[\hat\varphi(x_0,\ell_1),\hat p(x_0,\ell_1)]\right)
  \qquad
  Z_{\mathrm{P.B.}}=\Tr\!\big(e^{-\frac{i}{\hbar}\hat H T}\big),
  \label{eq:trace-hbar-field}
\end{equation}
where $\hat H$ is the (free) field Hamiltonian and $T$ is the temporal extent of the periodic box. Using Eq.~\eqref{eq:canonical-comm-rel-lat} at $\ell_1'=\ell_1$ gives
\begin{equation}
  c(\hbar)= i\hbar\,\frac{1}{a}\,,
  \qquad\Longleftrightarrow\qquad
  -i\,a\,c(\hbar)=\hbar,
  \label{eq:c-hbar-expect}
\end{equation}
which is the form we will directly compare to the numerical estimator.\par
To represent the operator ordering in the functional integral we proceed as in \cite{Giachello_Harmonic}: although the two operators in the commutator are labeled by the same coordinate time $x_0$, the rightmost operator acts infinitesimally earlier than the leftmost one. We therefore introduce $x_0^{(\pm)}$ with $x_0^{(+)}-x_0\ll 1$ and $x_0-x_0^{(-)}\ll 1$ and write the trace commutator as a difference of two insertions at slightly shifted times,
\begin{equation}
  c(\hbar)
  \;=\;
  \frac{1}{N_1\,Z_{\mathrm{P.B.}}}\sum_{\ell_1=0}^{N_1-1}
  \int_{\mathrm{P.B.}}\!\mathcal D\varphi\;
  e^{\frac{i}{\hbar}S[\varphi]}\,
  \Big[\varphi(x_0,\ell_1)\,p(x_0^{(-)},\ell_1)-\varphi(x_0,\ell_1)\,p(x_0^{(+)},\ell_1)\Big],
  \label{eq:c-hbar-pathint-field}
\end{equation}
where $p=\partial_{x_0}\varphi$ at the level of fields.\par
We now discretize Minkowskian time and space on an $N_0\times N_1$ periodic lattice with spacing $a$ and sites $(\ell_0,\ell_1)$, $\ell_0=0,\dots,N_0-1$, $\ell_1=0,\dots,N_1-1$. We denote $\varphi(\ell_0,\ell_1)\equiv \varphi(x_0^{(\ell_0)},x_1^{(\ell_1)})$ with $x_0^{(\ell_0)}=\ell_0 a$, $x_1^{(\ell_1)}=\ell_1 a$, and we implement the infinitesimal shifts by midpoint momenta. A convenient lattice definition of the conjugate momentum at half time-steps is
\begin{equation}
  p\!\left(x_0^{(\ell_0-\frac12)},\ell_1\right)
  \;\equiv\;
  \frac{\varphi(\ell_0,\ell_1)-\varphi(\ell_0-1,\ell_1)}{a},
  \qquad
  p\!\left(x_0^{(\ell_0+\frac12)},\ell_1\right)
  \;\equiv\;
  \frac{\varphi(\ell_0+1,\ell_1)-\varphi(\ell_0,\ell_1)}{a},
  \label{eq:pi-midpoint-field}
\end{equation}
with periodic indexing in $\ell_0$. Inserting Eq.~\eqref{eq:pi-midpoint-field} into Eq.~\eqref{eq:c-hbar-pathint-field} and using translation invariance yields a trace commutator entirely in terms of local and nearest-neighbour correlators in the $\ell_0$ direction,
\begin{align}
  c(\hbar)
  &=
  \frac{1}{N_0N_1}\sum_{\ell_0=0}^{N_0-1}\sum_{\ell_1=0}^{N_1-1}
  \Bigg\langle
    \varphi(\ell_0,\ell_1)\,
    \Big[
      \frac{\varphi(\ell_0,\ell_1)-\varphi(\ell_0-1,\ell_1)}{a}
      -
      \frac{\varphi(\ell_0+1,\ell_1)-\varphi(\ell_0,\ell_1)}{a}
    \Big]
  \Bigg\rangle_{\mathrm{P.B.}}
  \nonumber\\[1mm]
  &=
  \frac{1}{a}\Big[
    2\,\big\langle \varphi^2(\ell_0,\ell_1)\big\rangle_{\mathrm{P.B.}}
    -
    \big\langle \varphi(\ell_0,\ell_1)\varphi(\ell_0-1,\ell_1)\big\rangle_{\mathrm{P.B.}}
    -
    \big\langle \varphi(\ell_0,\ell_1)\varphi(\ell_0+1,\ell_1)\big\rangle_{\mathrm{P.B.}}
  \Big],
  \label{eq:comm-correlators-field}
\end{align}
where the final line is independent of the chosen site $(\ell_0,\ell_1)$ by translation invariance.\par
To express Eq.~\eqref{eq:comm-correlators-field} in terms of the variables that are actually evolved in our constrained dynamics, we use the stable-manifold parametrization introduced in Sec.~\ref{sec:numerical-algorithm}. In the numerical implementation, the complexified lattice field is \emph{not} stored as independent real and imaginary parts; rather, it is reconstructed at each intrinsic time $\tau$ from the even/odd stable-manifold components $\varphi_R^{E/O}$ as
\begin{equation}
  \varphi(\ell;\tau)
  \;=\;
  (1+i)\,\varphi_R^{E}(\ell;\tau)\;+\;(1-i)\,\varphi_R^{O}(\ell;\tau),
  \qquad \ell=(\ell_0,\ell_1),
  \label{eq:phi-reconstruct-EO}
\end{equation}
with the even/odd decomposition understood with respect to the temporal lattice index $\ell_0\mapsto -\ell_0$ (mod $N_0$). Equivalently, separating the real and imaginary parts of \eqref{eq:phi-reconstruct-EO} gives the explicit map
\begin{equation}
  \varphi_R(\ell,\tau)=\varphi_R^{E}(\ell,\tau)+\varphi_R^{O}(\ell,\tau),
  \qquad
  \varphi_I(\ell,\tau)=\varphi_R^{E}(\ell,\tau)-\varphi_R^{O}(\ell,\tau),
  \label{eq:phi-RI-from-EO}
\end{equation}
which is the relation used when rewriting the commutator estimator in the
$E/O$ language.
The constraints defining the stable integration manifold imply, for any two-point function,
\begin{equation}
  \Im\Big\langle \varphi(\ell_0,\ell_1)\,\varphi(\ell_0',\ell_1')\Big\rangle_{\mathrm{P.B.}}
  =
  2\Big[
    \Big\langle \varphi_R^{E}(\ell_0,\ell_1)\,\varphi_R^{E}(\ell_0',\ell_1')\Big\rangle_{\mathrm{P.B.}}
    -
    \Big\langle \varphi_R^{O}(\ell_0,\ell_1)\,\varphi_R^{O}(\ell_0',\ell_1')\Big\rangle_{\mathrm{P.B.}}
  \Big],
  \label{eq:Imphiphi-evenodd}
\end{equation}
so that, since the commutator trace is purely imaginary, \eqref{eq:comm-correlators-field} can be rewritten as
\begin{align}
  -i\,c(\hbar)
  =
  \frac{1}{a}\Big[&
    2\,\Im\big\langle \varphi(\ell_0,\ell_1)\varphi(\ell_0,\ell_1)\big\rangle_{\mathrm{P.B.}}
    -
    \Im\big\langle \varphi(\ell_0,\ell_1)\varphi(\ell_0-1,\ell_1)\big\rangle_{\mathrm{P.B.}}+\\\nonumber
    &-
    \Im\big\langle \varphi(\ell_0,\ell_1)\varphi(\ell_0+1,\ell_1)\big\rangle_{\mathrm{P.B.}}
  \Big],
  \label{eq:comm-correlators-field-im}
\end{align}
and substituting \eqref{eq:Imphiphi-evenodd} term by term yields an expression entirely in terms of correlators of $\varphi_R^{E}$ and $\varphi_R^{O}$,
\begin{equation}
\begin{aligned}
  -i\,c(\hbar)
  &= \frac{1}{a}\Bigg\{
      4\Big[
        \big\langle \varphi_R^{E}(\ell_0,\ell_1)\varphi_R^{E}(\ell_0,\ell_1)\big\rangle_{\mathrm{P.B.}}
        -
        \big\langle \varphi_R^{O}(\ell_0,\ell_1)\varphi_R^{O}(\ell_0,\ell_1)\big\rangle_{\mathrm{P.B.}}
      \Big]
\\
  &\hspace{1.25cm}
      -2\Big[
        \big\langle \varphi_R^{E}(\ell_0,\ell_1)\varphi_R^{E}(\ell_0-1,\ell_1)\big\rangle_{\mathrm{P.B.}}
        -
        \big\langle \varphi_R^{O}(\ell_0,\ell_1)\varphi_R^{O}(\ell_0-1,\ell_1)\big\rangle_{\mathrm{P.B.}}
      \Big]
\\
  &\hspace{1.25cm}
      -2\Big[
        \big\langle \varphi_R^{E}(\ell_0,\ell_1)\varphi_R^{E}(\ell_0+1,\ell_1)\big\rangle_{\mathrm{P.B.}}
        -
        \big\langle \varphi_R^{O}(\ell_0,\ell_1)\varphi_R^{O}(\ell_0+1,\ell_1)\big\rangle_{\mathrm{P.B.}}
      \Big]
    \Bigg\},
\end{aligned}
  \label{eq:c-hbar-phiREphiRO-final}
\end{equation}
which is the form directly implemented in the numerical estimator.\par
\begin{figure}[H]
  \centering
  \includegraphics[width=0.70\textwidth]{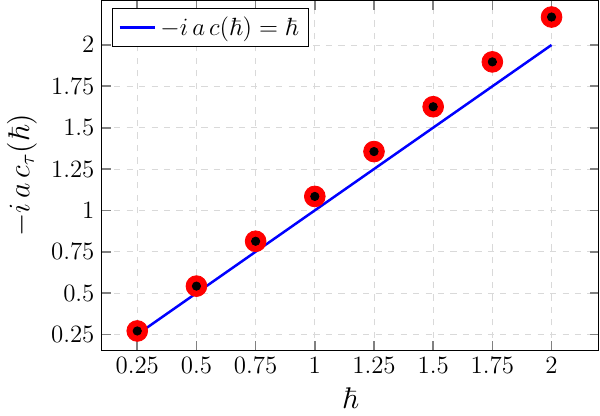}
    \caption{
Numerical test of the equal-time canonical commutator. 
The points show the quantity $-i\,a\,c_\tau(\hbar)$ measured from the intrinsic-time average in Eq.~\eqref{eq:comm-estimator-field} for different input values of $\hbar$. 
The solid line is the exact prediction $-i\,a\,c_\tau(\hbar)=\hbar$, with no fit parameters. 
The agreement shows that the constrained symplectic dynamics reproduces the canonical commutation relation. 
Simulation parameters: $m=2.5$, $a=0.1$, and $N_0\times N_1=4096$.
}
  \label{fig:comm-vs-hbar-field}
\end{figure}

Accordingly, in the simulations we measure the intrinsic-time average of the same combination of correlators along the stationary symplectic evolution and we average over all lattice sites,
\begin{align}
  c_\tau(\hbar)
  \;\equiv\;
  \frac{1}{N_0N_1\,\Delta\tau}
  &\sum_{\ell_0=0}^{N_0-1}\sum_{\ell_1=0}^{N_1-1}
  \int_{\tau_0}^{\tau_0+\Delta\tau}\! d\tau\;
  \frac{1}{a}\Bigg\{
  \,4\Big[
        \big(\varphi_R^{E}(\ell_0,\ell_1;\tau)\big)^2
        -
        \big(\varphi_R^{O}(\ell_0,\ell_1;\tau)\big)^2
      \Big]
\nonumber\\
  &-2\Big[
        \varphi_R^{E}(\ell_0,\ell_1;\tau)\,\varphi_R^{E}(\ell_0-1,\ell_1;\tau)
        -
        \varphi_R^{O}(\ell_0,\ell_1;\tau)\,\varphi_R^{O}(\ell_0-1,\ell_1;\tau)
      \Big]
\nonumber\\
  &-2\Big[
        \varphi_R^{E}(\ell_0,\ell_1;\tau)\,\varphi_R^{E}(\ell_0+1,\ell_1;\tau)
        -
        \varphi_R^{O}(\ell_0,\ell_1;\tau)\,\varphi_R^{O}(\ell_0+1,\ell_1;\tau)
      \Big]
  \Bigg\},
  \label{eq:comm-estimator-field}
\end{align}
with periodic indexing in $\ell_0$ and $\ell_1$. In the stationary regime and for large averaging windows, the equivalence between intrinsic-time dynamical averages and the periodic real-time path integral implies
\begin{equation}
  \lim_{\Delta\tau\to\infty} c_\tau(\hbar)= -i\,c(\hbar).
  \label{eq:comm-conv-field}
\end{equation}
Combining this with Eq.~\eqref{eq:c-hbar-expect} yields the parameter-free prediction
\begin{equation}
  \lim_{\Delta\tau\to\infty}\big(-i\,a\,c_\tau(\hbar)\big)=\hbar,
  \label{eq:comm-final-pred}
\end{equation}
namely $-i\,a\,c_\tau(\hbar)$ must fall on a straight line of unit slope as a function of $\hbar$. In Fig.~\ref{fig:comm-vs-hbar-field} we report the values of $-i\,a\,c_\tau(\hbar)$ extracted from Eq.~\eqref{eq:comm-estimator-field} for different choices of $\hbar$. The results are consistent with $-i\,a\,c_\tau(\hbar)=\hbar$ and with no adjustable parameters, confirming that the constrained symplectic dynamics reproduces the canonical field commutation relations and therefore the uncertainty principle.
%
%
%
%
\subsection{Dyson-Schwinger numerical verification (periodic boundaries)}
\label{sec:lat-dyson}
\begin{figure}[H]
  \centering
  \includegraphics[width=0.98\linewidth]{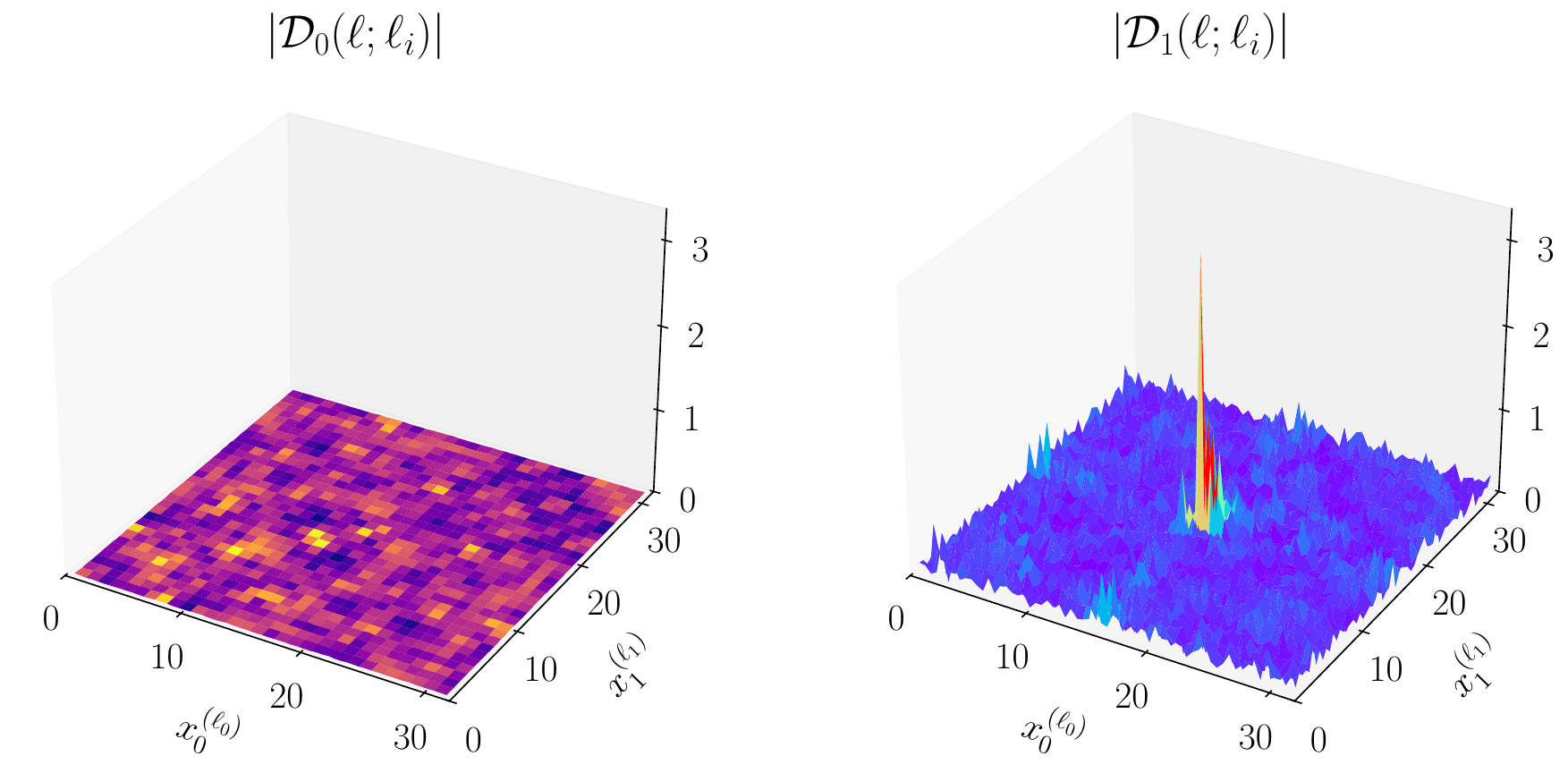}
  \caption{
Numerical verification of the Dyson--Schwinger identities on the periodic lattice. 
\textbf{Left}: the magnitude of the equation-of-motion estimator $|\mathcal D_0(\ell)|$, which is expected to vanish at every lattice site. 
\textbf{Right}: the magnitude of the one-insertion estimator $|\mathcal D_1(\ell;\ell_i)|$, which is expected to vanish everywhere except at the insertion point $\ell=\ell_i$. 
The localized peak is the lattice contact term predicted by Eq.~\eqref{eq:D1-pred}. 
Simulation parameters: $m=0.6$, $a=0.5$, and $N_0\times N_1=16384$.
}
  \label{fig:dyson-3d-periodic}
\end{figure}

A further non-trivial validation of the constrained symplectic quantization algorithm is obtained by \emph{numerically} testing the Dyson-Schwinger (DS) identities derived in Sec.~\ref{sec:ds}. In the fully periodic setup, the equivalence between stationary intrinsic-time averages along the constrained symplectic flow and the corresponding Minkowskian real-time path-integral expectation values implies that the DS relations must be reproduced directly by our long-$\tau$ measurements. For the free scalar field this yields, in particular, the continuum identities
\begin{equation}
  \Big\langle \frac{\delta S[\varphi]}{\delta \varphi(x)} \Big\rangle_{\mathrm{P.B.}} = 0,
  \label{eq:DS-1-cont}
\end{equation}
and, with a single field insertion,
\begin{equation}
  \Big\langle \frac{\delta S[\varphi]}{\delta \varphi(x)}\,\varphi(x_i) \Big\rangle_{\mathrm{P.B.}}
  = i\hbar\,\delta^{(2)}(x-x_i).
  \label{eq:DS-2-cont}
\end{equation}
We discretize $\mathcal V=[0,T]\times[0,L]$ on an $N_0\times N_1$ lattice with spacing $a$ and periodic indexing in both directions. Lattice sites are $\ell=(\ell_0,\ell_1)$ and we choose an insertion site $\ell_i=(\ell_{0,i},\ell_{1,i})$. The continuum delta distribution is represented as
\begin{equation}
  \delta^{(2)}(x-x_i)\;\longrightarrow\;\frac{1}{a^2}\,\delta_{\ell,\ell_i},
  \qquad
  \delta_{\ell,\ell_i}\equiv \delta_{\ell_0,\ell_{0,i}}\delta_{\ell_1,\ell_{1,i}}.
  \label{eq:delta2-lat}
\end{equation}
Accordingly, the DS identities become
\begin{align}
  \Big\langle \frac{\delta S[\varphi]}{\delta \varphi(\ell)} \Big\rangle_{\mathrm{P.B.}} &= 0,
  \label{eq:DS-1-lat}\\[2pt]
  \Big\langle \frac{\delta S[\varphi]}{\delta \varphi(\ell)}\,\varphi(\ell_i) \Big\rangle_{\mathrm{P.B.}}
  &= i\hbar\,\frac{1}{a^2}\,\delta_{\ell,\ell_i}.
  \label{eq:DS-2-lat}
\end{align}
For the free Minkowskian quadratic action, the lattice functional derivative is the local Klein--Gordon operator acting on $\varphi$,
\begin{equation}
\begin{aligned}
  \frac{\delta S[\varphi]}{\delta \varphi(\ell_0,\ell_1)}
  \;=\;
  -\Bigg[
  \frac{\varphi_{\ell_0+1,\ell_1}-2\varphi_{\ell_0,\ell_1}+\varphi_{\ell_0-1,\ell_1}}{a^2}
  \;-\;
  \frac{\varphi_{\ell_0,\ell_1+1}-2\varphi_{\ell_0,\ell_1}+\varphi_{\ell_0,\ell_1-1}}{a^2}
  \;+\;
  m^2\,\varphi_{\ell_0,\ell_1}
  \Bigg],
\end{aligned}
  \label{eq:dS-dphi-minkowski-lat-components}
\end{equation}
In our numerical implementation the complex field is reconstructed from the stable-manifold variables introduced in Sec.~\ref{sec:numerical-algorithm},
\begin{equation}
  \varphi(\ell;\tau)
  \;=\;
  (1+i)\,\varphi_R^{E}(\ell;\tau)\;+\;(1-i)\,\varphi_R^{O}(\ell;\tau),
  \label{eq:phi-reconstruct-EO-again}
\end{equation}
and we therefore define the pointwise $E/O$ force components
\begin{equation}
\begin{aligned}
  F^{E}(\ell;\tau)\;\equiv\;
  &\frac{\varphi_{R}^{E}(\ell_0+1,\ell_1;\tau)-2\varphi_{R}^{E}(\ell_0,\ell_1;\tau)+\varphi_{R}^{E}(\ell_0-1,\ell_1;\tau)}{a^2}
\\[-1mm]
  &-\frac{\varphi_{R}^{E}(\ell_0,\ell_1+1;\tau)-2\varphi_{R}^{E}(\ell_0,\ell_1;\tau)+\varphi_{R}^{E}(\ell_0,\ell_1-1;\tau)}{a^2}
  +m^2\,\varphi_{R}^{E}(\ell_0,\ell_1;\tau),
\end{aligned}
  \label{eq:FE-def}
\end{equation}
\begin{equation}
\begin{aligned}
  F^{O}(\ell;\tau)\;\equiv\;
  &\frac{\varphi_{R}^{O}(\ell_0+1,\ell_1;\tau)-2\varphi_{R}^{O}(\ell_0,\ell_1;\tau)+\varphi_{R}^{O}(\ell_0-1,\ell_1;\tau)}{a^2}
\\[-1mm]
  &-\frac{\varphi_{R}^{O}(\ell_0,\ell_1+1;\tau)-2\varphi_{R}^{O}(\ell_0,\ell_1;\tau)+\varphi_{R}^{O}(\ell_0,\ell_1-1;\tau)}{a^2}
  +m^2\,\varphi_{R}^{O}(\ell_0,\ell_1;\tau),
\end{aligned}
  \label{eq:FO-def}
\end{equation}
so that the lattice functional derivative admits the compact $E/O$ representation
\begin{equation}
  \frac{\delta S[\varphi]}{\delta\varphi(\ell;\tau)}
  \;=\;
  -(1+i)\,F^{E}(\ell;\tau)\;-\;(1-i)\,F^{O}(\ell;\tau).
  \label{eq:dS-dphi-EO-F}
\end{equation}
In symplectic quantization we evolve the complexified field $\varphi(\ell;\tau)$ and its conjugate momentum $\pi(\ell;\tau)$ in intrinsic time $\tau$ via
\begin{equation}
  \frac{d}{d\tau}\varphi(\ell;\tau)=\bar\pi(\ell;\tau),
  \qquad
  \frac{d}{d\tau}\pi(\ell;\tau)=i\,\frac{\delta S[\varphi]}{\delta\varphi(\ell;\tau)}
  \;=\;
  (1-i)\,F^{E}(\ell;\tau)\;-\;(1+i)\,F^{O}(\ell;\tau),
  \label{eq:sq-eom-lat-EO}
\end{equation}
and therefore, equivalently,
\begin{equation}
  \frac{\delta S[\varphi]}{\delta\varphi(\ell;\tau)}
  \;=\;
  -\,i\,\frac{d}{d\tau}\pi(\ell;\tau).
  \label{eq:force-from-pi}
\end{equation}
DS identities are probed through intrinsic--time averages over a stationary window $\Delta\tau=N_\tau\,\delta\tau$, with sampling times $\tau_n\equiv\tau_0+n\,\delta\tau$. For a chosen insertion site $\ell_i$ we define the $E/O$ pointwise estimators
\begin{align}
  \mathcal D_0(\ell)
  &\;\equiv\;
  \frac{1}{N_\tau}\sum_{n=0}^{N_\tau-1}\frac{1}{\hbar}\,
  \frac{\delta S[\varphi]}{\delta\varphi(\ell;\tau_n)}
  \;=\;
  -\,\frac{1}{\hbar\,N_\tau}\sum_{n=0}^{N_\tau-1}
  \Big[(1+i)\,F^{E}(\ell;\tau_n)+(1-i)\,F^{O}(\ell;\tau_n)\Big],
  \label{eq:D0-est-EO}\\[2mm]
  \mathcal D_1(\ell;\ell_i)
  &\;\equiv\;
  \frac{1}{N_\tau}\sum_{n=0}^{N_\tau-1}\frac{1}{\hbar}\,
  \frac{\delta S[\varphi]}{\delta\varphi(\ell;\tau_n)}\,\varphi(\ell_i;\tau_n)
  \nonumber\\
  &\;=\;
  -\,\frac{1}{\hbar\,N_\tau}\sum_{n=0}^{N_\tau-1}
  \Big[(1+i)\,F^{E}(\ell;\tau_n)+(1-i)\,F^{O}(\ell;\tau_n)\Big]
  \Big[(1+i)\,\varphi_R^{E}(\ell_i;\tau_n)+(1-i)\,\varphi_R^{O}(\ell_i;\tau_n)\Big],
  \label{eq:D1-est-EO}
\end{align}
where the complex insertion field is reconstructed as in Eq.~\eqref{eq:phi-reconstruct-EO}. In the stationary regime and for large averaging windows, equivalence between intrinsic--time averages and the periodic real--time path integral implies the predictions
\begin{align}
  \lim_{\Delta\tau\to\infty}\mathcal D_0(\ell) &= 0,
  \label{eq:D0-pred}\\
  \lim_{\Delta\tau\to\infty}\mathcal D_1(\ell;\ell_i) &= i\frac{1}{a^2}\,\delta_{\ell,\ell_i},
  \label{eq:D1-pred}
\end{align}
namely: the equation of motion estimator vanishes pointwise on the torus, while the estimator with one insertion reproduces a purely local contact term concentrated at the insertion site. This is exactly what we observe numerically, as visible in Fig.~\ref{fig:dyson-3d-periodic}.
%
%
%
%
\subsection{Extracting physical observables with initial-value dynamics (mixed temporal boundaries)}
\label{sec:initial-value-dynamics}

We now turn to a second class of observables, tailored to an \emph{initial-value} formulation in Minkowski time.  The key difference with respect to the periodic setup is that we do not compute a trace.  Rather, we condition the path integral on a prescribed initial field configuration and read out the ensuing real-time response through correlators anchored at the fixed initial slice.  The data at $x_0=0$ selects the sector of field histories being sampled, while the end of the interval at $x_0=T$ sets the observation window.  In our simulations we implement this philosophy in two closely related ways, depending on the observable: for spectroscopy (mass gap and dispersion) we adopt mixed temporal boundaries with an unconstrained final slice, while for light-cone visualizations we replace the unconstrained endpoint by a fixed terminal slice chosen to coincide with the exact free evolution of the preparation.  In both cases the spatial boundaries are taken \emph{free} in the Neumann sense,
\begin{equation}
  \partial_{x_1}\varphi(x_0,0)=0,\qquad \partial_{x_1}\varphi(x_0,L)=0,
  \label{eq:mixed-spatial-Neumann}
\end{equation}
for all $x_0\in[0,T]$.\par
We begin with the mixed temporal prescription in which the initial time-slice is fixed and the final time-slice is left free in the variational sense. This setup provides a direct spectroscopy tool: after projecting onto the spatial Neumann modes, the resulting real-time correlators resolve the discrete oscillation frequencies of the free theory.  Concretely, fixing $\varphi$ at $x_0=0$ and allowing arbitrary variations at $x_0=T$ implies the mixed temporal conditions
\begin{equation}
  \varphi(0,x_1)=f(x_1),
  \qquad
  \partial_{x_0}\varphi(T,x_1)=0,
  \label{eq:mixed-time-bc-main}
\end{equation}
while Eq.~\eqref{eq:mixed-spatial-Neumann} enforces Neumann reflection in space.  If we vary the action $S[\varphi]=\int_{\mathcal V} d^2x\,\mathcal L(\varphi,\partial\varphi)$ and integrate by parts in $x_0$, independently of the bulk equations of motion, the variation contains an endpoint term
\begin{equation}
  \delta S\Big|_{\text{endpoints in }x_0}
  =
  \int_{0}^{L}\!dx_1\;
  \frac{\partial\mathcal L}{\partial(\partial_{x_0}\varphi)}(T,x_1)\,\delta\varphi(T,x_1)
  \;-\;
  \int_{0}^{L}\!dx_1\;
  \frac{\partial\mathcal L}{\partial(\partial_{x_0}\varphi)}(0,x_1)\,\delta\varphi(0,x_1).
  \label{eq:endpoint-variation}
\end{equation}
Fixing the initial configuration means $\delta\varphi(0,x_1)=0$, so the second term vanishes.  If the final configuration is left free, then $\delta\varphi(T,x_1)$ is arbitrary. Stationarity therefore requires
\begin{equation}
  \frac{\partial\mathcal L}{\partial(\partial_{x_0}\varphi)}(T,x_1)=0
  \qquad \forall~x_1\in[0,L].
  \label{eq:natural-bc-general}
\end{equation}
For the free scalar with $\mathcal L=\frac12(\partial_{x_0}\varphi)^2-\frac12(\partial_{x_1}\varphi)^2-\frac12 m^2\varphi^2$ one has $\frac{\partial\mathcal L}{\partial(\partial_{x_0}\varphi)}=\partial_{x_0}\varphi$, so Eq.~\eqref{eq:natural-bc-general} becomes precisely the Neumann endpoint condition in Eq.~\eqref{eq:mixed-time-bc-main}.  In this precise sense the surface $x_0=T$ marks the end of the observation window without fixing the value of $\varphi(T,\cdot)$.\par
We now connect the mixed temporal boundary problem to the canonical formulation.  Let $\hat\varphi(x_1)$ denote the field operator at fixed time, with field eigenstates $|\varphi\rangle$ defined by
\begin{equation}
  \hat\varphi(x_1)\,|\varphi\rangle=\varphi(x_1)\,|\varphi\rangle.
\end{equation}
The real-time kernel with fixed initial and final field configurations is the matrix element of the time-evolution operator,
\begin{equation}
  \big\langle \varphi_T, T \big|\,\widehat{\mathcal O}(x_0)\,\big| f,0 \big\rangle
  =
  \langle \varphi_T|\,e^{-\frac{i}{\hbar}\hat H (T-x_0)}\,
  \widehat{\mathcal O}\,
  e^{-\frac{i}{\hbar}\hat H x_0}\,|f\rangle,
  \label{eq:kernel-fixed-fixed-op}
\end{equation}
and the standard time-slicing construction gives the equivalent path-integral representation
\begin{equation}
  \big\langle \varphi_T, T \big|\,\widehat{\mathcal O}(x_0)\,\big| f,0 \big\rangle
  =
  \int_{\substack{\varphi(0,x_1)=f(x_1)\\ \varphi(T,x_1)=\varphi_T(x_1)}}
  \!\!\mathcal D\varphi\;
  \mathcal O[\varphi]\;
  \exp\!\left(\frac{i}{\hbar}S[\varphi]\right),
  \label{eq:kernel-fixed-fixed}
\end{equation}
with $\mathcal O[\varphi]$ the c-number functional corresponding to $\widehat{\mathcal O}$ in the functional integral.  The probability amplitude relevant for the initial-value setup is obtained by summing over all final configurations,
\begin{align}
  \mathcal K[\widehat{\mathcal O};f]
  \;\equiv\;
  \int \mathcal D\varphi_T\;
  \big\langle \varphi_T, T \big|\,\widehat{\mathcal O}(x_0)\,\big| f,0 \big\rangle
  &=
  \int_{\varphi(0,\cdot)=f}
  \!\!\mathcal D\varphi\;
  \mathcal O[\varphi]\;
  \exp\!\left(\frac{i}{\hbar}S[\varphi]\right),
  \label{eq:Kmix-def}
\end{align}
where in the last expression the final slice $\varphi(T,\cdot)$ is unrestricted.  This unrestricted endpoint is precisely the implementation of the natural variational condition discussed above.  In the Schr\"odinger picture the integration over $\varphi_T$ inserts a distinguished boundary bra state,
\begin{equation}
  \langle N|
  \;\equiv\;
  \int \mathcal D\varphi_T\;\langle \varphi_T|,
  \qquad\text{so that}\qquad
  \langle \varphi_T|N\rangle=1,
  \label{eq:Neumann-boundary-bra}
\end{equation}
and therefore
\begin{equation}
  \mathcal K[\widehat{\mathcal O};f]
  =
  \langle N|\,e^{-\frac{i}{\hbar}\hat H (T-x_0)}\,
  \widehat{\mathcal O}\,
  e^{-\frac{i}{\hbar}\hat H x_0}\,|f\rangle,
  \qquad
  \mathcal K[\mathbbm 1;f]
  =
  \langle N|e^{-\frac{i}{\hbar}\hat H T}|f\rangle.
  \label{eq:Kmix-boundary-state}
\end{equation}
From Eq.~\eqref{eq:Kmix-def} we define the normalized mixed expectation value
\begin{equation}
  \big\langle \widehat{\mathcal O}(x_0)\big\rangle_f
  \;\equiv\;
  \frac{\mathcal K[\widehat{\mathcal O};f]}
       {\mathcal K[\mathbbm 1;f]},
  \qquad 0\le x_0\le T,
  \label{eq:mixed-def-normalised}
\end{equation}
which depends only on the chosen initial profile $f$ and on the insertion: it probes the quantum spread generated by the dynamics starting from the field eigenstate $|f\rangle$, without conditioning on a final configuration.\par
We now express the free theory in a form that makes the harmonic content manifest.  With the Neumann spatial boundaries of Eq.~\eqref{eq:mixed-spatial-Neumann}, the spatial Laplacian is diagonalized by the Neumann cosine basis
\begin{equation}
  u^{(0)}(x_1)=\frac{1}{\sqrt{L}},
  \qquad
  u^{(n_1)}(x_1)=\sqrt{\frac{2}{L}}\cos\!\big(k_1^{(n_1)}x_1\big)\quad(n_1\ge1),
  \qquad
  k_1^{(n_1)}\equiv\frac{n_1\pi}{L},
  \label{eq:neumann-cos-modes}
\end{equation}
satisfying $\int_0^L dx_1\,u^{(n_1)}(x_1)u^{(n_1')}(x_1)=\delta_{n_1n_1'}$.  Expanding the field as
\begin{equation}
  \varphi(x_0,x_1)=\sum_{n_1=0}^{\infty}u^{(n_1)}(x_1)\,\varphi^{(n_1)}(x_0),
  \qquad
  \varphi^{(n_1)}(x_0)\equiv\int_0^L\!dx_1\,u^{(n_1)}(x_1)\,\varphi(x_0,x_1),
  \label{eq:mixed-neumann-mode-expansion}
\end{equation}
diagonalizes the spatial integral of the quadratic action,
\begin{equation}
  S_0[\varphi]
  =
  \sum_{n_1\ge0}\frac12\int_0^T\!dx_0\,
  \Big[\big(\partial_{x_0}\varphi^{(n_1)}(x_0)\big)^2-\Omega_{n_1}^2\,\varphi^{(n_1)}(x_0)^2\Big],
  \qquad
  \Omega_{n_1}\equiv\sqrt{m^2+\big(k_1^{(n_1)}\big)^2}.
  \label{eq:neumann-action-diag}
\end{equation}
In other words, the free scalar field on $[0,L]$ is an infinite tower of \emph{decoupled} harmonic oscillators evolving in coordinate time $x_0$, with discrete frequencies $\Omega_{n_1}$ fixed by the spatial Neumann spectrum.  The mixed temporal boundary conditions act mode by mode: the fixed initial profile $\varphi(0,x_1)=f(x_1)$ implies
\begin{equation}
  \varphi^{(n_1)}(0)=f^{(n_1)},
  \qquad
  f^{(n_1)}\equiv\int_0^L\!dx_1\,u^{(n_1)}(x_1)\,f(x_1),
  \label{eq:fn-def}
\end{equation}
while the natural endpoint condition $\partial_{x_0}\varphi(T,x_1)=0$ gives $\partial_{x_0}\varphi^{(n_1)}(T)=0$ for all $n_1$.  Thus the initial data selects which oscillators are initially displaced, and the subsequent real-time signal is a superposition of harmonics at the discrete set of frequencies $\{\Omega_{n_1}\}$.\par
In the mixed temporal setup the freedom to choose the initial profile $f$ is precisely what allows us to turn the real-time evolution into a spectroscopy tool: fixing $\varphi(0,\cdot)=f$ selects the initial displacements of the spatial normal modes, while leaving the final slice unconstrained lets the system evolve and reveals its harmonic content through time-dependent correlators.  We implement this directly on the lattice and define the relevant observables on the field variable $\varphi(\ell_0,\ell_1;\tau)$.  We work on an $N_0\times N_1$ grid with sites $x_0^{(\ell_0)}=\ell_0 a$, $x_1^{(\ell_1)}=\ell_1 a$, and we fix the initial slice by prescribing
\begin{equation}
  \varphi(0,\ell_1;\tau)=f_{\ell_1},
  \qquad \ell_1=0,\dots,N_1-1,
  \label{eq:mixed-lat-IC}
\end{equation}
while the unconstrained endpoint is implemented by the lattice Neumann condition in the temporal direction,
\begin{equation}
  \varphi(N_0,\ell_1;\tau)=\varphi(N_0-1,\ell_1;\tau),
  \qquad \ell_1=0,\dots,N_1-1,
  \label{eq:mixed-lat-final-Neumann}
\end{equation}
and analogously in space, as discussed in App.~\ref{app:mixed-lattice-implementation}.  To read out the subsequent dynamics we consider temporal correlators anchored at the fixed slice,
\begin{equation}
  C_f(\ell_0;\ell_1,\ell_1')
  \;\equiv\;
  \Big\langle \varphi(\ell_0,\ell_1)\,\varphi(0,\ell_1')\Big\rangle_f,
  \qquad \ell_0=0,\dots,N_0-1,
  \label{eq:mixed-lat-C-def}
\end{equation}
where $\langle\cdot\rangle_f$ denotes the normalized mixed expectation value of Eq.~\eqref{eq:mixed-def-normalised} evaluated as a large-$\tau$ dynamical average of the constrained symplectic flow.  For spectral information it is convenient to project onto the spatial Neumann eigenmodes of the lattice Laplacian.  These are the discrete cosine functions
\begin{equation}
  u^{(0)}_{\ell_1}\equiv \frac{1}{\sqrt{N_1}},\qquad
  u^{(n_1)}_{\ell_1}\equiv
  \sqrt{\frac{2}{N_1}}\cos\!\big(\varphi_{n_1}\ell_1\big)\quad (n_1=1,\dots,N_1-1),
  \qquad
  \varphi_{n_1}\equiv \frac{\pi n_1}{N_1-1},
  \label{eq:mixed-lat-neumann-basis}
\end{equation}
which satisfy the orthonormality relation
\begin{equation}
  \sum_{\ell_1=0}^{N_1-1}u^{(n_1)}_{\ell_1}\,u^{(n_1')}_{\ell_1}=\delta_{n_1n_1'}.
  \label{eq:mixed-lat-neumann-orth}
\end{equation}
We define the corresponding mode amplitudes at time-slice $\ell_0$ by the discrete cosine transform
\begin{equation}
  \varphi^{(n_1)}(\ell_0)
  \;\equiv\;
  \sum_{\ell_1=0}^{N_1-1}u^{(n_1)}_{\ell_1}\,\varphi(\ell_0,\ell_1),
  \qquad
  f^{(n_1)}
  \;\equiv\;
  \sum_{\ell_1=0}^{N_1-1}u^{(n_1)}_{\ell_1}\,f_{\ell_1},
  \label{eq:mixed-lat-mode-def}
\end{equation}
so that the mixed Dirichlet condition of Eq.~\eqref{eq:mixed-lat-IC} is equivalent to $\varphi^{(n_1)}(0)=f^{(n_1)}$ for all $n_1$.  In the free theory each spatial mode behaves as a harmonic oscillator in coordinate time with lattice frequency
\begin{equation}
  \Omega_{n_1}^2
  \;\equiv\;
  m^2+\widehat{k}_1^{\,2}(n_1),
  \qquad
  \widehat{k}_1(n_1)\equiv \frac{2}{a}\sin\!\Big(\frac{\varphi_{n_1}}{2}\Big),
  \label{eq:mixed-lat-Omega}
\end{equation}
so the spectrum of temporal oscillations is discrete and labeled by $n_1$.  We extract these frequencies from the mode autocorrelators
\begin{equation}
  C_f^{(n_1)}(\ell_0)
  \;\equiv\;
  \Big\langle \varphi^{(n_1)}(\ell_0)\,\varphi^{(n_1)}(0)\Big\rangle_f,
  \label{eq:mixed-lat-mode-corr}
\end{equation}
whose discrete Fourier transform in $\ell_0$ exhibits peaks at $\Omega_{n_1}$ whenever $f^{(n_1)}\neq 0$.\par
The three preparations used in our simulations correspond to three progressively less selective choices of the initial overlaps $f^{(n_1)}$.

\begin{enumerate}
%
	\item \emph{Zero-mode preparation (mass gap).}  Choosing a spatially uniform initial slice,
\begin{equation}
  f_{\ell_1}=C,
  \label{eq:mixed-lat-f-zeromode}
\end{equation}
implies $f^{(0)}\neq0$ and $f^{(n_1)}=0$ for all $n_1\ge1$, so only the Neumann zero mode contributes.  The correlator $C_f^{(0)}(\ell_0)$ therefore contains a single dominant harmonic at $\Omega_0=m$, which provides a direct determination of the mass gap.  A representative example is shown in Fig.~\ref{fig:mass_gap_extraction}.
\begin{figure}[H]
  \centering
  \includegraphics[width=0.7\linewidth]{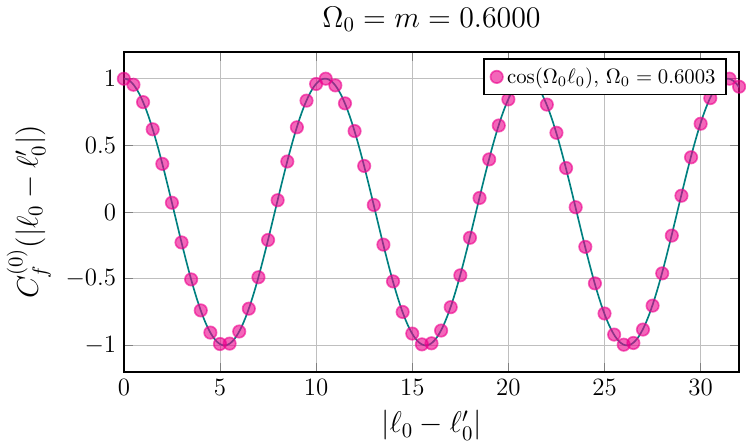}
  \caption{
Mass-gap extraction from a uniform initial field profile. 
A constant initial slice excites only the Neumann zero mode, so the projected correlator $C_f^{(0)}(\ell_0)$ in Eq. \eqref{eq:mixed-lat-f-zeromode} contains a single dominant frequency. 
That frequency is $\Omega_0=m$, allowing a direct determination of the mass gap from the real-time oscillation. 
Simulation parameters: $m=0.6$, $a=0.5$, $d\tau=10^{-2}$, and $N_0\times N_1=256\times256$.
}
  \label{fig:mass_gap_extraction}
\end{figure}
%
\item \emph{Single-mode preparation (dispersion).}  Choosing the initial slice to coincide with a single cosine mode,
\begin{equation}
  f_{\ell_1}=C\,u^{(n_1)}_{\ell_1},
  \qquad n_1\ge 1,
  \label{eq:mixed-lat-f-singlemode}
\end{equation}
sets $f^{(n_1')}\propto\delta_{n_1'n_1}$ and isolates a single oscillator frequency $\Omega_{n_1}$.  The corresponding correlator $C_f^{(n_1)}(\ell_0)$ oscillates with $\Omega_{n_1}$, and repeating the extraction for several values of $n_1$ reconstructs the lattice dispersion relation of Eq.~\eqref{eq:mixed-lat-Omega}.  Two representative single-mode runs are shown in Fig.~\ref{fig:dispersion_relation}.
\begin{figure}[H]
  \centering
  \includegraphics[width=1\linewidth]{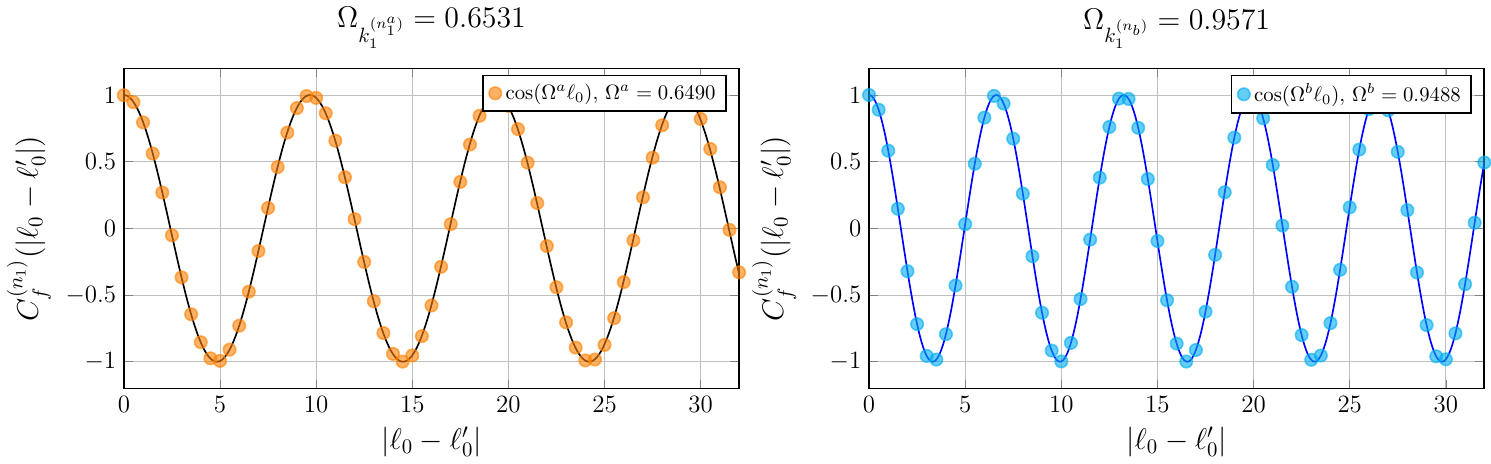}
  \caption{
Mode-by-mode extraction of the lattice dispersion relation obtained from  Eq.~\eqref{eq:mixed-lat-f-singlemode}. 
The initial field profile is chosen to be a single Neumann spatial mode, so the correlator contains one dominant temporal frequency $\Omega_{n_1}$. 
Repeating the measurement for different mode numbers $n_1$ gives the dispersion relation $\Omega_{n_1}^2=m^2+\widehat{k}_1^{\,2}(n_1)$. 
The plotted correlators show representative single-mode measurements.
}
  \label{fig:dispersion_relation}
\end{figure}
%
\item \emph{All-modes preparation (localized pulse).}
A Kronecker-localized initial slice has overlap with the entire Neumann cosine basis and therefore excites all spatial frequencies.  For this preparation we keep the \emph{same} Neumann spatial boundaries of Eq.~\eqref{eq:mixed-spatial-Neumann}, but we modify the \emph{temporal} implementation in order to obtain a clean causal profile on a finite slab: rather than leaving the final slice unconstrained, we impose a fixed terminal time-slice equal to the exact free evolution of the preparation. This allows us to avoid all kinds of boundary reflection, and for the purpose of this work, which is essentially a proof of concept, allows for clear visualization. Concretely, we fix the initial configuration at $x_0=0$ to
\begin{equation}
  f_{\ell_1}=\delta_{\ell_1,\ell_{1,\star}},
  \qquad\Rightarrow\qquad
  \varphi_R(0,\ell_{1,\star})=1,\quad \varphi_I(0,\ell_{1,\star})=0,
  \label{eq:mixed-lat-f-delta}
\end{equation}
with $f_{\ell_1}=0$ for $\ell_1\neq \ell_{1,\star}$.  Denoting coordinate time by $t=a\,\ell_0$ and $T\equiv (N_0-1)a$, we define the \emph{terminal} Dirichlet profile to be the free lattice evolution evaluated at $t=T$ for the same preparation released from rest, as reviewed in App.~\ref{app:lightcone-bessel-pulse} and implemented in App.~\ref{app:fixedfixed-terminal-implementation}.  In other words, for light-cone measurements the temporal boundaries are
\begin{equation}
  \varphi(0,\ell_1;\tau)=f_{\ell_1},
  \qquad
  \varphi(N_0-1,\ell_1;\tau)=g_{\ell_1}\equiv \varphi_{\mathrm{free}}(T,\ell_1),
  \label{eq:lightcone-fixedfixed-slices}
\end{equation}
while the spatial direction remains Neumann. The corresponding space-time observable is the mixed correlator sourced at the preparation site,
\begin{equation}
  C_f(\ell_0,\ell_1)
  \;\equiv\;
  \Big\langle \varphi(\ell_0,\ell_1)\,\varphi(0,\ell_{1,\star})\Big\rangle_f,
  \label{eq:mixed-lat-lightcone}
\end{equation}
evaluated with the fixed-fixed temporal implementation of Eq~\eqref{eq:lightcone-fixedfixed-slices} and Neumann in space.  Because the initial slice is conditioned by Eq.~\eqref{eq:mixed-lat-f-delta}, $C_f$ is not a translationally invariant vacuum two-point function; it measures the response to a localized initial displacement.  In the continuum, the corresponding initial-value problem admits an explicit causal solution in terms of Bessel functions (App.~\ref{app:lightcone-bessel-pulse}).  We first visualize the full two-dimensional profile $C_f(\ell_0,\ell_1)$ as a heatmap in the $(\ell_0,\ell_1)$ plane.  A representative result is shown in Fig.~\ref{fig:light_cone}, where the dominant signal propagates away from $\ell_{1,\star}$ within a causal wedge.
\begin{figure}[H]
  \centering
  \includegraphics[width=0.85\linewidth]{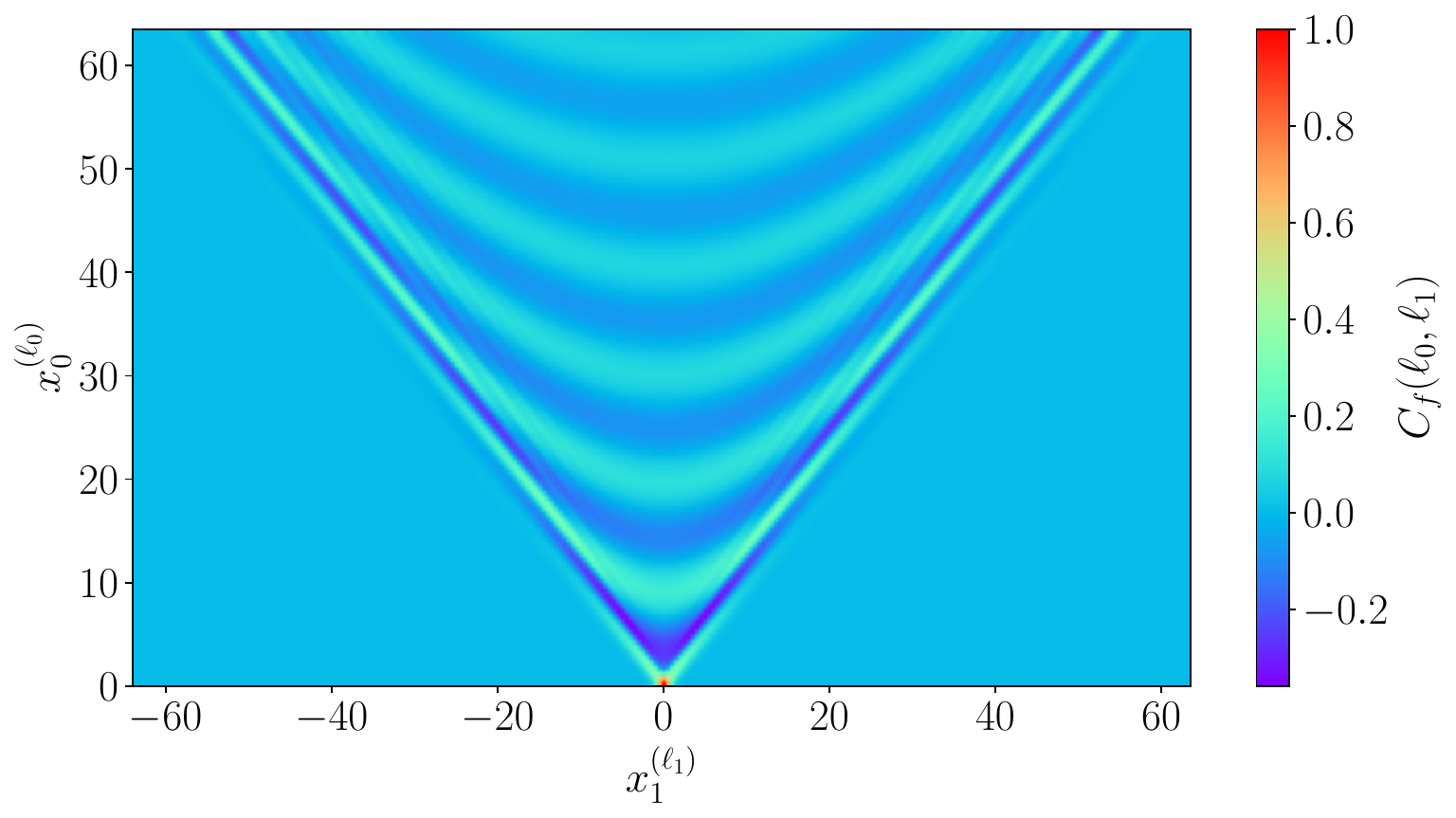}
\caption{
Real-time response to a localized initial excitation. 
The initial field is nonzero only at the source site $\ell_{1,\star}$, so it excites all Neumann spatial modes. 
The heatmap shows the resulting correlator $C_f(\ell_0,\ell_1)$ as the excitation propagates away from the source. 
The correlator is largest inside the causal region $t\gtrsim r$, where $t=a\ell_0$ and $r=a|\ell_1-\ell_{1,\star}|$. 
Simulation parameters: $m=0.6$, $a=0.5$, $d\tau=10^{-2}$, and $N_0\times N_1=128\times256$.
}
  \label{fig:light_cone}
\end{figure}
To compare with the continuum causal prediction we reduce $C_f(\ell_0,\ell_1)$ to an invariant one-dimensional profile.  Introducing
\begin{equation}
  t \equiv a\,\ell_0,
  \qquad
  r \equiv a\,|\ell_1-\ell_{1,\star}|,
  \qquad
  s(t,r)\equiv \sqrt{t^2-r^2},
  \label{eq:mixed-lat-invariants}
\end{equation}
we restrict to points strictly inside the cone ($t>r$) and define a binned average over invariant shells,
\begin{equation}
  \overline{C}_f(s)
  \;\equiv\;
  \frac{1}{|\mathcal S_s|}
  \sum_{(\ell_0,\ell_1)\in\mathcal S_s}
  C_f(\ell_0,\ell_1),
  \qquad
  \mathcal S_s:\ s(t_{\ell_0},r_{\ell_1})\in[s,s+\Delta s].
  \label{eq:mixed-lat-radial-average}
\end{equation}
In the continuum, a localized displacement at $t=0$ implies (App.~\ref{app:lightcone-bessel-pulse})
\begin{equation}
  \Big\langle \varphi(t,x_1)\,\varphi(0,x_1^\star)\Big\rangle_f
  =
  \frac{1}{2}\,\partial_t\!\left[
    \Theta(t^2-r^2)\,J_0\!\big(m\sqrt{t^2-r^2}\big)
  \right],
  \qquad r=|x_1-x_1^\star|,
  \label{eq:mixed-bessel-master}
\end{equation}
and therefore, for $t>r>0$,
\begin{equation}
  \Big\langle \varphi(t,x_1)\,\varphi(0,x_1^\star)\Big\rangle_f
  =
  -\,\frac{m\,t}{2\sqrt{t^2-r^2}}\,
  J_1\!\big(m\sqrt{t^2-r^2}\big),
  \qquad (t>r),
  \label{eq:mixed-bessel-inside-explicit}
\end{equation}
using $J_0'(z)=-J_1(z)$.  We fit the collapsed lattice data $\overline{C}_f(s)$ to Eq.~\eqref{eq:mixed-bessel-inside-explicit}, with overall normalization and the mass parameter treated as fit inputs,
\begin{equation}
  \overline{C}_f(s)
  \;\overset{\textrm{fit}}{\longrightarrow}\;
  -
  \frac{t}{s}\,
  J_1(m s),
  \qquad s=\sqrt{t^2-r^2},
  \label{eq:mixed-bessel-fit}
\end{equation}
restricting the fit to $t>r$.  The collapse of points with different $(t,r)$ but identical $s$ onto a single curve, together with the best-fit value of the mass parameter, provides a direct check of the expected dependence on the invariant $t^2-r^2$ in the all-modes initial-value measurement. In particular, treating $m$ as a free fit parameter, we recover a value consistent with the input mass used in the simulation, showing that the light-cone profile can be used as an independent determination of the mass scale. Temporal boundary reflections are suppressed by the fixed-fixed implementation of App.~\ref{app:fixedfixed-terminal-implementation}.
\begin{figure}[H]
  \centering
  \includegraphics[width=0.85\linewidth]{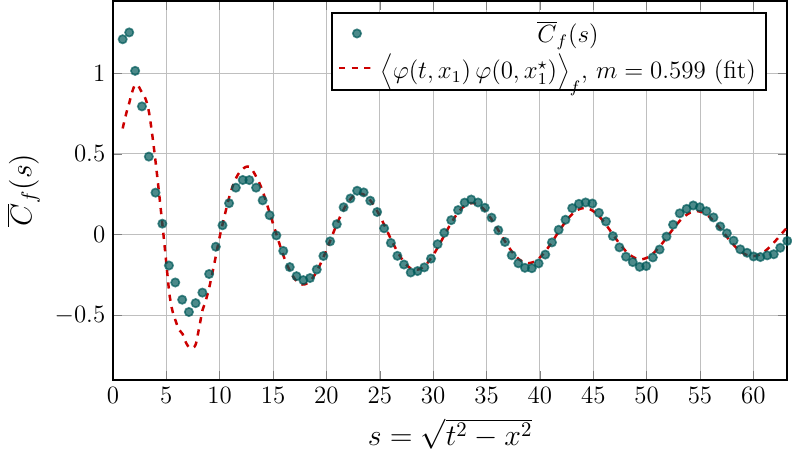}

\caption{
Comparison of the localized-pulse correlator with the continuum massive-field prediction. 
Starting from the heatmap in Fig.~\ref{fig:light_cone}, we keep only points inside the light cone, where the time separation is larger than the spatial separation: $t>r$. 
For each such point we compute the light-cone distance $s=\sqrt{t^2-r^2}$, with $t=a\ell_0$ and $r=a|\ell_1-\ell_{1,\star}|$. 
The plotted points are the lattice correlator averaged over small intervals of $s$. 
The solid curve is the expected continuum Bessel-function form for a massive field. 
The agreement shows that the lattice response follows the continuum light-cone prediction and gives a fitted mass compatible with the input value $m=0.6$. 
Simulation parameters: $a=0.5$, $d\tau=10^{-2}$, and $N_0\times N_1=128\times256$.
}
  \label{fig:bessel_fit}
\end{figure}
\end{enumerate}
%
%
%
\section{Conclusions}

In this paper we extended constrained symplectic quantization to a relativistic scalar field in Minkowski space-time. The construction is based on the analytic continuation of fields and action from $\mathbb{R}$ to $\mathbb{C}$ together with constraints that select stable intrinsic-time trajectories and define the corresponding integration contours in the complexified field space.

We have proved in full generality that the CSQ microcanonical generating functional at fixed generalized action $\mA=\hbar M$ reproduces, in the continuum limit, the Feynman generating functional of scalar quantum field theory with interactions, provided that its renormalizable. For the same class of models We also showed that the intrinsic-time Hamiltonian flow encodes the quantum equations of motion and, more generally, the Dyson--Schwinger hierarchy, including the correct contact terms.

Then, in order to provide a first numerical validation of the above correspondence, we specialized our study to the case of a free scalar field in $1+1$ dimensions, deriving the constrained equations of motion, identifying the corresponding stable manifold, and implementing the resulting dynamics numerically. The simulations reproduce the expected real-time two-point function, the equal-time commutator structure, and the Dyson--Schwinger identities. These results provide concrete evidence that constrained symplectic quantization can access genuinely Minkowskian observables through deterministic intrinsic-time evolution. Moreover we have shown full control over various types of boundary conditions suitable for studying different observables.

The next natural step is to extend the same strategy beyond the free theory and to devise how to implement the constraints that guaranty the stability of the dynamics in the presence of interactions. From the perspective of any functional approach to quantum field theory and quantum mechanics the simplest interacting theory is the anharmonic oscillator, which represents the already in progress next chapter of our investigation~\cite{CSQ_Anharmonic_inprep}. Then we can foresee a bifurcation of the investigation paths: one the one hand an appealing direction will be the extension of constrained symplectic quantization to the non-relativistic quantum field theories of interest in condensed matter, for which the functional approaches aimed at studying non-equilibrium dynamics are typically plagued by the sign problem and an efficient numerical strategy to sample the non-equilibrium evolution in Minkowskian time, which is crucial for many relaxation and transport phenomena, is still missing. On the other hand, a natural development of this investigation leads to the design of methods to handle relativistic quantum field theories with interactions on the lattice. For instance, in order of increasing difficulty: the prototypical model $\lambda \varphi^4$, pure gauge theories and theories with fermions. For such systems the constrained symplectic framework may offer a practical route to real-time quantum dynamics beyond the reach of Euclidean importance sampling, providing a new strategy to attack problems notoriously difficult in numerical particle physics.
Possible future applications include spectral functions and transport coefficients, scattering amplitudes and resonance physics in finite volume~\cite{Luscher:1990ux,Luscher:1991cf}, and, more broadly, real-time processes in strongly coupled gauge theories.

\newpage
\appendix
\section*{\texorpdfstring{\Huge\bfseries Appendices}{Appendices}}
\addcontentsline{toc}{section}{Appendices}
\bigskip
%
%
%
%
\section{Microcanonical expectation values}
\label{app:alpha}
In this appendix we prove, using a space-time lattice regularization, that in the continuum limit and at fixed generalized action $\mA_*=\hbar M$,
\begin{align}\label{eq:expect}
&\frac{1}{\Omega[\mA_*]}~\sum_{i=1}^n\,\int \mD_M\varphi\,\mD_M\bar \varphi\,\mD_M\pi\,\mD_M\bar\pi~
\delta\left(\mA_*-\mathbb{H}_{\rm SQ}[\varphi,\bar{\varphi},\pi,\bar{\pi}]\right)
&\nonumber\\
&\pi(x)\bar{\pi}(x_i)\varphi(x_1)\ldots\varphi(x_{i-1})\varphi(x_{i+1})\ldots\varphi(x_n)\nonumber\\
&=\hbar \sum_{i=1}^n\delta(x-x_i)\left\langle\varphi(x_1)\ldots\varphi(x_{i-1})\varphi(x_{i+1})\ldots\varphi(x_n)\right\rangle\,.
\end{align}
The right-hand side is the usual QFT expectation value obtained after taking the continuum limit $M\to\infty$.

The momentum integrals can be separated from the field integrals. By symmetry, the momentum integral vanishes unless $x_i=x$. We are therefore led to evaluate
\begin{align}
I = \frac{1}{\Omega[\mA_*]}\,\int\mD_M\pi\,\mD_M\bar{\pi}\;
\delta\left(\mA_*-\mathbb{H}_{\rm SQ}[\varphi,\bar{\varphi},\pi,\bar{\pi}]\right)\pi(x_i)\bar{\pi}(x_i)\,.
\end{align}
Since the lattice measure is finite-dimensional, it is enough to evaluate the corresponding finite-dimensional model integral
\begin{align}
I=&\int_{-\infty}^{\infty} (x_1^2+y_1^2)\,
\delta\left(c - x_1^2 - x_2^2 - \cdots - x_n^2 - y_1^2 - y_2^2 - \cdots - y_n^2\right)
\,dx_1\cdots dx_n\,dy_1\cdots dy_n\,.
\end{align}
Writing $\pi\cdot\bar{\pi}=x\cdot x+y\cdot y$ and using rotational symmetry, this becomes
\begin{align}
I=2 \int_{-\infty}^{\infty} x_1^2\,
\delta\left(c - x_1^2 - x_2^2 - \cdots - x_{2n}^2\right)
\,dx_1\,dx_2\cdots dx_{2n}\,.
\end{align}
We now perform a hyperspherical change of variables,
\begin{align}
	x_1 &= r \cos \theta, \nonumber\\
	x_2 &= r \sin \theta \cos \theta_2,\nonumber \\
	&\vdots \nonumber\\
	x_{2n} &= r \sin \theta \sin \theta_2 \cdots \sin \theta_{2n-1},
\end{align}
for which the volume element is
\begin{equation}
dV = r^{2n-1} \, dr \, d\Omega\,,
\end{equation}
with $d\Omega$ the solid-angle measure on the unit sphere in $2n$ dimensions. Substituting into the integral, we obtain
\begin{align}
	I= 2\int x_1^2 \, \delta\left(c - r^2\right) dV
	&=2 \int (r \cos \theta)^2 \, \delta\left(c - r^2\right) r^{2n-1} \, dr \, d\Omega \nonumber\\
	&=2 \int r^{2n+1} \cos^2 \theta \, \delta\left(c-r^2\right) \, dr \, d\Omega.
\end{align}
Using
\[
\delta\left(c - r^2\right)=\frac{\delta\left(r-\sqrt{c}\right)}{2r}\,,
\]
we find
\begin{align}
	I
	&= 2\int r^{2n+1} \cos^2 \theta \,\frac{\delta\left(r - \sqrt{c}\right)}{2r} \, dr \, d\Omega\nonumber \\
	&= c^{n}\int \cos^2 \theta \, d\Omega.
\end{align}
By rotational symmetry on the unit sphere,
\begin{equation}
\int \cos^2 \theta \, d\Omega = \frac{S_{2n}}{2n}\,,
\end{equation}
where
\begin{equation}
S_{2n} = \frac{2 \pi^{n}}{\Gamma\left(n\right)}
\end{equation}
is the surface area of the unit sphere in $2n$ dimensions. Hence
\begin{align}
	\int_{-\infty}^{\infty} x_1^2 \, \delta\left(c - x^2_1 - \cdots - x^2_{2n}\right) \, dx_1 \cdots dx_{2n}
	&= \frac{c^{n}}{2}\cdot \frac{S_{2n}}{2n} \nonumber\\
	&= \frac{c^{n}}{2} \cdot \frac{2 \pi^{n}}{2n \Gamma\left(n\right)}\nonumber \\
	&= \frac{ \left(\pi c\right)^{n}}{2 n \Gamma\left(n\right)}.
\end{align}
Therefore,
\begin{align}
	I = \frac{1}{\Omega[\mA_*]}\, \frac{ \left(\pi\right)^{M}}{M \Gamma\left(M\right)}(\mA_*-2\im S[\varphi,\bar{\varphi}])^{M}\,.
\end{align}
Substituting this result into Eq.~\eqref{eq:expect}, we obtain
\begin{align}
\frac{1}{\Omega[\mA_*]}\, \frac{\left(\pi\right)^{M}}{M \Gamma\left(M\right)}\sum_{i=1}^n\delta(x-x_i)\int \mD_M\varphi\,\mD_M\bar\varphi\,(\mA_*-2\im S[\varphi,\bar{\varphi}])^{M}\varphi(x_1)\ldots\varphi(x_{i-1})\varphi(x_{i+1})\ldots\varphi(x_n)\,,
\end{align}
while
\begin{equation}
\Omega[\mA_*] = \frac{(2\pi)^{M}}{\Gamma\left(M\right)}\int \mD_M\varphi\,\mD_M\bar\varphi\,(\mA_*-2\im S[\varphi,\bar{\varphi}])^{M-1}\,.
\end{equation}
This proves Eq.~\eqref{eq:expect} on the lattice.

Having fixed $\mA_{*} =  \hbar M$, we can now take the continuum limit. Dividing numerator and denominator by $\mA_*$, we obtain
\begin{align}
\frac{\hbar\sum_{i=1}^n\delta(x-x_i)\int \mD_M\varphi\,\mD_M\bar\varphi\,\left(1-\frac{2\im S[\varphi,\bar{\varphi}]}{\mA_*}\right)^{M}\varphi(x_1)\ldots\varphi(x_{i-1})\varphi(x_{i+1})\ldots\varphi(x_n)}{\int \mD_M\varphi\,\mD_M\bar\varphi\,\left(1-\frac{2\im S[\varphi,\bar{\varphi}]}{\mA_{*}}\right)^{M-1}}\,.
\end{align}
Following the same steps as in Sec.~\ref{sec:equivalence}, this expression converges in the continuum limit to
\begin{align}
&\hbar\sum_{i=1}^n\delta(x-x_i)\,\frac{ \int_{\Gamma_{\varphi}} \mathcal{D}\varphi~e^{\frac{i}{\hbar}S[\varphi]}\varphi(x_1)\ldots\varphi(x_{i-1})\varphi(x_{i+1})\ldots\varphi(x_n)}{\int_{\Gamma_{\varphi}} \mathcal{D}\varphi~e^{\frac{i}{\hbar}S[\varphi]}}\nonumber\\
&=\hbar\sum_{i=1}^n\delta(x-x_i)\left\langle\varphi(x_1)\ldots\varphi(x_{i-1})\varphi(x_{i+1})\ldots\varphi(x_n)\right\rangle\,.
\end{align}
This completes the proof of Eq.~\eqref{eq:expect}.

\section{Mixed temporal boundaries: mode decomposition and lattice discretization}
\label{app:mixed-lattice-implementation}

We give a precise implementation of the mixed temporal boundary conditions of Sec.~\ref{sec:initial-value-dynamics}---Dirichlet on the initial time-slice and Neumann on the final time-slice---together with Neumann spatial boundaries, first in the continuum and then on the lattice.\par
 In the continuum we consider the free real scalar in $1+1$ dimensions on $\mathcal V=[0,T]\times[0,L]$ with Minkowskian action
\begin{equation}
  S[\varphi]
  =
  \frac12\int_0^T\!dx_0\int_0^L\!dx_1\;
  \Big[(\partial_{x_0}\varphi)^2-(\partial_{x_1}\varphi)^2-m^2\varphi^2\Big],
  \label{eq:app-mix-action}
\end{equation}
and mixed temporal/spatial Neumann boundary conditions
\begin{equation}
  \varphi(0,x_1)=f(x_1),\qquad
  \partial_{x_0}\varphi(T,x_1)=0,\qquad
  \partial_{x_1}\varphi(x_0,0)=\partial_{x_1}\varphi(x_0,L)=0.
  \label{eq:app-mix-bc}
\end{equation}
For numerical implementation, it is convenient to remove the inhomogeneous Dirichlet datum by a background-field split
\begin{equation}
  \varphi(x_0,x_1)=\varphi_{\mathrm{cl}}(x_0,x_1)+\eta(x_0,x_1),
  \label{eq:app-mix-split}
\end{equation}
where $\varphi_{\mathrm{cl}}$ is chosen to solve the bulk Klein--Gordon equation $(\partial_{x_0}^2-\partial_{x_1}^2+m^2)\varphi_{\mathrm{cl}}(x_0,x_1)=0$ together with the same mixed boundary conditions as $\varphi$,
\begin{equation}
  \varphi_{\mathrm{cl}}(0,x_1)=f(x_1),\qquad
  \partial_{x_0}\varphi_{\mathrm{cl}}(T,x_1)=0,\qquad
  \partial_{x_1}\varphi_{\mathrm{cl}}(x_0,0)=\partial_{x_1}\varphi_{\mathrm{cl}}(x_0,L)=0,
  \label{eq:app-mix-phicl-bc}
\end{equation}
so that the fluctuation field satisfies homogeneous mixed conditions,
\begin{equation}
  \eta(0,x_1)=0,\qquad
  \partial_{x_0}\eta(T,x_1)=0,\qquad
  \partial_{x_1}\eta(x_0,0)=\partial_{x_1}\eta(x_0,L)=0.
  \label{eq:app-mix-eta-bc}
\end{equation}
With this choice the action splits exactly
\begin{equation}
  S[\varphi_{\mathrm{cl}}+\eta]=S[\varphi_{\mathrm{cl}}]+S[\eta],
  \label{eq:app-mix-action-split}
\end{equation}
because the cross term reduces to boundary contributions that vanish by \eqref{eq:app-mix-phicl-bc} and \eqref{eq:app-mix-eta-bc}. Consequently the mixed-boundary path integral factorizes into a classical phase $\exp(\frac{i}{\hbar}S[\varphi_{\mathrm{cl}}])$ times a Gaussian functional integral over $\eta$ with homogeneous mixed boundary conditions.\par
In practice $\varphi_{\mathrm{cl}}$ is obtained by separating variables in the spatial Neumann eigenbasis. One first expands the prescribed initial profile as
\begin{equation}
  f(x_1)=\sum_{n_1=0}^{\infty} f^{(n_1)}\,v^{(n_1)}(x_1),
  \qquad
  f^{(n_1)}\equiv \int_{0}^{L}\!dx_1\,v^{(n_1)}(x_1)\,f(x_1),
\end{equation}
where the normalized Neumann modes on $[0,L]$ are
\begin{equation}
  v^{(0)}(x_1)=\frac{1}{\sqrt{L}},
  \qquad
  v^{(n_1)}(x_1)=\sqrt{\frac{2}{L}}\cos\!\big(k^{(n_1)}_1 x_1\big)\quad (n_1\ge 1),
  \qquad
  k^{(n_1)}_1\equiv \frac{\pi n_1}{L}.
\end{equation}
and satisfy $\int_0^L dx_1\,v^{(n_1)}(x_1)v^{(n_1')}(x_1)=\delta_{n_1n_1'}$ and $\partial_{x_1}v^{(n_1)}(0)=\partial_{x_1}v^{(n_1)}(L)=0$.  Seeking the classical background in separable form
\begin{equation}
  \varphi_{\mathrm{cl}}(x_0,x_1)=\sum_{n_1=0}^{\infty}\varphi^{(n_1)}(x_0)\,v^{(n_1)}(x_1).
\end{equation}
the bulk equation $(\partial_{x_0}^2-\partial_{x_1}^2+m^2)\varphi_{\mathrm{cl}}=0$ reduces mode-by-mode to the ordinary differential equation
\begin{equation}
  \partial^2_{x_0}\varphi^{(n_1)}(x_0)+\Omega_{n_1}^2\,\varphi^{(n_1)}(x_0)=0,
  \qquad
  \Omega_{n_1}\equiv \sqrt{m^2+\big(k^{(n_1)}_1\big)^2}.
\end{equation}
The mixed temporal boundary conditions $\varphi_{\mathrm{cl}}(0,x_1)=f(x_1)$ and $\partial_{x_0}\varphi_{\mathrm{cl}}(T,x_1)=0$ become the mode conditions
\begin{equation}
  \varphi^{(n_1)}(0)=f^{(n_1)},
  \qquad
  \partial_{x_0}\varphi^{(n_1)}(T)=0,
\end{equation}
whose unique solution for generic $T$ is
\begin{equation}
  \varphi^{(n_1)}(x_0)
  =
  f^{(n_1)}\,\frac{\cos\!\big(\Omega_{n_1}(T-x_0)\big)}{\cos(\Omega_{n_1} T)},
  \qquad
  \cos(\Omega_{n_1} T)\neq 0.
\end{equation}
Therefore the explicit classical background that carries the inhomogeneous initial Dirichlet data and satisfies the Neumann endpoint condition is
\begin{equation}
  \varphi_{\mathrm{cl}}(x_0,x_1)
  =
  \sum_{n_1=0}^{\infty}
  f^{(n_1)}\,\frac{\cos\!\big(\Omega_{n_1}(T-x_0)\big)}{\cos(\Omega_{n_1} T)}\,v^{(n_1)}(x_1).
\end{equation}
The exceptional values $\cos(\Omega_{n_1} T)=0$ correspond to mixed Dirichlet/Neumann temporal resonances for that spatial mode; they are avoided by working at generic $T$, or treated with the discretization of the time direction. \par
On the lattice we discretize $\mathcal V=[0,T]\times[0,L]$ on a $N_0\times N_1$ grid,
\begin{equation}
  x_0^{(\ell_0)}\equiv \ell_0 a,\qquad \ell_0=0,\dots,N_0-1,\qquad
  x_1^{(\ell_1)}\equiv \ell_1 a,\qquad \ell_1=0,\dots,N_1-1.
\end{equation}
and we denote the field by $\varphi(\ell_0,\ell_1;\tau)$, with real/imaginary parts $\varphi=\varphi_R+i\varphi_I$.  The mixed temporal boundary conditions together with free spatial boundaries~\eqref{eq:app-mix-bc} are implemented by
\begin{align} 
  &\varphi(0,\ell_1;\tau)=f_{\ell_1},\qquad
  \varphi(N_0,\ell_1;\tau)=\varphi(N_0-1,\ell_1;\tau),\\\nonumber
  &\varphi(\ell_0,-1;\tau)=\varphi(\ell_0,0;\tau),\qquad
  \varphi(\ell_0,N_1;\tau)=\varphi(\ell_0,N_1-1;\tau),
  \label{eq:app-mix-lat-bc-ellj}
\end{align}
where $\ell_0=N_0$ and $j=-1,N_1$ are ghost indices used to impose vanishing normal derivatives by reflection (e.g.\ $(\varphi(N_0,\ell_1)-\varphi(N_0-1,\ell_1))/a=0$ and $(\varphi(\ell_0,0)-\varphi(\ell_0,-1))/a=0$).  As in the continuum  we remove the inhomogeneous Dirichlet slice by shifting
\begin{equation}
  \varphi(\ell_0,\ell_1;\tau)=\varphi_{\rm cl}(\ell_0,\ell_1)+\eta(\ell_0,\ell_1;\tau),
  \label{eq:app-mix-lat-shift-ellj}
\end{equation}
choosing $\varphi_{\rm cl}$ to satisfy the same lattice boundary conditions and, for the cleanest decoupling, to solve the discrete bulk equation associated with the free quadratic kernel; then the fluctuation obeys homogeneous mixed boundaries,
\begin{align}
  &\eta(0,\ell_1;\tau)=0,\qquad
  \eta(N_0,\ell_1;\tau)=\eta(N_0-1,\ell_1;\tau),\\\nonumber
  &\eta(\ell_0,-1;\tau)=\eta(\ell_0,0;\tau),\qquad
  \eta(\ell_0,N_1;\tau)=\eta(\ell_0,N_1-1;\tau),
  \label{eq:app-mix-lat-eta-bc-ellj}
\end{align}
so that the number of dynamical fluctuation variables is $M=(N_0-1)N_1$.  We define the second-difference operators (with boundary-adjacent values reduced using \eqref{eq:app-mix-lat-eta-bc-ellj})
\begin{align}
  (\Delta_0 \eta)(\ell_0,\ell_1)\equiv \frac{\eta(\ell_0+1,\ell_1)-2\eta(\ell_0,\ell_1)+\eta(\ell_0-1,\ell_1)}{a^2},
  \\\nonumber
  (\Delta_1 \eta)(\ell_0,\ell_1)\equiv \frac{\eta(\ell_0,\ell_1+1)-2\eta(\ell_0,\ell_1)+\eta(\ell_0,\ell_1-1)}{a^2},
  \label{eq:app-mix-lat-deltas-ellj}
\end{align}
and the lattice Klein--Gordon operator acting on fluctuations,
\begin{equation}
  (\mathcal K \eta)(\ell_0,\ell_1)\equiv (\Delta_0\eta)(\ell_0,\ell_1)-(\Delta_1\eta)(\ell_0,\ell_1)+m^2\,\eta(\ell_0,\ell_1).
  \label{eq:app-mix-lat-K-ellj}
\end{equation}
Because the domain and boundary conditions separate, the eigenmodes of $\mathcal K$ factorize into a temporal mixed (Dirichlet/Neumann) basis times a spatial Neumann basis.  Accordingly, the correct lattice analogue of a Fourier transform is a finite sine--cosine transform. Concretely, the spatial Neumann sector is labeled by $n_1=0,1,\dots,N_1-1$ with angles and momenta
\begin{equation}
  \varphi_{n_1}\equiv \frac{\pi n_1}{N_1-1},
  \qquad
  k_1^{(n_1)}\equiv \frac{\varphi_{n_1}}{a},
  \qquad
  w_{n_1}(\ell_1)\equiv \mathcal N_1^{(n_1)}\cos\!\big(k_1^{(n_1)}x_1^{(\ell_1)}\big)
  =\mathcal N_1^{(n_1)}\cos(\varphi_{n_1} \ell_1),
  \label{eq:app-mix-lat-space-basis-ellj}
\end{equation}
while the temporal mixed Dirichlet/Neumann sector is labeled by $n_0=0,1,\dots,N_0-2$ with angles and momenta
\begin{equation}
  \theta_{n_0}\equiv \frac{(2n_0+1)\pi}{2N_0-1},
  \qquad
  k_0^{(n_0)}\equiv \frac{\theta_{n_0}}{a},
  \qquad
  s_{n_0}(\ell_0)\equiv \mathcal N_0^{(n_0)}\sin\!\big(k_0^{(n_0)}x_0^{(\ell_0)}\big)
  =\mathcal N_0^{(n_0)}\sin(\theta_{n_0}\ell_0),
  \label{eq:app-mix-lat-time-basis-ellj}
\end{equation}
so that $s_{n_0}(0)=0$ and the Neumann ghost condition at the final slice holds, $s_{n_0}(N_0)=s_{n_0}(N_0-1)$.  The normalization factors $\mathcal N^{(n_0)}_0$ and $\mathcal N^{(n_1)}_1$ are chosen so that
\begin{equation}
  \sum_{\ell_0=0}^{N_0-1} s_{n_0}(\ell_0)\,s_{n_0'}(\ell_0)=\delta_{n_0 n_0'},
  \qquad
  \sum_{\ell_1=0}^{N_1-1} w_{n_1}(\ell_1)\,w_{n_1'}(\ell_1)=\delta_{n_1 n_1'}.
  \label{eq:app-mix-lat-orth-ellj}
\end{equation}
The 2D orthonormal eigenbasis is then $u_{n_0,n_1}(\ell_0,\ell_1)\equiv s_{n_0}(\ell_0)\,w_{n_1}(\ell_1)$, and the homogeneous fluctuation admits the exact finite mode expansion 
\begin{align}
  \eta_R(\ell_0,\ell_1;\tau)&=\sum_{n_0=0}^{N_0-2}\sum_{n_1=0}^{N_1-1}
  s_{n_0}(\ell_0)\,w_{n_1}(\ell_1)\,\widehat{\eta}_R(n_0,n_1;\tau),
  \\\nonumber
  \eta_I(\ell_0,\ell_1;\tau)&=\sum_{n_0=0}^{N_0-2}\sum_{n_1=0}^{N_1-1}
  s_{n_0}(\ell_0)\,w_{n_1}(\ell_1)\,\widehat{\eta}_I(n_0,n_1;\tau),
  \label{eq:app-mix-lat-forward-ellj}
\end{align}
with inverse transform
\begin{equation}
  \widehat{\eta}_{R,I}(n_0,n_1;\tau)=
  \sum_{\ell_0=0}^{N_0-1}\sum_{\ell_1=0}^{N_1-1} s_{n_0}(\ell_0)\,w_{n_1}(\ell_1)\,\eta_{R,I}(\ell_0,\ell_1;\tau),
  \label{eq:app-mix-lat-inverse-ellj}
\end{equation}
where the orthonormality \eqref{eq:app-mix-lat-orth-ellj} ensures exact inversion.  Acting with $\mathcal K$ on a basis mode yields a diagonal eigenvalue,
\begin{equation}
  \big(\mathcal K u_{n_0,n_1}\big)(\ell_0,\ell_1)=\omega(n_0,n_1)\,u_{n_0,n_1}(\ell_0,\ell_1),
  \qquad
  \omega(n_0,n_1)=-\widehat{k}_0^{\,2}(n_0)+\widehat{k}_1^{\,2}(n_1)+m^2,
  \label{eq:app-mix-lat-eigs-ellj}
\end{equation}
with one-dimensional eigenvalues
\begin{align}
  \widehat{k}_0^{\,2}(n_0)&=-\frac{2}{a^2}\big(\cos\theta_{n_0}-1\big)
  =\frac{4}{a^2}\sin^2\!\Big(\frac{\theta_{n_0}}{2}\Big)\\\nonumber
  \widehat{k}_1^{\,2}(n_1)&=-\frac{2}{a^2}\big(\cos\varphi_{n_1}-1\big)
  =\frac{4}{a^2}\sin^2\!\Big(\frac{\varphi_{n_1}}{2}\Big),
  \label{eq:app-mix-lat-1d-eigs-ellj}
\end{align}
where, in direct analogy with the periodic lattice dispersion, we introduced the standard lattice momenta
\begin{equation}
  \widehat{k}_0(n_0)\equiv \frac{2}{a}\sin\!\Big(\frac{\theta_{n_0}}{2}\Big),
  \qquad
  \widehat{k}_1(n_1)\equiv \frac{2}{a}\sin\!\Big(\frac{\varphi_{n_1}}{2}\Big).
  \label{eq:app-mix-lat-khat-ellj}
\end{equation}
Thus
\begin{equation}
  \lambda(n_0,n_1)= -\widehat{k}_0^{\,2}(n_0)+\widehat{k}_1^{\,2}(n_1)+m^2.
\end{equation}
To match the convention used in the main text, we identify
\begin{equation}
  \omega^2(n_0,n_1)\equiv \lambda(n_0,n_1)
  =
  -\widehat{k}_0^{\,2}(n_0)+\widehat{k}_1^{\,2}(n_1)+m^2.
  \label{eq:app-mix-lat-omega-ellj}
\end{equation}
Projecting the complexified constrained symplectic flow onto the orthonormal eigenbasis \eqref{eq:app-mix-lat-inverse-ellj} yields completely decoupled intrinsic-time evolution for each mode $(n_0,n_1)$: each complex amplitude $\widehat{\eta}(n_0,n_1;\tau)$ evolves with its own parameter $\omega^2(n_0,n_1)$, and any residual mismatch from a non-on-shell choice of $\varphi_{\rm cl}$ appears only as a constant driving term given by the mode projection of $J_{\rm cl}\equiv\mathcal K\varphi_{\rm cl}$.  Finally, stability is enforced mode-by-mode by restricting each mode to the stable linear subspace determined by the sign of $\omega^2(n_0,n_1)$: writing $\widehat{\eta}(n_0,n_1;\tau)=\widehat{\eta}_R(n_0,n_1;\tau)+i\,\widehat{\eta}_I(n_0,n_1;\tau)$ and similarly for the conjugate momentum mode $\widehat{\pi}(n_0,n_1;\tau)$, one imposes
\begin{align}
  \widehat{\eta}_I(n_0,n_1;\tau)&=-\sgn\!\big[\omega^2(n_0,n_1)\big]\;\widehat{\eta}_R(n_0,n_1;\tau),\\
  \widehat{\pi}_I(n_0,n_1;\tau)&=-\sgn\!\big[\omega^2(n_0,n_1)\big]\;\widehat{\pi}_R(n_0,n_1;\tau),
  \label{eq:app-mix-lat-stable-ellj}
\end{align}
i.e.\ a $\pm\pi/4$ rotation in the complex plane for each lattice mode.
%
%
%
%
\section{Light-cone Bessel profile from a localized initial pulse}
\label{app:lightcone-bessel-pulse}

This appendix provides the continuum prediction used in the main text to interpret the ``all-modes'' (localized) preparation.  Fixing a sharply localized profile on the initial time-slice excites the full tower of spatial modes.  In a relativistic free theory this initial disturbance propagates causally, and the resulting space--time signal depends only on the Minkowski invariant separation from the preparation point.  In $1+1$ dimensions the causal kernel can be written in closed form in terms of Bessel functions; this is the origin of the Bessel light-cone profile used for invariant collapse and fitting.\par
We consider the free real scalar field in $1+1$ dimensions in infinite volume, with Minkowskian action
\begin{equation}
  S[\varphi]
  =
  \frac12\int_{\mathbb R^{1,1}}\!d^2x\;
  \Big[(\partial_{x_0}\varphi)^2-(\partial_{x_1}\varphi)^2-m^2\varphi^2\Big].
  \label{eq:app-pulse-action}
\end{equation}
The Klein--Gordon equation
\begin{equation}
  \big(\partial_{x_0}^2-\partial_{x_1}^2+m^2\big)\varphi(x_0,x_1)=0,
  \qquad (x_0>0),
  \label{eq:app-pulse-KG}
\end{equation}
and an initial-value problem is specified by prescribing the field and its canonical momentum on the initial time-slice.  In the setup relevant for the main text we fix an initial \emph{displacement} profile and ``release from rest'',
\begin{equation}
  \varphi(0,x_1)=f(x_1),
  \qquad
  \partial_{x_0}\varphi(0,x_1)=0,
  \label{eq:app-pulse-IC}
\end{equation}
so that the ensuing dynamics is entirely generated by the initial displacement. The solution may be written in a manifestly causal form by introducing the retarded Green kernel $G_R$, defined as the unique distribution solving
\begin{equation}
  \big(\partial_{x_0}^2-\partial_{x_1}^2+m^2\big)\,G_R(x_0,x_1)
  =
  \delta(x_0)\,\delta(x_1),
  \qquad
  G_R(x_0,x_1)=0\ \ (x_0<0).
  \label{eq:app-pulse-GR-def}
\end{equation}
The condition $G_R=0$ for $x_0<0$ enforces causality: disturbances propagate only forward in time.  For generic initial data $(\varphi,\partial_{x_0}\varphi)$, the causal representation reads
\begin{equation}
  \varphi(x_0,x_1)
  =
  \int_{-\infty}^{\infty}\!dy_1\;
  G_R(x_0,x_1-y_1)\,\partial_{x_0}\varphi(0,y_1)
  \;+\;
  \int_{-\infty}^{\infty}\!dy_1\;
  \partial_{x_0}G_R(x_0,x_1-y_1)\,\varphi(0,y_1),
  \qquad x_0>0,
  \label{eq:app-pulse-general-solution}
\end{equation}
which separates the response to an initial \emph{velocity} profile $\partial_{x_0}\varphi(0,\cdot)$ from the response to an initial \emph{displacement} profile $\varphi(0,\cdot)$.  In the ``released from rest'' case \eqref{eq:app-pulse-IC} the first term vanishes and one obtains
\begin{equation}
  \varphi(x_0,x_1)
  =
  \int_{-\infty}^{\infty}\!dy_1\;
  \partial_{x_0}G_R(x_0,x_1-y_1)\,f(y_1),
  \qquad x_0>0.
  \label{eq:app-pulse-duhamel}
\end{equation}
Equation \eqref{eq:app-pulse-duhamel} makes explicit what is being tested in the main text: once $G_R$ is known, the space--time profile produced by any initial preparation $f$ is fixed, and in particular it must vanish outside the causal future of the support of $f$.\par
For the \emph{localized pulse} centered at $x_1^\star$ we take
\begin{equation}
  f(x_1)=A\,\delta(x_1-x_1^\star),
  \qquad
  \partial_{x_0}\varphi(0,x_1)=0,
  \label{eq:app-pulse-deltaIC}
\end{equation}
with real amplitude $A$.  Inserting \eqref{eq:app-pulse-deltaIC} into \eqref{eq:app-pulse-duhamel} yields the particularly simple form
\begin{equation}
  \varphi(x_0,x_1)
  =
  A\,\partial_{x_0}G_R\!\big(x_0,x_1-x_1^\star\big),
  \qquad x_0>0,
  \label{eq:app-pulse-solution-from-GR}
\end{equation}
so the full prediction reduces to an explicit expression for the retarded kernel.  It is convenient to introduce the invariant separations between the observation point $x=(x_0,x_1)$ and the preparation point $(0,x_1^\star)$,
\begin{equation}
  t\equiv x_0,
  \qquad
  r\equiv |x_1-x_1^\star|,
  \qquad
  s(t,r)\equiv \sqrt{t^2-r^2}.
  \label{eq:app-pulse-separations}
\end{equation}
In $1+1$ dimensions the retarded Green kernel admits the explicit light-cone Bessel form
\begin{equation}
  G_R(t,r)
  =
  \frac12\,\Theta(t)\,\Theta(t^2-r^2)\,
  J_0\!\big(m\,s(t,r)\big),
  \qquad
  s(t,r)=\sqrt{t^2-r^2}.
  \label{eq:app-pulse-GR-bessel}
\end{equation}
The double step function in \eqref{eq:app-pulse-GR-bessel} encodes causal support: for space-like separations ($t^2<r^2$) one has $G_R(t,r)=0$, i.e.\ no signal can be observed outside the light cone, while for time-like separations ($t^2>r^2$) the kernel is an oscillatory function of the invariant distance $s=\sqrt{t^2-r^2}$.  The special function appearing in \eqref{eq:app-pulse-GR-bessel} is the Bessel function of the first kind, order zero, which we define explicitly as
\begin{equation}
  J_0(z)
  \equiv
  \sum_{p=0}^{\infty}\frac{(-1)^p}{(p!)^2}\Big(\frac{z}{2}\Big)^{2p}
  \;=\;
  \frac{1}{\pi}\int_0^\pi\!d\theta\;\cos\!\big(z\sin\theta\big).
  \label{eq:app-pulse-J0-def}
\end{equation}
Combining \eqref{eq:app-pulse-solution-from-GR} and \eqref{eq:app-pulse-GR-bessel} gives the explicit causal prediction for the field generated by the localized displacement \eqref{eq:app-pulse-deltaIC},
\begin{equation}
  \varphi(t,x_1)
  =
  \frac{A}{2}\,\Theta(t)\,
  \partial_t\!\left[
    \Theta(t^2-r^2)\,J_0\!\big(m\sqrt{t^2-r^2}\big)
  \right],
  \qquad
  r=|x_1-x_1^\star|.
  \label{eq:app-pulse-field-explicit}
\end{equation}
Equation \eqref{eq:app-pulse-field-explicit} makes two physically distinct contributions manifest.  For points strictly inside the cone ($t>r>0$) the derivative acts only on the smooth Bessel factor; using $J_0'(z)=-J_1(z)$ one obtains the bulk profile
\begin{equation}
  \varphi(t,x_1)
  =
  -\frac{A m t}{2\sqrt{t^2-r^2}}\,
  J_1\!\big(m\sqrt{t^2-r^2}\big),
  \qquad (t>r),
  \label{eq:app-pulse-field-inside}
\end{equation}
where $J_1$ is the Bessel function of the first kind, order one.  On the light cone $t=r$ the derivative of $\Theta(t^2-r^2)$ generates a distributional wave-front term supported on the cone.  In the main text this sharp front is effectively regularized because the lattice preparation \eqref{eq:mixed-lat-f-delta} is of finite resolution and the measured correlator is binned/averaged.\par
The connection to the mixed correlator measured in the initial-value setup is as follows. In the localized preparation one fixes the initial slice at the insertion point, so that $\varphi(0,x_1^\star)=A$ is not fluctuating.  For $t>0$ this implies the identity
\begin{equation}
  \big\langle \varphi(t,x_1)\,\varphi(0,x_1^\star)\big\rangle_f
  =
  A\,\big\langle \varphi(t,x_1)\big\rangle_f,
  \label{eq:app-pulse-corr-identity}
\end{equation}
and in the free theory $\langle\varphi(t,x_1)\rangle_f$ is precisely the classical solution $\varphi(t,x_1)$ determined by the initial data \eqref{eq:app-pulse-deltaIC}.  Therefore the continuum prediction for the measured correlator is directly given by \eqref{eq:app-pulse-field-explicit}, or, for points strictly inside the cone, by \eqref{eq:app-pulse-field-inside}.  In particular, for space-like separations ($t<r$) one obtains the strict causal prediction
\begin{equation}
  \big\langle \varphi(t,x_1)\,\varphi(0,x_1^\star)\big\rangle_f
  =
  0,
  \qquad (t<r),
  \label{eq:app-pulse-corr-spacelike}
\end{equation}
while for time-like separations ($t>r$) the correlator is governed by the universal Bessel light-cone profile in \eqref{eq:app-pulse-field-inside}, i.e.\ it depends on $(t,x_1)$ only through the invariant combination $\sqrt{t^2-r^2}$ and oscillates with scale set by the mass parameter $m$.\par
We now give the lattice analogue of the continuum construction above.  At finite lattice spacing and finite volume the causal response to a localized initial displacement is still determined by a retarded Green kernel, but no closed Bessel form exists in general; the exact answer is instead a discrete normal--mode sum. We discretize the spatial interval $[0,L]$ with spacing $a$ and $N_1$ sites $\ell_1=0,\dots,N_1-1$ (so $L\equiv (N_1-1)a$ for Neumann endpoints), and we consider the free lattice Hamiltonian evolution in coordinate time $t$ for the real field $\varphi(t,\ell_1)$. The discrete Neumann Laplacian $\Delta_1$ acts as
\begin{equation}
  (\Delta_1 \varphi)(\ell_1)\equiv \frac{\varphi(\ell_1+1)-2\varphi(\ell_1)+\varphi(\ell_1-1)}{a^2},
  \label{eq:app-pulse-lat-Delta1}
\end{equation}
with Neumann reflection implemented at the boundaries by ghost values $\varphi(-1)=\varphi(0)$ and $\varphi(N_1)=\varphi(N_1-1)$.  The lattice Klein--Gordon equation for $t>0$ then reads
\begin{equation}
  \partial_t^2 \varphi(t,\ell_1)\;-\;(\Delta_1\varphi)(t,\ell_1)\;+\;m^2\,\varphi(t,\ell_1)=0,
  \qquad (t>0),
  \label{eq:app-pulse-lat-KG}
\end{equation}
and we impose the initial-value preparation 
\begin{equation}
  \varphi(0,\ell_1)=f_{\ell_1},
  \qquad
  \partial_t\varphi(0,\ell_1)=0.
  \label{eq:app-pulse-lat-IC}
\end{equation}
The lattice retarded Green kernel $G_R^{\rm lat}$ is defined as the causal fundamental solution of the lattice operator,
\begin{equation}
  \big(\partial_t^2-\Delta_1+m^2\big)\,G_R^{\rm lat}(t;\ell_1,\ell_1')
  =
  \delta(t)\,\frac{\delta_{\ell_1,\ell_1'}}{a},
  \qquad
  G_R^{\rm lat}(t;\ell_1,\ell_1')=0\ \ (t<0),
  \label{eq:app-pulse-lat-GR-def}
\end{equation}
where the factor $1/a$ matches the continuum normalization $\delta(x_1-y_1)$ in the limit $a\to0$ (any consistent convention may be used provided it is kept fixed throughout).\par
In terms of $G_R^{\rm lat}$ the solution of \eqref{eq:app-pulse-lat-KG} with initial data \eqref{eq:app-pulse-lat-IC} is the exact lattice analogue of \eqref{eq:app-pulse-duhamel},
\begin{equation}
  \varphi(t,\ell_1)
  =
  \sum_{\ell_1'=0}^{N_1-1} a\;
  \partial_t G_R^{\rm lat}(t;\ell_1,\ell_1')\,f_{\ell_1'},
  \qquad (t>0).
  \label{eq:app-pulse-lat-solution}
\end{equation}
For a Kronecker-localized displacement at $\ell_{1,\star}$,
\begin{equation}
  f_{\ell_1}=A\,\delta_{\ell_1,\ell_{1,\star}},
  \label{eq:app-pulse-lat-deltaIC}
\end{equation}
this reduces to
\begin{equation}
  \varphi(t,\ell_1)=A\,\partial_t G_R^{\rm lat}(t;\ell_1,\ell_{1,\star}),
  \qquad (t>0),
  \label{eq:app-pulse-lat-solution-delta}
\end{equation}
in direct analogy with \eqref{eq:app-pulse-solution-from-GR}.\par
To write $G_R^{\rm lat}$ explicitly one diagonalizes the spatial operator $(-\Delta_1)$ in the Neumann eigenbasis.  Let $u^{(n_1)}_{\ell_1}$ be the orthonormal eigenvectors of the discrete Neumann Laplacian,
\begin{equation}
  \sum_{\ell_1'=0}^{N_1-1} a\;
  u^{(n_1)}_{\ell_1}\,u^{(n_1')}_{\ell_1}
  =\delta_{n_1 n_1'},
  \qquad
  -(\Delta_1 u^{(n_1)})(\ell_1)=\widehat{k}_1^{\,2}(n_1)\,u^{(n_1)}_{\ell_1},
  \label{eq:app-pulse-lat-eig}
\end{equation}
with $n_1=0,1,\dots,N_1-1$. Concretely, for Neumann boundaries one may take the standard discrete cosine basis
\begin{equation}
  u^{(0)}_{\ell_1}=\frac{1}{\sqrt{L}}\;,
  \qquad
  u^{(n_1)}_{\ell_1}=\sqrt{\frac{2}{L}}\cos\!\Big(\varphi_{n_1}\,\ell_1\Big),
  \qquad
  \varphi_{n_1}\equiv \frac{\pi n_1}{N_1-1}\quad(n_1\ge 1),
  \label{eq:app-pulse-lat-cos-basis}
\end{equation}
which yields the usual lattice momenta
\begin{equation}
  \widehat{k}_1(n_1)\equiv \frac{2}{a}\sin\!\Big(\frac{\varphi_{n_1}}{2}\Big),
  \qquad
  \Omega_{n_1}\equiv \sqrt{m^2+\widehat{k}_1^{\,2}(n_1)}.
  \label{eq:app-pulse-lat-dispersion}
\end{equation}
Expanding $G_R^{\rm lat}$ in this basis and solving the resulting decoupled oscillator equations in time gives the exact mode-sum representation
\begin{equation}
  G_R^{\rm lat}(t;\ell_1,\ell_1')
  =
  \Theta(t)\,
  \sum_{n_1=0}^{N_1-1}
  u^{(n_1)}_{\ell_1}\,u^{(n_1)}_{\ell_1'}\,
  \frac{\sin(\Omega_{n_1} t)}{\Omega_{n_1}},
  \label{eq:app-pulse-lat-GR-modesum}
\end{equation}
and therefore, for a localized displacement,
\begin{equation}
  \varphi(t,\ell_1)
  =
  A\,\Theta(t)\,
  \sum_{n_1=0}^{N_1-1}
  u^{(n_1)}_{\ell_1}\,u^{(n_1)}_{\ell_{1,\star}}\,
  \cos(\Omega_{n_1} t),
  \label{eq:app-pulse-lat-field-modesum}
\end{equation}
since $\partial_t\sin(\Omega t)=\Omega\cos(\Omega t)$. Equations \eqref{eq:app-pulse-lat-GR-modesum} and \eqref{eq:app-pulse-lat-field-modesum} are the precise lattice analogues of the continuum Bessel expressions: at finite $a$ they define the full causal profile, while in the joint limit $a\to0$ and $L\to\infty$ the mode sum approaches the continuum Fourier integral and reproduces the Bessel light-cone form of $G_R$ in \eqref{eq:app-pulse-GR-bessel}).\par
Finally, in the mixed initial-value measurement of the main text, the correlator sourced at $\ell_{1,\star}$ satisfies the same factorization as in the continuum: since the preparation fixes $\varphi(0,\ell_{1,\star})=A$, for $t>0$ one has
\begin{equation}
  \Big\langle \varphi(t,\ell_1)\,\varphi(0,\ell_{1,\star})\Big\rangle_f
  =
  A\,\Big\langle \varphi(t,\ell_1)\Big\rangle_f,
  \label{eq:app-pulse-lat-corr-identity}
\end{equation}
and in the free theory $\langle\varphi(t,\ell_1)\rangle_f$ is given exactly by the mode sum \eqref{eq:app-pulse-lat-field-modesum}.  
\section{Fixed temporal boundaries with exact terminal slice: implementation for light-cone measurements}
\label{app:fixedfixed-terminal-implementation}

For the light-cone measurements on a finite Minkowskian volume
$\mathcal V=[0,T]\times[0,L]$, temporal boundary reflections must be
controlled explicitly. We therefore use a fixed--fixed temporal
implementation: the initial slice contains the localized preparation,
while the final slice is fixed to the exact free evolution of the same
initial profile. This choice suppresses spurious reflections from the
temporal edges and yields a clean causal signal in the bulk. The setup is
used here as a proof-of-concept implementation: the bulk degrees of
freedom evolve dynamically according to the constrained symplectic flow,
whereas the terminal slice is imposed analytically. In this appendix we
spell out the corresponding continuum construction and its lattice
implementation, including the background/fluctuation split used in the
simulations. References to the mixed-boundary implementation are to App.~\ref{app:mixed-lattice-implementation}, while the continuum/light-cone prediction for a localized pulse is reviewed in App.~\ref{app:lightcone-bessel-pulse}.\par
We consider the free real scalar field in $1+1$ dimensions on $\mathcal V=[0,T]\times[0,L]$ with Minkowskian action
\begin{equation}
  S[\varphi]
  =
  \frac12\int_0^T\!dx_0\int_0^L\!dx_1\;
  \Big[(\partial_{x_0}\varphi)^2-(\partial_{x_1}\varphi)^2-m^2\varphi^2\Big].
  \label{eq:app-ff-action}
\end{equation}
We impose Neumann spatial boundaries and Dirichlet temporal boundaries,
\begin{equation}
  \varphi(0,x_1)=f(x_1),\qquad
  \varphi(T,x_1)=g(x_1),\qquad
  \partial_{x_1}\varphi(x_0,0)=\partial_{x_1}\varphi(x_0,L)=0.
  \label{eq:app-ff-bc}
\end{equation}
The initial profile $f$ is the chosen preparation, e.g.\ the localized displacement $f(x_1)=A\,\delta(x_1-x_1^\star)$ as in App.~\ref{app:lightcone-bessel-pulse}. The final profile $g$ is fixed \emph{not} arbitrarily, but to the value obtained by the exact free evolution from $f$ released from rest,
\begin{equation}
  \partial_{x_0}\varphi(0,x_1)=0,
  \label{eq:app-ff-release}
\end{equation}
so that the bulk signal coincides with the causal prediction derived in App.~\ref{app:lightcone-bessel-pulse}.  Expanding $f$ in the spatial Neumann basis $v^{(n_1)}(x_1)$ on $[0,L]$,
\begin{equation}
  f(x_1)=\sum_{n_1=0}^{\infty} f^{(n_1)}\,v^{(n_1)}(x_1),
  \qquad
  f^{(n_1)}\equiv \int_{0}^{L}\!dx_1\,v^{(n_1)}(x_1)\,f(x_1),
  \label{eq:app-ff-f-expand}
\end{equation}
the bulk Klein-Gordon equation $(\partial_{x_0}^2-\partial_{x_1}^2+m^2)\varphi=0$ reduces mode-by-mode to
\begin{equation}
  \partial_{x_0}^2\varphi^{(n_1)}(x_0)+\Omega_{n_1}^2\,\varphi^{(n_1)}(x_0)=0,
  \qquad
  \Omega_{n_1}\equiv\sqrt{m^2+\big(k^{(n_1)}_1\big)^2}.
  \label{eq:app-ff-ode}
\end{equation}
With initial conditions $\varphi^{(n_1)}(0)=f^{(n_1)}$ and $\partial_{x_0}\varphi^{(n_1)}(0)=0$ one obtains the unique free evolution
\begin{equation}
  \varphi_{\mathrm{free}}(x_0,x_1)
  =
  \sum_{n_1=0}^{\infty} f^{(n_1)}\,v^{(n_1)}(x_1)\,\cos(\Omega_{n_1}x_0),
  \label{eq:app-ff-freeevol-cont}
\end{equation}
and we define the terminal Dirichlet profile to be
\begin{equation}
  g(x_1)\equiv \varphi_{\mathrm{free}}(T,x_1).
  \label{eq:app-ff-g-def}
\end{equation}
This choice is tailored to the proof-of-concept goal: by matching the field at $x_0=T$ to the exact causal free prediction, it suppresses spurious reflections from the final temporal boundary in the light-cone analysis.

As in App.~\ref{app:mixed-lattice-implementation}, it is convenient for numerical purposes to remove the inhomogeneous Dirichlet data by a background-field split,
\begin{equation}
  \varphi(x_0,x_1)=\varphi_{\mathrm{cl}}(x_0,x_1)+\eta(x_0,x_1),
  \label{eq:app-ff-split}
\end{equation}
where $\varphi_{\mathrm{cl}}$ solves the bulk Klein--Gordon equation
\[
(\partial_{x_0}^2-\partial_{x_1}^2+m^2)\varphi_{\mathrm{cl}}=0
\]
together with the same fixed temporal and Neumann spatial boundary conditions as $\varphi$,
\begin{equation}
  \varphi_{\mathrm{cl}}(0,x_1)=f(x_1),\qquad
  \varphi_{\mathrm{cl}}(T,x_1)=g(x_1),\qquad
  \partial_{x_1}\varphi_{\mathrm{cl}}(x_0,0)=\partial_{x_1}\varphi_{\mathrm{cl}}(x_0,L)=0,
\label{eq:app-ff-phicl-bc}
\end{equation}
so that the fluctuation satisfies homogeneous fixed boundaries,
\begin{equation}
  \eta(0,x_1)=0,\qquad
  \eta(T,x_1)=0,\qquad
  \partial_{x_1}\eta(x_0,0)=\partial_{x_1}\eta(x_0,L)=0.
  \label{eq:app-ff-eta-bc}
\end{equation}
With these choices the action splits as
\begin{equation}
  S[\varphi_{\mathrm{cl}}+\eta]=S[\varphi_{\mathrm{cl}}]+S[\eta],
  \label{eq:app-ff-action-split}
\end{equation}
since the cross term reduces to boundary contributions that vanish by \eqref{eq:app-ff-phicl-bc} and \eqref{eq:app-ff-eta-bc}.  In the light-cone setup we take $\varphi_{\mathrm{cl}}$ to be precisely the free evolution \eqref{eq:app-ff-freeevol-cont}, so that $g$ is fixed by \eqref{eq:app-ff-g-def} and the dynamical variables evolved by the constrained symplectic flow are the homogeneous quantum fluctuations $\eta$.\par
We discretize $\mathcal V=[0,T]\times[0,L]$ on a $N_0\times N_1$ lattice with spacing $a$,
\begin{equation}
  x_0^{(\ell_0)}\equiv \ell_0 a,\qquad \ell_0=0,\dots,N_0-1,\qquad
  x_1^{(\ell_1)}\equiv \ell_1 a,\qquad \ell_1=0,\dots,N_1-1,
\end{equation}
and denote the complexified lattice field by $\varphi(\ell_0,\ell_1;\tau)$, $\varphi=\varphi_R+i\varphi_I$.  Neumann spatial boundaries are implemented by reflection through ghost values,
\begin{equation}
  \varphi(\ell_0,-1;\tau)=\varphi(\ell_0,0;\tau),\qquad
  \varphi(\ell_0,N_1;\tau)=\varphi(\ell_0,N_1-1;\tau),
  \label{eq:app-ff-lat-space-ghost}
\end{equation}
while fixed temporal boundaries are imposed directly on the endpoint slices,
\begin{equation}
  \varphi(0,\ell_1;\tau)=f_{\ell_1},\qquad
  \varphi(N_0-1,\ell_1;\tau)=g_{\ell_1}.
  \label{eq:app-ff-lat-time-fixed}
\end{equation}
As in the continuum we shift
\begin{equation}
  \varphi(\ell_0,\ell_1;\tau)=\varphi_{\rm cl}(\ell_0,\ell_1)+\eta(\ell_0,\ell_1;\tau),
  \label{eq:app-ff-lat-split}
\end{equation}
choosing $\varphi_{\rm cl}$ to satisfy the same lattice boundary conditions as $\varphi$ and, for exact decoupling, to solve the discrete bulk equation associated with the free quadratic kernel.  Then the fluctuation obeys homogeneous fixed boundaries in time and Neumann in space,
\begin{align}
  &\eta(0,\ell_1;\tau)=0,\qquad
  \eta(N_0-1,\ell_1;\tau)=0,\\
  &\eta(\ell_0,-1;\tau)=\eta(\ell_0,0;\tau),\qquad
  \eta(\ell_0,N_1;\tau)=\eta(\ell_0,N_1-1;\tau).
  \label{eq:app-ff-lat-eta-bc}
\end{align}
Accordingly, the number of dynamical fluctuation variables is
$M=(N_0-2)N_1$, i.e.\ the interior time-slices $\ell_0=1,\dots,N_0-2$ for each
spatial site.\par
We define the second-difference operators (with boundary-adjacent values reduced using \eqref{eq:app-ff-lat-eta-bc})
\begin{align}
  (\Delta_0 \eta)(\ell_0,\ell_1) &\equiv \frac{\eta(\ell_0+1,\ell_1)-2\eta(\ell_0,\ell_1)+\eta(\ell_0-1,\ell_1)}{a^2},  \\\nonumber
  (\Delta_1 \eta)(\ell_0,\ell_1)
  &\equiv \frac{\eta(\ell_0,\ell_1+1)-2\eta(\ell_0,\ell_1)+\eta(\ell_0,\ell_1-1)}{a^2},
  \label{eq:app-ff-lat-deltas}
\end{align}
and the lattice Klein--Gordon operator acting on fluctuations,
\begin{equation}
  (\mathcal K \eta)(\ell_0,\ell_1)
  \equiv
  (\Delta_0\eta)(\ell_0,\ell_1)-(\Delta_1\eta)(\ell_0,\ell_1)+m^2\,\eta(\ell_0,\ell_1).
  \label{eq:app-ff-lat-K}
\end{equation}
In the fixed--fixed setup the temporal difference $\Delta_0$ is understood with Dirichlet endpoints: for interior sites $\ell_0=1,\dots,N_0-2$ it uses the nearest neighbours at $\ell_0\pm1$, while the boundary values at $\ell_0=0$ and $\ell_0=N_0-1$ are fixed to zero for $\eta$.\par
Because the spatial direction is Neumann and the temporal direction is Dirichlet--Dirichlet for fluctuations, the eigenmodes of $\mathcal K$ factorize into a temporal sine basis times a spatial cosine basis. The spatial Neumann sector is labeled by $n_1=0,1,\dots,N_1-1$ with angles
\begin{equation}
  \varphi_{n_1}\equiv \frac{\pi n_1}{N_1-1},
  \qquad
  w_{n_1}(\ell_1)\equiv \mathcal N_1^{(n_1)}\cos(\varphi_{n_1}\ell_1),
  \label{eq:app-ff-lat-space-basis}
\end{equation}
and lattice momenta
\begin{equation}
  \widehat{k}_1(n_1)\equiv \frac{2}{a}\sin\!\Big(\frac{\varphi_{n_1}}{2}\Big).
  \label{eq:app-ff-lat-k1hat}
\end{equation}
The temporal Dirichlet--Dirichlet sector is naturally labeled by $n_0=0,1,\dots,N_0-3$, with angles
\begin{equation}
  \theta_{n_0}\equiv \frac{(n_0+1)\pi}{N_0-1},
  \qquad
  s_{n_0}(\ell_0)\equiv \mathcal N_0^{(n_0)}\sin(\theta_{n_0}\ell_0),
  \qquad \ell_0=0,\dots,N_0-1,
  \label{eq:app-ff-lat-time-basis}
\end{equation}
so that $s_{n_0}(0)=s_{n_0}(N_0-1)=0$.  The corresponding lattice temporal momenta are
\begin{equation}
  \widehat{k}_0(n_0)\equiv \frac{2}{a}\sin\!\Big(\frac{\theta_{n_0}}{2}\Big)
  =
  \frac{2}{a}\sin\!\Big(\frac{(n_0+1)\pi}{2(N_0-1)}\Big).
  \label{eq:app-ff-lat-k0hat}
\end{equation}
The orthonormal 2D eigenbasis is $u_{n_0,n_1}(\ell_0,\ell_1)\equiv s_{n_0}(\ell_0)\,w_{n_1}(\ell_1)$, and fluctuations admit the finite mode expansion
\begin{equation}
  \eta_{R,I}(\ell_0,\ell_1;\tau)
  =
  \sum_{n_0=0}^{N_0-3}\sum_{n_1=0}^{N_1-1}
  s_{n_0}(\ell_0)\,w_{n_1}(\ell_1)\,\widehat{\eta}_{R,I}(n_0,n_1;\tau),
  \label{eq:app-ff-lat-forward}
\end{equation}
with inverse
\begin{equation}
  \widehat{\eta}_{R,I}(n_0,n_1;\tau)=
  \sum_{\ell_0=0}^{N_0-1}\sum_{\ell_1=0}^{N_1-1}
  s_{n_0}(\ell_0)\,w_{n_1}(\ell_1)\,\eta_{R,I}(\ell_0,\ell_1;\tau).
  \label{eq:app-ff-lat-inverse}
\end{equation}
Acting with $\mathcal K$ on a basis mode yields a diagonal eigenvalue,
\begin{equation}
  (\mathcal K u_{n_0,n_1})(\ell_0,\ell_1)
  =
  \omega^2(n_0,n_1)\,u_{n_0,n_1}(\ell_0,\ell_1),
  \qquad
  \omega^2(n_0,n_1)= -\widehat{k}_0^{\,2}(n_0)+\widehat{k}_1^{\,2}(n_1)+m^2.
  \label{eq:app-ff-lat-eigs}
\end{equation}
so that the sign of $\omega^2$ determines the stable subspace for each mode in the constrained symplectic flow. In the light-cone measurements discussed in the main text, the initial profile is taken to be a Kronecker-localized displacement on the spatial lattice,
\begin{equation}
  f_{\ell_1}=A\,\delta_{\ell_1,\ell_{1,\star}},
  \label{eq:app-ff-lat-deltaIC}
\end{equation}
with ``release from rest'' (zero initial coordinate-time velocity), the lattice analogue of \eqref{eq:app-ff-release}.  The exact free lattice evolution of this preparation is the discrete normal-mode sum (App.~\ref{app:lightcone-bessel-pulse}),
\begin{equation}
  \varphi_{\mathrm{free}}(t,\ell_1)
  =
  A\,\Theta(t)\,
  \sum_{n_1=0}^{N_1-1}
  u^{(n_1)}_{\ell_1}\,u^{(n_1)}_{\ell_{1,\star}}\,
  \cos(\Omega_{n_1} t),
  \qquad
  \Omega_{n_1}\equiv \sqrt{m^2+\widehat{k}_1^{\,2}(n_1)},
  \label{eq:app-ff-lat-field-modesum}
\end{equation}
with $u^{(n_1)}_{\ell_1}$ the discrete Neumann eigenvectors and $\widehat{k}_1(n_1)=\tfrac{2}{a}\sin(\tfrac{\varphi_{n_1}}{2})$ as in \eqref{eq:app-ff-lat-k1hat}.  In our fixed--fixed setup we define the terminal Dirichlet slice by evaluating \eqref{eq:app-ff-lat-field-modesum} at the final coordinate time $t=T$,
\begin{equation}
  g_{\ell_1}\equiv \varphi_{\mathrm{free}}(T,\ell_1),
  \qquad T\equiv (N_0-1)a,
  \label{eq:app-ff-lat-g-def}
\end{equation}
and we choose the background field on the full lattice to be precisely this free solution at each discrete time,
\begin{equation}
  \varphi_{\rm cl}(\ell_0,\ell_1)\equiv \varphi_{\mathrm{free}}(x_0^{(\ell_0)},\ell_1),
  \qquad x_0^{(\ell_0)}=\ell_0 a.
  \label{eq:app-ff-lat-phicl-def}
\end{equation}
With this choice $\varphi_{\rm cl}$ satisfies the fixed temporal slices \eqref{eq:app-ff-lat-time-fixed} by construction and solves the free lattice equation in the bulk. The evolved dynamical variables in the code are the homogeneous fluctuations $\eta$ with boundary conditions \eqref{eq:app-ff-lat-eta-bc}. The full measured field is reconstructed as
\begin{equation}
  \varphi(\ell_0,\ell_1;\tau)=\varphi_{\rm cl}(\ell_0,\ell_1)+\eta(\ell_0,\ell_1;\tau),
  \label{eq:app-ff-lat-reconstruct}
\end{equation}
so that the imposed terminal slice is respected automatically by enforcing $\eta(N_0-1,\ell_1;\tau)=0$ at all intrinsic times $\tau$.

\vskip 0.4cm
\newpage
\bibliographystyle{JHEP}
\bibliography{biblio}
\end{document}